\documentclass[aps,prl,reprint,groupedaddress,floatfix,longbibliography]{revtex4-2}

\usepackage[T1]{fontenc}
\usepackage{newtxtext}
\usepackage{newtxmath}

\usepackage{amsmath}
\usepackage{mathtools}
\usepackage{amsfonts}
\usepackage{bbold}
\usepackage{braket}

\usepackage{color}
\usepackage{xcolor}
\usepackage[colorlinks,citecolor=blue,linkcolor=blue,urlcolor=blue]{hyperref}

\usepackage{graphicx}
\usepackage[all]{hypcap}

\begin{document}

\title{Parisi Energy before Parisi Geometry in the\\Quantum Approximate Optimization Algorithm}

\author{Maxime Dupont}
\email[]{mdupont@rigetti.com}
\affiliation{Rigetti Computing, 775 Heinz Avenue, Berkeley, California 94710, USA}

\author{Bhuvanesh Sundar}
\affiliation{Rigetti Computing, 775 Heinz Avenue, Berkeley, California 94710, USA}

\begin{abstract}
    The fractured low-energy landscape of spin glasses presents a notorious barrier to optimization. While high-depth calculations indicate that the Quantum Approximate Optimization Algorithm (QAOA) can approach the Parisi ground-state energy of the Sherrington--Kirkpatrick spin glass, how it does so remains elusive. Here, we treat the QAOA distribution as a statistical-physics object and probe its geometry directly in the thermodynamic limit. We draw two independent bitstrings from the same problem instance and couple them through their macroscopic similarity. We find that typical QAOA samples remain macroscopically uncorrelated, even as their sensitivity to this coupling grows with circuit depth $p$. By contrast, the low-energy tail shows a much stronger tendency to overlap and reveals a Parisi-like phase with the same critical behavior as the thermal spin-glass transition. The onset of this organization moves toward the typical output as the circuit deepens, consistent with the separation vanishing only as $p\to\infty$. Together, these findings suggest a route distinct from structure-exploiting classical algorithms and thermal sampling, one in which the QAOA approaches the Parisi energy before Parisi geometry becomes typical.
\end{abstract}

\maketitle

The modern understanding of spin glasses emerged from decades of effort culminating in the replica framework and the Parisi solution of the Sherrington--Kirkpatrick (SK) model~\cite{EdwardsAnderson1975,SherringtonKirkpatrick1975,Parisi1979,MezardParisiVirasoro1987,BinderYoung1986,CastellaniCavagna2005,Talagrand2006ParisiFormula,Panchenko2013Book,AuffingerChenZeng2020}. This paradigm demonstrated that spin-glass structure is captured not by a conventional local observable, such as magnetization, but by the global statistics of overlaps between independent replicas. In frustrated Ising models, low-energy configurations cannot satisfy all interactions simultaneously and instead form amorphous states whose mutual organization is encoded by their overlaps. This overlap structure is where much of the geometric and algorithmic intractability of a low-energy spin-glass phase resides~\cite{MezardParisiVirasoro1987,CastellaniCavagna2005,Panchenko2013Ultrametricity,Panchenko2013Book,Gamarnik2021OGP,ElAlaouiMontanariSellke2021}.

This geometry creates a direct challenge for reaching low-energy states. The low-energy landscape is not a smooth basin reached by steady descent, but an archipelago of valleys separated by large barriers and organized by mutual overlaps. Local search can become trapped within one valley, while moving between low-energy regions may require crossing large barriers or navigating a highly nontrivial overlap landscape. Successful algorithms must therefore either exploit detailed spin-glass structure or avoid following a localized path through the fractured landscape.

This obstruction appears directly in thermal classical algorithms. Monte Carlo and annealing-based methods use temperature as a control parameter to search for low-energy configurations~\cite{Metropolis1953,Kirkpatrick1983,MarinariParisi1992,HukushimaNemoto1996,WangMachtaKatzgraber2015a,WangMachtaKatzgraber2015b}. At low temperature, however, the SK model enters a spin-glass phase in which configuration space fractures into many competing low-free-energy regions~\cite{MezardParisiVirasoro1987,BinderYoung1986,CastellaniCavagna2005,Sellke2025SlowMixing}.

A contrasting classical route is provided by algorithms that turn the analytic structure of the SK spin glass into efficient procedures. This structure is described by the cavity method, the Thouless--Anderson--Palmer (TAP) equations, and the Parisi solution, which characterize the glassy organization of the low-energy landscape~\cite{ThoulessAndersonPalmer1977,MezardParisiVirasoro1987,Bolthausen2014}. Approximate message passing (AMP) uses this structure to construct configurations arbitrarily close to the SK optimum in polynomial time~\cite{Montanari2025,ElAlaouiMontanariSellke2021,ElAlaouiMontanari2020}, while recent Hessian-ascent methods provide a complementary route through the TAP landscape~\cite{JekelSandhuShi2025}.

Against this classical backdrop, quantum computing offers a distinct way to search the landscape. Instead of defining an explicit trajectory through configuration space, a quantum circuit prepares a state whose measurement produces a distribution over candidate solutions. Among quantum algorithms, the Quantum Approximate Optimization Algorithm (QAOA)~\cite{Farhi2014,BLEKOS20241} provides a canonical variational ansatz, built from $p$ alternating cost and mixer layers. While the QAOA is widely studied heuristically, the SK model provides a rare setting where its layer-by-layer approach to the optimum can be analyzed directly in the thermodynamic limit. Analytical evaluations have advanced from calculations at depths $p\le 20$~\cite{FarhiGoldstoneGutmannZhou2022,BassoFarhiMarwahaVillalongaZhou2022} to a recent spin-boson mapping reaching $p=160$, providing evidence for an empirical depth scaling $p\sim\varepsilon^{-1.13}$ toward a $(1-\varepsilon)$ approximation to the SK optimum~\cite{BoulebnaneKhanLiuLarsonHermanShaydulinPistoia2026,BoulebnaneKhanSubrahmanyaHermanFarhiVillalongaShaydulin2026,OAI:QAOA}. At the level of energy, these results place the high-depth QAOA alongside classical methods such as AMP and Hessian ascent~\cite{Montanari2025,ElAlaouiMontanariSellke2021,JekelSandhuShi2025}, but along a route that is not explicitly built from the special analytic structure of SK.

A complementary perspective on how the QAOA approaches the optimum comes from its $p\to\infty$ connection to quantum annealing, in which the alternating circuit approaches a continuous interpolation from the transverse-field initial Hamiltonian to the final SK Hamiltonian~\cite{Farhi2014,KadowakiNishimori1998,BoulebnaneSudShaydulinPistoia2025}. In the ideal adiabatic limit, the system follows the instantaneous ground state of the transverse-field SK Hamiltonian, whose low-field phase develops glassy overlap structure as the field is reduced~\cite{KissZarandLovas2024}. This makes the central question geometric: when and how does Parisi geometry emerge in the QAOA output as the circuit depth increases?

\begin{figure}[!t]
    \centering
    \includegraphics[width=1\columnwidth]{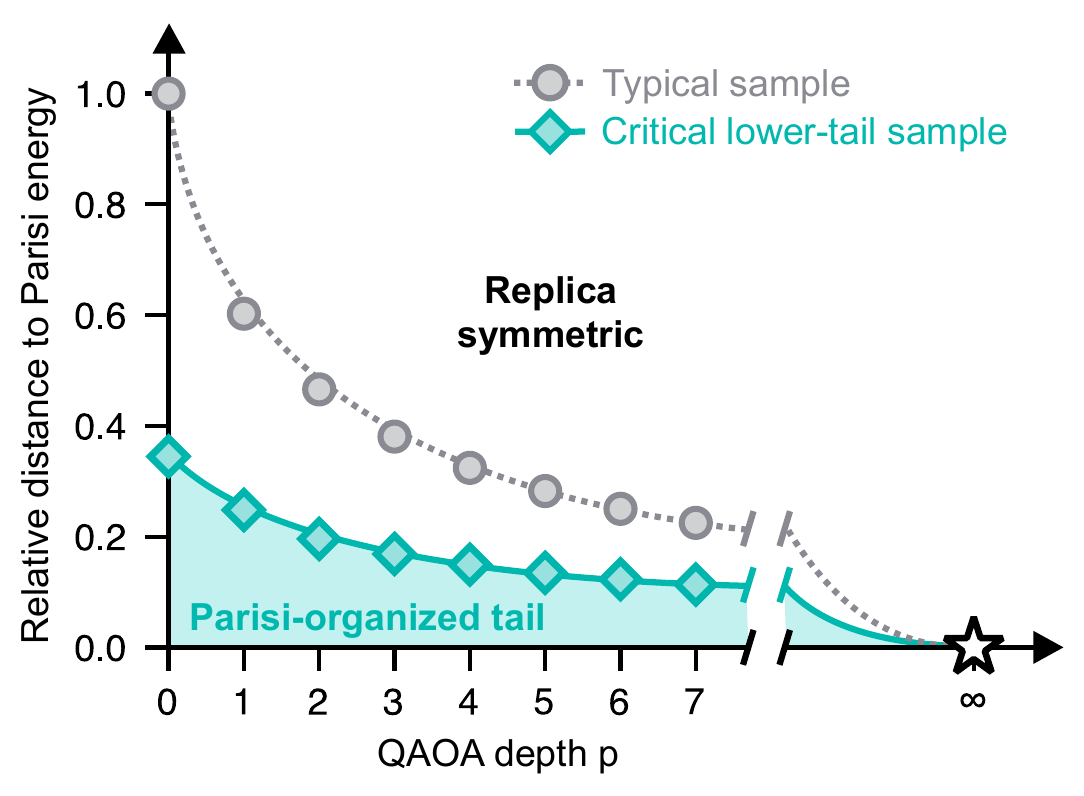}
    \caption{Schematic phase diagram of the QAOA overlap geometry versus circuit depth $p$ and relative distance to the Parisi energy $1-|E/E_{\mathrm{Parisi}}|$. Typical samples remain macroscopically uncorrelated, while Parisi organization appears first in the low-energy tail and is consistent with becoming typical only asymptotically as $p\to\infty$. Thus, the QAOA approaches the Parisi energy before Parisi overlap geometry becomes typical.}
    \label{fig:phase_diagram}
\end{figure}

Here, we address this question with a coupled-replica overlap probe evaluated in the thermodynamic limit, in the spirit of the Franz--Parisi construction~\cite{FranzParisi1995,FranzParisi1997,FranzParisi1998EffectivePotential}. We find that typical QAOA samples remain macroscopically uncorrelated, even as their sensitivity to the coupling grows with depth. The low-energy tail, by contrast, shows a much stronger tendency for independently drawn samples to overlap. At $p=1$, we characterize this low-energy overlap organization and find the same Parisi structure as in the equilibrium SK Gibbs distribution near its spin-glass transition. As $p$ increases, the energetic and statistical separation between typical samples and the onset of this organization narrows, as summarized in Fig.~\ref{fig:phase_diagram}. Finite-size simulations connect the growing response of typical samples to a depth-dependent crossover with system size.

The SK Hamiltonian for $N$ Ising spins $\boldsymbol z=(z_1,\ldots,z_N)\in\{\pm1\}^N$ is~\cite{SherringtonKirkpatrick1975,MezardParisiVirasoro1987,BinderYoung1986}
\begin{equation}
    C_J(\boldsymbol z)=\frac{1}{\sqrt N}\sum\nolimits_{j<k}J_{jk}z_jz_k,
    \label{eq:sk_cost}
\end{equation}
where the couplings $J_{jk}$ are independent standard Gaussian random variables and $C_J$ is minimized. We denote the thermodynamic ground-state energy density by $E_{\mathrm{Parisi}}=\lim_{N\to\infty}N^{-1}\mathbb E_J[\min_{\boldsymbol z}C_J(\boldsymbol z)]$~\cite{Parisi1979,MezardParisiVirasoro1987,Talagrand2006ParisiFormula,Panchenko2013Book} and the average over SK disorder couplings by $\mathbb E_J$. Promoting $C_J$ to the diagonal operator $\hat C_J|\boldsymbol z\rangle=C_J(\boldsymbol z)|\boldsymbol z\rangle$, the depth-$p$ QAOA state is~\cite{Farhi2014,BLEKOS20241}
\begin{equation}
    \bigl|\Psi_{J,p}\bigr\rangle=\prod\nolimits_{\ell=1}^{p}e^{-i\beta_\ell\sum_j\hat X_j}e^{-i\gamma_\ell\hat C_J}|+\rangle^{\otimes N},
    \label{eq:qaoa_state}
\end{equation}
where $|+\rangle^{\otimes N}$ is an equal superposition of all basis states, $\hat{X}_j$ is the Pauli-$X$ operator on qubit $j$, and $\ell$ is increasing from right to left. The angles $(\boldsymbol\gamma,\boldsymbol\beta)\in\mathbb R^{2p}$ are chosen to minimize the thermodynamic disorder-averaged energy density $E_p=\lim_{N\to\infty}\mathbb E_J\langle\hat{C}_J\rangle/N$~\cite{supplemental}. After measurement in the computational basis, the state defines a probability distribution over spin configurations
\begin{equation}
    P_{J,p}(\boldsymbol z) =\left|\left\langle \boldsymbol z \middle|\Psi_{J,p}\right\rangle\right|^2,\qquad\sum\nolimits_{\boldsymbol z}P_{J,p}(\boldsymbol z)=1.
    \label{eq:qaoa_dist}
\end{equation}
We probe the geometry of this distribution by drawing two independent configurations, $\boldsymbol z^A$ and $\boldsymbol z^B$, from the same $P_{J,p}$ and measuring their overlap~\cite{EdwardsAnderson1975,Parisi1979,MezardParisiVirasoro1987,Panchenko2013Book}
\begin{equation}
    q_{AB}=\frac{1}{N}\sum\nolimits_{j=1}^N z^A_j z^B_j.
    \label{eq:overlap}
\end{equation}
Collective structure in $P_{J,p}$ can be probed by coupling two QAOA outputs through a source field $\lambda$ conjugate to their extensive overlap $Nq_{AB}$, with positive $\lambda$ favoring larger overlap~\cite{FranzParisi1995,FranzParisi1997,FranzParisi1998EffectivePotential}. For a disordered system, the relevant thermodynamic quantity is the typical free-entropy density at fixed $\lambda$~\cite{MezardParisiVirasoro1987,CastellaniCavagna2005}
\begin{equation}
    \begin{aligned}
        \phi_p(\lambda)=\lim_{N\to\infty}\frac{1}{N}\mathbb E_J\ln\sum_{\boldsymbol z^A,\boldsymbol z^B}&P_{J,p}\bigl(\boldsymbol z^A\bigr)P_{J,p}\bigl(\boldsymbol z^B\bigr)e^{\lambda N q_{AB}}.
    \end{aligned}
    \label{eq:quenched_source_free_entropy}
\end{equation}
A Legendre transform from $\lambda$ to the overlap $q$ defines a Franz--Parisi-type potential associated with this source-biased construction~\cite{FranzParisi1995,FranzParisi1997,FranzParisi1998EffectivePotential}
\begin{equation}
    V_p(q)=\max\nolimits_\lambda\left[\lambda q-\phi_p(\lambda)\right],
    \label{eq:franz_parisi_potential}
\end{equation}
which quantifies the cost of selecting two QAOA samples with mutual overlap $q$.

At fixed depth, these quantities can be evaluated directly in the thermodynamic limit. Expanding the QAOA amplitudes into forward and backward histories and averaging over the Gaussian disorder yields a finite set of saddle-point equations whose size is independent of $N$. As in previous infinite-size SK QAOA calculations, the remaining history sum grows exponentially with $p$~\cite{FarhiGoldstoneGutmannZhou2022,BassoFarhiMarwahaVillalongaZhou2022,supplemental}.

Operationally, Eq.~\eqref{eq:quenched_source_free_entropy} is computed with the replica trick~\cite{MezardParisiVirasoro1987,CastellaniCavagna2005}, which introduces auxiliary replicas for the disorder average, distinct from the two QAOA samples whose overlap is probed. We use a replica-symmetric ansatz with vanishing overlaps between distinct auxiliary replicas, while allowing the two physical samples to correlate. The resulting equations are solved by continuation in $\lambda$ from the uncoupled distribution at $\lambda=0$, yielding $\phi_p(\lambda)$ and $V_p(q)$~\cite{supplemental}. Within this construction, searches from multiple initial conditions find no lower-potential solution, and the branch continued from $\lambda=0$ remains locally stable against correlations between auxiliary replicas over the depths and source ranges studied~\cite{supplemental}.

\begin{figure}[!t]
    \centering
    \includegraphics[width=1\columnwidth]{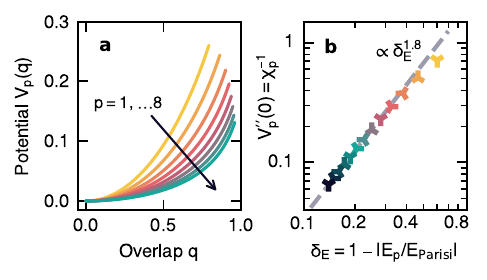}
    \caption{(a) Overlap potential from Eq.~\eqref{eq:franz_parisi_potential} plotted against the overlap $q$ for circuit depths $p=1,2,\dots,8$. (b) Local potential curvature versus the relative distance to the Parisi ground-state energy, $\delta_E=1-|E_p/E_{\mathrm{Parisi}}|$. Data up to $p=13$ are compatible with an empirical power law $\sim \delta_E^{1.8}$ (dashed line).}
    \label{fig:typical_distribution}
\end{figure}

Figure~\ref{fig:typical_distribution}a shows that the overlap potential progressively flattens as $p$ increases, making nonzero overlap less costly while $q=0$ remains the preferred value. Thus, across the depths studied, independently drawn samples remain macroscopically uncorrelated even as they become progressively easier to bias toward mutual overlap.

The response near $\lambda=0$ provides a compact measure of this softening. Global spin-flip symmetry makes the unbiased overlap distribution symmetric and hence $\phi_p(\lambda)$ even. Since $\phi_p(0)=0$, its small-$\lambda$ expansion is $\phi_p(\lambda)=\chi_p\lambda^2/2+O(\lambda^4)$, where $\chi_p=\lim_{N\to\infty}N\mathbb E_J\langle q_{AB}^2\rangle_{J,p}$ is the overlap susceptibility and $V_p''(0)=\chi_p^{-1}$~\cite{supplemental}. The susceptibility can equivalently be written as~\cite{supplemental}
\begin{equation}
    \chi_p=\lim_{N\to\infty}\frac{1}{N}\mathbb E_J\left[\sum\nolimits_{j,k}\left\langle \hat Z_j\hat Z_k\right\rangle_{J,p}^2\right].
    \label{eq:chi_correlation_power}
\end{equation}
The flattening of the potential therefore reflects an increase in the overall strength of the two-spin correlations. At $p=1$, the thermodynamic calculation can be carried out analytically, allowing the small-$\lambda$ response to be evaluated exactly at the optimized SK angles $\gamma_1=1/2$ and $\beta_1=\pi/8$~\cite{FarhiGoldstoneGutmannZhou2022,supplemental}.
\begin{equation}
    1/\chi_1=V_1''(0)=\left(1+\frac{1}{2e}\right)^{-2}\approx 0.7134.
    \label{eq:p1_exact_susceptibility}
\end{equation}
For higher depths, we solve the thermodynamic equations numerically~\cite{supplemental}. Figure~\ref{fig:typical_distribution}b connects the overlap susceptibility to the energetic benchmark by plotting $V_p''(0)=\chi_p^{-1}$ against the relative distance to the Parisi ground-state energy, $\delta_E=1-|E_p/E_{\mathrm{Parisi}}|$. As $E_p$ approaches the Parisi value, the local curvature decreases, with the data compatible with $V_p''(0)\sim \delta_E^{1.8}$ over the studied depths. Already at $p=4$, the QAOA energy density $E_4=-0.51568$~\cite{supplemental} lies below the equilibrium SK transition energy density $-1/2$~\cite{SherringtonKirkpatrick1975,BinderYoung1986,MezardParisiVirasoro1987}. An equilibrium SK Gibbs distribution at the same mean energy would therefore be in the spin-glass phase, with independently drawn thermal samples exhibiting nontrivial overlap structure, whereas QAOA samples remain macroscopically uncorrelated.

\begin{figure}[!t]
    \centering
    \includegraphics[width=1\columnwidth]{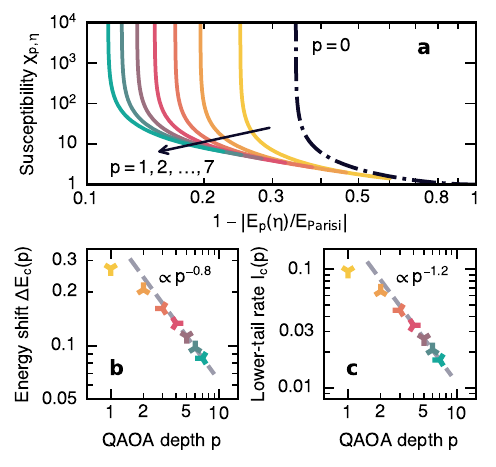}
    \caption{(a) Overlap susceptibility $\chi_{p,\eta}$ versus the relative distance $1-|E_p(\eta)/E_{\mathrm{Parisi}}|$ from the Parisi ground-state energy. Solid curves show QAOA depths $p=1,2,\ldots,7$, and the dash-dotted curve shows the exact $p=0$ Gibbs result. (b) Energy shift $\Delta E_c(p)=E_p(0)-E_p(\eta_c(p))$ from the typical distribution to the lower-tail threshold at $\eta_c(p)$. (c) Lower-tail rate $I_c(p)$ defined in Eq.~\eqref{eq:critical_lower_tail_rate}. Dashed lines show power-law fits to the largest-depth points, giving approximately $\Delta E_c(p)\sim p^{-0.8}$ and $I_c(p)\sim p^{-1.2}$.}
    \label{fig:energy_tilted_response}
\end{figure}

We next examine the overlap properties of the low-energy tail of the QAOA distribution. To probe this tail without changing the circuit, we introduce an energy tilt $\eta$ that reweights sampled bitstrings by their SK energy~\cite{supplemental}
\begin{equation}
    P_{J,p}^{(\eta)}(\boldsymbol z)\propto P_{J,p}(\boldsymbol z)e^{-\eta C_J(\boldsymbol z)},\quad\sum\nolimits_{\boldsymbol z}P_{J,p}^{(\eta)}(\boldsymbol z)=1,
    \label{eq:energy_tilted_distribution}
\end{equation}
where $\eta>0$ favors lower-energy configurations. We denote the mean energy density of this distribution by $E_p(\eta)$ and its overlap susceptibility by $\chi_{p,\eta}$. At $p=0$, the QAOA distribution is uniform, so Eq.~\eqref{eq:energy_tilted_distribution} reduces exactly to the classical SK Gibbs distribution at inverse temperature $\eta$.

Figure~\ref{fig:energy_tilted_response}a shows that $\chi_{p,\eta}$ grows as lower energies are selected and approaches a divergence at a depth-dependent threshold $\eta_c(p)$. At $p=0$, the exact Gibbs result $\chi_{0,\eta}=(1-\eta^2)^{-1}$ has $\eta_c(0)=1$, marking the SK transition from a paramagnet to the Parisi spin-glass phase~\cite{SherringtonKirkpatrick1975,deAlmeidaThouless1978,Parisi1979,MezardParisiVirasoro1987,supplemental}. At $p=1$, we characterize the phase emerging at $\eta_c(1)$ and find the same Parisi organization as in the equilibrium SK Gibbs distribution near this transition, with matching critical exponents but renormalized nonuniversal coefficients~\cite{supplemental}. For $p\geq2$, we anticipate that the analogous instabilities mark the onset of the same Parisi-organized phase.

We quantify in Fig.~\ref{fig:energy_tilted_response}b the energetic separation between the typical QAOA distribution and this lower-tail threshold through $\Delta E_c(p)=E_p(0)-E_p(\eta_c(p))$. In addition, we quantify its sampling cost under the untilted QAOA distribution through the probability of drawing a sample with energy density at or below $E_p(\eta_c(p))$. Within the standard large-deviation description of energy fluctuations~\cite{Touchette2009LargeDeviations,supplemental}, this probability is characterized, for a typical disorder instance, by
\begin{equation}
    \lim_{N\to\infty}\frac{1}{N}\ln\Pr\left[\frac{C_J(\boldsymbol z)}{N}\leq E_p(\eta_c(p))\right]=-I_c(p),
    \label{eq:critical_lower_tail_rate}
\end{equation}
where the probability $\Pr$ is over untilted QAOA samples drawn from Eq.~\eqref{eq:qaoa_dist}. Figure~\ref{fig:energy_tilted_response}c shows that $I_c(p)$ remains positive at every depth studied but decreases steadily with $p$. The tail therefore remains exponentially rare at fixed depth while gaining statistical weight as the circuit deepens. The depth-zero Gibbs benchmark gives $\Delta E_c(0)=1/2$ and $I_c(0)=1/4$ exactly at the SK transition~\cite{supplemental}.

Over the largest accessible depths, the decreases are consistent with effective finite-depth scalings $\Delta E_c(p)\sim p^{-0.8}$ and $I_c(p)\sim p^{-1.2}$. The former must cross over asymptotically, since $\Delta E_c(p)\leq E_p(0)-E_{\mathrm{Parisi}}$ requires decay at least as fast as the known $p^{-0.88}$ energy convergence~\cite{BassoFarhiMarwahaVillalongaZhou2022,BoulebnaneKhanLiuLarsonHermanShaydulinPistoia2026}. The faster decrease of $I_c(p)$ suggests that the threshold becomes less rare more rapidly than it approaches the typical energy.

\begin{figure}[!t]
    \centering
    \includegraphics[width=1\columnwidth]{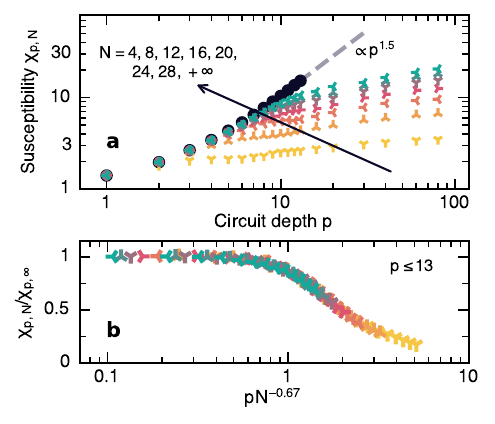}
    \caption{(a) Finite-$N$ state-vector estimates of $\chi_{p,N}$ compared with the thermodynamic value $\chi_{p,\infty}$ as a function of the QAOA depth $p$. Finite-$N$ data are averaged over $10^3$--$10^4$ random problem instances; statistical error bars are smaller than the markers. The dashed line is a $\sim p^{1.5}$ guide to the eye. (b) Rescaled susceptibility $\chi_{p,N}/\chi_{p,\infty}$ as a function of $pN^{-0.67}$ for the available $N\leq 28$ and $p\leq 13$ data.}
    \label{fig:finite_size}
\end{figure}

Finally, to examine how the softening of the typical QAOA distribution emerges in finite systems, we evaluate the finite-$N$ counterpart of Eq.~\eqref{eq:chi_correlation_power}, denoted by $\chi_{p,N}$, using exact QAOA state-vector simulation. Figure~\ref{fig:finite_size}a compares these estimates with the thermodynamic susceptibility $\chi_{p,\infty}\equiv\chi_p$ from Fig.~\ref{fig:typical_distribution}b. Over the largest accessible depths, $\chi_{p,\infty}$ grows approximately as $p^{1.5}$. At small $p$, the finite-size data follow the thermodynamic prediction, while at larger depths they bend toward an $N$-dependent cutoff. This is consistent with the ideal adiabatic limit, in which sampling the ground-state manifold gives $\chi_{p,N}\to N$ following Eq.~\eqref{eq:chi_correlation_power}. The available data are compatible with the empirical finite-size scaling form
\begin{equation}
    \chi_{p,N}/\chi_{p,\infty}=\mathfrak F\bigl(pN^{-0.67}\bigr).
    \label{eq:finite_size_scaling}
\end{equation}
As shown in Fig.~\ref{fig:finite_size}b, $\mathfrak F(x)\to1$ as $x\to0$, while the crossover at $x=O(1)$ corresponds to $N^\star(p)\sim p^{1/0.67}\simeq p^{1.5}$. Since $\chi_{p,N}=N\mathbb E_J\langle q_{AB}^2\rangle_{J,p,N}$, the crossover has a direct geometric meaning. For $N\gg N^\star(p)$, the overlap remains concentrated near zero, whereas for $N\lesssim N^\star(p)$ the response becomes system-spanning and $O(1)$ overlaps can carry substantial, strongly sample-dependent weight~\cite{supplemental,AspelmeierBilloireMarinariMoore2008}. This parallels the finite-size growth of overlap fluctuations across the Gibbs SK transition~\cite{BinderYoung1986,AspelmeierKatzgraberLarsonMooreWittmannYeo2016}. A parameter-free analysis as a function of $\chi_{p,\infty}/N$ gives a comparable collapse, with the largest deviations confined to the smallest sizes~\cite{supplemental}.

To conclude, the QAOA approaches the Parisi energy~\cite{FarhiGoldstoneGutmannZhou2022,BassoFarhiMarwahaVillalongaZhou2022,BoulebnaneKhanLiuLarsonHermanShaydulinPistoia2026} before Parisi geometry becomes typical. Across the depths studied, typical samples remain macroscopically uncorrelated, while signatures of Parisi organization appear first in the low-energy tail. At $p=1$, we characterize its onset as a continuous transition with the same critical behavior as the equilibrium SK transition. As summarized in Fig.~\ref{fig:phase_diagram}, the energetic and statistical gap separating this organization from typical samples narrows with $p$, consistent with Parisi geometry becoming typical only asymptotically as $p\to\infty$.

This trajectory differs from structure-informed classical algorithms, which explicitly exploit spin-glass overlap structure, and from thermal searches, whose low-temperature distributions already carry it~\cite{ElAlaouiMontanariSellke2021,Montanari2025,ElAlaouiMontanari2020,JekelSandhuShi2025,Kirkpatrick1983,MarinariParisi1992,HukushimaNemoto1996}. These results also qualify pseudo-Boltzmann descriptions of QAOA output probabilities~\cite{DiezValle2023PseudoBoltzmann,Lotshaw2023ApproximateBoltzmann,DiezValle2024Interferometry}. Although QAOA output probabilities may appear thermal when organized by energy, the overlap geometry of typical samples need not match that of an equilibrium Gibbs distribution. A recent semiclassical analysis finds that truncated-Wigner dynamics can slightly outperform the QAOA energetically while approaching the Parisi value~\cite{SelsMorone2026}. Determining the full asymptotic overlap geometry of the QAOA, and whether the semiclassical output follows the same energy--geometry trajectory, would offer a sharper test of what is specifically quantum in glassy optimization.

\let\oldaddcontentsline\addcontentsline
\renewcommand{\addcontentsline}[3]{}
\begin{acknowledgments}
    This work was partly supported by the U.S. Department of Energy, Office of Science, National Quantum Information Science Research Centers, Superconducting Quantum Materials and Systems Center (SQMS), under Contract No. 89243024CSC000002. Fermilab is operated by Fermi Forward Discovery Group, LLC under Contract No. 89243024CSC000002 with the U.S. Department of Energy, Office of Science, Office of High Energy Physics. The authors used ChatGPT 5.5 (OpenAI) to assist with selected analytical calculations reported in the SM. The authors declare no competing interests.
\end{acknowledgments}
\let\addcontentsline\oldaddcontentsline

% Remove the bibliography entry from the table of contents
\let\oldaddcontentsline\addcontentsline
\renewcommand{\addcontentsline}[3]{}
\bibliography{bibliography}
\let\addcontentsline\oldaddcontentsline

\onecolumngrid
\setcounter{secnumdepth}{3}
\setcounter{figure}{0}
\setcounter{equation}{0}
\renewcommand\thefigure{S\arabic{figure}}
\renewcommand\theequation{S\arabic{equation}}
\renewcommand\theHfigure{S\arabic{figure}}
\renewcommand\theHequation{S\arabic{equation}}
\newpage\clearpage

\setlength{\belowcaptionskip}{0pt}

\begin{center}
    \large\textbf{Supplemental Material for ``\textit{Parisi Energy before Parisi Geometry\\in the Quantum Approximate Optimization Algorithm}''}
\end{center}

{\hypersetup{linkcolor=black}\tableofcontents}

\section{Definitions}
\label{sec:sm_definitions}

This section fixes the conventions used throughout the Supplemental Material, restating some of the definitions briefly introduced in the main text.

\subsection{The Sherrington--Kirkpatrick spin glass}
\label{subsec:sm_sk}

We consider $N$ Ising spins $\boldsymbol z=(z_1,\ldots,z_N)\in\{\pm1\}^N$ and independent Gaussian couplings $J_{jk}\sim\mathcal N(0,1)$ for $j<k$. Throughout, $\mathbb E_J$ denotes expectation over these Sherrington--Kirkpatrick (SK) couplings. For a fixed disorder realization $J$, the SK Hamiltonian is~\cite{SherringtonKirkpatrick1975,MezardParisiVirasoro1987,BinderYoung1986}
\begin{equation}
    C_J(\boldsymbol z)=\frac{1}{\sqrt N}\sum_{j=1}^N\sum_{k=j+1}^N J_{jk} z_j z_k,
    \label{eq:sm_sk_cost}
\end{equation}
where $C_J$ is the energy to be minimized. With this convention, the ground-state energy density is negative, and the corresponding thermodynamic Parisi value is $E_{\mathrm{Parisi}}=-0.763166726566547\ldots$~\cite{Parisi1979,Schmidt2008}.

\subsection{The Quantum Approximate Optimization Algorithm}
\label{subsec:sm_quantum}

We map one Ising spin to one qubit and label computational-basis states by their Pauli-$Z$ eigenvalues, $\boldsymbol z\in\{\pm1\}^N$. Thus $|z_j=+1\rangle=|0\rangle$, $|z_j=-1\rangle=|1\rangle$, or equivalently $z_j=(-1)^{b_j}$ for a binary bit $b_j\in\{0,1\}$. The quantum cost operator corresponding to the Hamiltonian~\eqref{eq:sm_sk_cost} is diagonal in this basis
\begin{equation}
    \hat C_J |\boldsymbol z\rangle = C_J(\boldsymbol z)|\boldsymbol z\rangle.
    \label{eq:sm_quantum_cost}
\end{equation}
We also define the transverse mixer operator
\begin{equation}
    \hat B = \sum_{j=1}^N \hat X_j,
    \label{eq:sm_mixer}
\end{equation}
where $\hat X_j$ is the Pauli-$X$ operator on qubit $j$. For a depth-$p$ QAOA circuit with angles $\boldsymbol\gamma=(\gamma_1,\ldots,\gamma_p)$ and $\boldsymbol\beta=(\beta_1,\ldots,\beta_p)$, known as the phase separator and mixer angles, respectively, the quantum state is~\cite{Farhi2014,BLEKOS20241}
\begin{equation}
    |\Psi_{J,p}(\boldsymbol\gamma,\boldsymbol\beta)\rangle=e^{-i\beta_p\hat B}e^{-i\gamma_p\hat C_J}\cdots e^{-i\beta_1\hat B}e^{-i\gamma_1\hat C_J}|+\rangle^{\otimes N},
\label{eq:sm_qaoa_state}
\end{equation}
where $|+\rangle=(|0\rangle+|1\rangle)/\sqrt2$. The angles are chosen to minimize the QAOA energy expectation at the corresponding depth $p$
\begin{equation}
    E_p=\lim_{N\to\infty}\frac{1}{N}\mathbb E_J\Bigl[\langle\Psi_{J,p}|\hat C_J|\Psi_{J,p}\rangle\Bigr].
    \label{eq:sm_energy_density}
\end{equation}
Throughout this work, we use the depth-dependent angles from Refs.~\cite{FarhiGoldstoneGutmannZhou2022,BassoFarhiMarwahaVillalongaZhou2022,BoulebnaneKhanLiuLarsonHermanShaydulinPistoia2026}, listed in Table~\ref{tab:sm_qaoa_angles_benchmarks}. Measuring the QAOA state in the computational basis defines the instance-dependent distribution
\begin{equation}
    P_{J,p}(\boldsymbol z)=\left|\langle \boldsymbol z|\Psi_{J,p}(\boldsymbol\gamma,\boldsymbol\beta)\rangle\right|^2,\qquad\sum_{\boldsymbol z} P_{J,p}(\boldsymbol z)=1.
    \label{eq:sm_qaoa_distribution}
\end{equation}
The QAOA distribution has an exact global spin-flip symmetry since the operator $\prod_{j=1}^N \hat X_j$ leaves $|+\rangle^{\otimes N}$ invariant and commutes with both $\hat C_J$, which contains only products $\hat Z_j\hat Z_k$, and the mixer $\hat B$ from Eq.~\eqref{eq:sm_mixer}. Hence $P_{J,p}(\boldsymbol z)=P_{J,p}(-\boldsymbol z)$ for every finite $N$, disorder realization $J$, depth $p$, and angle sequence $\{\boldsymbol\gamma,\boldsymbol\beta\}$. The subscript $p$ always refers to the chosen depth and optimized angle sequence. As such, we suppress the explicit angle dependence and simply refer to the quantum state and resulting quantities by $p$. For any diagonal observable $O(\boldsymbol z)$ of a single measured bitstring, we write
\begin{equation}
    \bigl\langle\hat O\bigr\rangle_{J,p}=\langle\Psi_{J,p}|\hat O|\Psi_{J,p}\rangle=\sum_{\boldsymbol z}P_{J,p}(\boldsymbol z) O(\boldsymbol z).
    \label{eq:sm_single_replica_expectation}
\end{equation}
To probe the geometry of the distribution, we draw several independent samples from the same distribution and the same disorder realization, which are called physical replicas. For $k$ such replicas, a diagonal observable in the computational basis reads
\begin{equation}
    \left\langle\hat{O}(\boldsymbol z^1,\ldots,\boldsymbol z^k)\right\rangle_{J,p}^{\otimes k}=\sum_{\boldsymbol z^1,\ldots,\boldsymbol z^k}\left[\prod_{a=1}^k P_{J,p}(\boldsymbol z^a)\right]O(\boldsymbol z^1,\ldots,\boldsymbol z^k).
    \label{eq:sm_k_copy_expectation}
\end{equation}
The superscript $\otimes k$ denotes $k$ independent samples at fixed disorder instance $J$, i.e., all replicas share the same Hamiltonian with couplings $J_{jk}$ and the same angle sequence. For the standard single-sample case $k=1$ and whenever unambiguous, we typically omit the superscript. For two physical replicas $A$ and $B$, the normalized overlap is~\cite{EdwardsAnderson1975,Parisi1979,MezardParisiVirasoro1987,Panchenko2013Book}
\begin{equation}
    q_{AB}=\frac{1}{N}\sum_{j=1}^N z_j^A z_j^B.
    \label{eq:sm_two_replica_overlap}
\end{equation}
This two-replica overlap is the basic geometric coordinate used throughout this work. It probes how independently sampled bitstrings from the same distribution are organized relative to one another in Hamming space. Because the product $P_{J,p}(\boldsymbol z^A)P_{J,p}(\boldsymbol z^B)$ is invariant under flipping only one replica, while $q_{AB}\to -q_{AB}$ under that flip, the two-replica overlap distribution is even.

\subsection{Overlap source and quenched free-entropy density}
\label{subsec:sm_overlap_source}

The two-replica overlap $q_{AB}$ defined in Eq.~\eqref{eq:sm_two_replica_overlap} can be probed by introducing a field $\lambda$ conjugate to the extensive overlap $Nq_{AB}$. For a fixed disorder instance $J$, using the two-sample expectation notation of Eq.~\eqref{eq:sm_k_copy_expectation}, we define the two-replica source partition function
\begin{equation}
    Z^{(2)}_{J,p}(\lambda)=\left\langle e^{\lambda N q_{AB}}\right\rangle_{J,p}^{\otimes 2}=\sum_{\boldsymbol z^A,\boldsymbol z^B}P_{J,p}\bigl(\boldsymbol z^A\bigr)P_{J,p}\bigl(\boldsymbol z^B\bigr)e^{\lambda\sum_{j=1}^N z_j^A z_j^B}.
    \label{eq:sm_two_replica_source}
\end{equation}
The normalization of the QAOA distribution from Eq.~\eqref{eq:sm_qaoa_distribution} gives $Z^{(2)}_{J,p}(0)=1$. In particular, $\lambda$ defines a source-biased two-replica distribution
\begin{equation}
    P^{(2)}_{J,p,\lambda}\left(\boldsymbol z^A,\boldsymbol z^B\right)=\frac{P_{J,p}\bigl(\boldsymbol z^A\bigr)P_{J,p}\bigl(\boldsymbol z^B\bigr)e^{\lambda N q_{AB}}}{Z^{(2)}_{J,p}(\lambda)},
    \label{eq:sm_source_biased_pair_measure}
\end{equation}
where positive $\lambda$ favors pairs of QAOA outputs with larger mutual overlap, while negative $\lambda$ favors negative mutual overlap. This construction does not change the QAOA circuit and only reweights pairs of independently drawn samples from the same QAOA distribution. The corresponding thermodynamic free-entropy density is
\begin{equation}
    \phi_p(\lambda)=\lim_{N\to\infty}\frac{1}{N}\mathbb E_J\left[\ln Z^{(2)}_{J,p}(\lambda)\right].
    \label{eq:sm_quenched_source_density}
\end{equation}
The logarithm is taken before the disorder average, which is known as the quenched convention. Thus $\phi_p(\lambda)$ describes the typical overlap response across SK disorder realizations. At finite $N$, differentiating the logarithm gives
\begin{equation}
    \frac{1}{N}\frac{\partial}{\partial\lambda}\ln Z^{(2)}_{J,p}(\lambda)=\bigl\langle q_{AB}\bigr\rangle^{(2)}_{J,p,\lambda},
    \label{eq:sm_finite_size_sourced_overlap}
\end{equation}
where the expectation is taken with respect to the source-biased distribution of Eq.~\eqref{eq:sm_source_biased_pair_measure}. If the thermodynamic limit is differentiable at $\lambda$, the source-selected overlap is therefore
\begin{equation}
    q_p(\lambda)=\partial_\lambda \phi_p(\lambda).
    \label{eq:sm_sourced_overlap}
\end{equation}
Similarly, the finite-$N$ second derivative is
\begin{equation}
    \frac{1}{N}\frac{\partial^2}{\partial\lambda^2}\ln Z^{(2)}_{J,p}(\lambda)=N\Biggl[\bigl\langle q_{AB}^2\bigr\rangle^{(2)}_{J,p,\lambda}-\Bigl(\bigl\langle q_{AB}\bigr\rangle^{(2)}_{J,p,\lambda}\Bigr)^2\Biggr],
    \label{eq:sm_source_second_derivative}
\end{equation}
so the local curvature of $\phi_p(\lambda)$ measures the extensive overlap fluctuations under the source-biased pair distribution. It is also useful to introduce the finite-instance overlap distribution induced by two independent QAOA samples
\begin{equation}
    P^{(2)}_{J,p,N}(q)=\bigl\langle\delta\left(q-q_{AB}\right)\bigr\rangle_{J,p}^{\otimes 2},
    \label{eq:sm_overlap_distribution}
\end{equation}
where $\delta$ denotes a delta mass on the finite-$N$ overlap grid where $q$ takes values in $\{-1,-1+2/N,\ldots,1\}$. The source partition function is the exact discrete Laplace transform
of this overlap distribution
\begin{equation}
    Z^{(2)}_{J,p}(\lambda)=\sum_qP^{(2)}_{J,p,N}(q)e^{\lambda N q}.
    \label{eq:sm_source_as_laplace_transform}
\end{equation}
Thus $\phi_p(\lambda)$ in Eq.~\eqref{eq:sm_quenched_source_density} is the cumulant-generating function density for the two-replica overlap in a typical disorder instance.

\subsection{Overlap potential from the source response}
\label{subsec:sm_field_selected_potential}

Starting from the cumulant-generating function of Eq.~\eqref{eq:sm_quenched_source_density}, the associated overlap potential is obtained by Legendre transforming from the source field $\lambda$ to the overlap $q$
\begin{equation}
    V_p(q)=\max_\lambda\left[\lambda q-\phi_p(\lambda)\right].
    \label{eq:sm_field_selected_potential}
\end{equation}
We refer to this object as the Franz--Parisi-type overlap potential associated with the QAOA distribution~\cite{FranzParisi1995,FranzParisi1997,FranzParisi1998EffectivePotential}. The $\lambda=0$ point corresponds to the unbiased two-sample QAOA distribution. Since $Z_{J,p}^{(2)}(0)=1$, one has $\phi_p(0)=0$. Moreover, the two-replica overlap distribution is even by the global spin-flip symmetry discussed above, so the source-selected overlap vanishes at the origin, i.e., $q_p(0)=0$. Thus the source-biased construction probes how the unbiased two-sample QAOA distribution responds when a field favors mutual similarity. When $\phi_p$ is differentiable, the maximizer in Eq.~\eqref{eq:sm_field_selected_potential} satisfies the stationarity condition $q=\partial_\lambda\phi_p(\lambda)$. Equivalently, tuning the source selects the overlap $q_p(\lambda)$ defined in Eq.~\eqref{eq:sm_sourced_overlap}. At points reached in this way, the Legendre transform is evaluated parametrically as
\begin{equation}
    V_p\bigl(q_p(\lambda)\bigr)=\lambda q_p(\lambda)-\phi_p(\lambda).
    \label{eq:sm_parametric_potential}
\end{equation}
This is the construction used in the main text. It assumes only that the source-overlap map is smooth and locally invertible on the solution being followed. Differentiating Eq.~\eqref{eq:sm_parametric_potential} gives the standard conjugacy relations
\begin{equation}
    \frac{d V_p(q)}{dq}=\lambda,\quad\frac{d^2 V_p(q)}{dq^2}=\left[\frac{d q_p(\lambda)}{d\lambda}\right]^{-1}=\left[\partial_\lambda^2\phi_p(\lambda)\right]^{-1}.
    \label{eq:sm_legendre_conjugacy}
\end{equation}
This source construction should be distinguished from a fixed-overlap Franz--Parisi calculation discussed in Sec.~\ref{subsec:sm_constrained_overlap_branch_search}.

\section{Finite-depth Keldysh-history formalism}
\label{sec:sm_keldysh_formalism}

We derive the finite-depth representation of $Z^{(2)}_{J,p}(\lambda)$ defined in Eq.~\eqref{eq:sm_two_replica_source}. The construction has three layers and extends the established forward--backward, or Keldysh, history representation of the QAOA~\cite{FarhiGoldstoneGutmannZhou2022,BassoFarhiMarwahaVillalongaZhou2022,BassoGamarnikMeiZhou2022,BoulebnaneKhanLiuLarsonHermanShaydulinPistoia2026}. First, each probability $P_{J,p}(\boldsymbol z)$ is written as a history sum containing one forward history for the amplitude and one backward history for its complex conjugate, joined at the measured terminal spin. Second, two physical replicas $A$ and $B$ are introduced, with $\lambda$ coupling only their terminal spins. Third, the quenched logarithm $\mathbb E_J[\ln Z^{(2)}_{J,p}(\lambda)]$ is evaluated by employing the replica trick~\cite{MezardParisiVirasoro1987,CastellaniCavagna2005}: we introduce $n$ auxiliary replicas $\alpha=1,\ldots,n$ and take $n\to 0$. These auxiliary replicas are a device for the disorder average and should not be confused with the two physical replicas whose overlap is being probed. After the Gaussian SK disorder average, the microscopic histories enter only through empirical overlaps between Keldysh histories. Introducing collective variables for these overlaps factorizes the $N$-site sum into identical one-site traces, namely sums over the history variables associated with a single spin. The thermodynamic limit is then obtained by a saddle point over a finite set of history-overlap variables. The resulting equations have size independent of $N$ at fixed $p$, while the one-site trace involves sums over the two one-site QAOA histories associated with the physical replicas in $Z^{(2)}_{J,p}(\lambda)$.

\subsection{One-sample history expansion}
\label{subsec:sm_single_replica_path_expansion}

We begin from the fixed-instance probability
\begin{equation}
    P_{J,p}(\boldsymbol z)=\left|\langle \boldsymbol z|\Psi_{J,p}\rangle\right|^2=\langle \boldsymbol z|\Psi_{J,p}\rangle\langle \Psi_{J,p}|\boldsymbol z\rangle.
    \label{eq:sm_probability_as_squared_amplitude}
\end{equation}
Since this quantity is a squared amplitude, its path expansion contains two time tracks: one from the amplitude and one from its complex conjugate. We obtain the expansion by inserting computational-basis resolutions of the identity $\hat{\mathbb{1}}_{2^N}=\sum_{\boldsymbol z}|\boldsymbol z\rangle\langle\boldsymbol z|$ between every cost and mixer layer on both tracks. The cost operator is diagonal in this basis, while the transverse mixer contributes single-qubit matrix elements. Thus, for each site, the squared amplitude contains $p$ internal spins on one branch, $p$ internal spins on the conjugate branch, and the terminal measured spin. We now package these $2p+1$ variables into the standard QAOA history notation. We define the history index set
\begin{equation}
    \mathcal I_p=\bigl\{1,2,\ldots,p,0,-p,\ldots,-2,-1\bigr\}.
    \label{eq:sm_history_index_set}
\end{equation}
A one-site history is
\begin{equation}
    \boldsymbol a=(a_1,\ldots,a_p,a_0,a_{-p},\ldots,a_{-1})\in\{\pm1\}^{2p+1}.
    \label{eq:sm_history_vector}
\end{equation}
The central entry $a_0$ is the terminal measured spin. With the convention used here, the positive indices carry the complex-conjugate phases and the negative indices carry the forward phases. This convention is encoded by the entries of the signed phase vector $\boldsymbol{\Gamma}=(\Gamma_r)_{r\in\mathcal I_p}$
\begin{equation}
    \Gamma_r=
    \begin{cases}
        \gamma_r, & r>0,\\
        0, & r=0,\\
        -\gamma_{-r}, & r<0.
    \end{cases}
    \label{eq:sm_gamma_signed}
\end{equation}
The terminal slice has $\Gamma_0=0$ because the measured spin is not acted on by the diagonal cost phase; it is the point where the forward and backward branches are sewn together. The identity $\sum_{r\in\mathcal I_p}\Gamma_r=0$ is the cancellation between the two branches of the contour. The single-qubit mixer matrix element is
\begin{equation}
    \langle a|e^{-i\beta \hat X}|b\rangle=\cos\beta\delta_{a,b}-i\sin\beta\delta_{a,-b},\quad a,b\in\{\pm1\}.
    \label{eq:sm_mixer_matrix_element}
\end{equation}
Collecting all mixer factors associated with a one-site history gives the one-site weight
\begin{equation}
    f_{\boldsymbol\beta}(\boldsymbol a)=\frac{1}{2}\left[\prod_{\ell=1}^{p-1}\langle a_\ell|e^{i\beta_\ell \hat X}|a_{\ell+1}\rangle\right]\langle a_p|e^{i\beta_p \hat X}|a_0\rangle\langle a_0|e^{-i\beta_p \hat X}|a_{-p}\rangle\left[\prod_{\ell=1}^{p-1}\langle a_{-\ell-1}|e^{-i\beta_\ell \hat X}|a_{-\ell}\rangle\right].
    \label{eq:sm_history_weight}
\end{equation}
The prefactor $1/2$ comes from the initial $|+\rangle$ state on the two branches at a single site. The factors with $e^{+i\beta_\ell \hat X}$ belong to the complex-conjugate branch, while those with $e^{-i\beta_\ell \hat X}$ belong to the forward branch. For a fixed disorder realization $J$, the probability of measuring $\boldsymbol z$ can then be written as an exact sum over one history per site
\begin{equation}
    P_{J,p}(\boldsymbol z)
    =\sum_{\boldsymbol a_1,\ldots,\boldsymbol a_N}\prod_{j=1}^Nf_{\boldsymbol\beta}(\boldsymbol a_j)\delta_{a_{0,j},z_j}\exp\left[\frac{i}{\sqrt N}\sum_{j<k}J_{jk}\sum_{r\in\mathcal I_p}\Gamma_ra_{r,j}a_{r,k}\right].
    \label{eq:sm_single_replica_probability_path}
\end{equation}
Here the sum runs over $\boldsymbol a_j\in\{\pm1\}^{2p+1}$ independently for every site $j$, and the Kronecker delta fixes the terminal history spin to the measured output. Equation~\eqref{eq:sm_single_replica_probability_path} is an exact finite-$N$ identity.

\subsection{Exact two-replica overlap source}
\label{subsec:sm_exact_two_replica_source}

We now insert Eq.~\eqref{eq:sm_single_replica_probability_path} into the definition of Eq.~\eqref{eq:sm_two_replica_source}. For each site $j$ we introduce two independent histories, $\boldsymbol a_j^A$ and $\boldsymbol a_j^B$, corresponding to the two physical replicas $A$ and $B$. Their terminal spins are $a_{0,j}^A=z_j^A$ and $a_{0,j}^B=z_j^B$. The overlap source couples only these terminal entries. Using the history weight $f_{\boldsymbol\beta}$ from Eq.~\eqref{eq:sm_history_weight} and the signed phases $\Gamma_r$ from Eq.~\eqref{eq:sm_gamma_signed}, this gives the exact finite-$N$ expression
\begin{equation}
    Z^{(2)}_{J,p}(\lambda)=\sum_{\{\boldsymbol a_j^A,\boldsymbol a_j^B\}_{j=1}^N}\prod_{j=1}^Nf_{\boldsymbol\beta}\bigl(\boldsymbol a_j^A\bigr)f_{\boldsymbol\beta}\bigl(\boldsymbol a_j^B\bigr)\exp\left[\lambda\sum_{j=1}^Na_{0,j}^A a_{0,j}^B\right]\exp\left[\frac{i}{\sqrt N}\sum_{j<k}J_{jk}\sum_{\sigma=A,B}\sum_{r\in\mathcal I_p}\Gamma_ra_{r,j}^{\sigma}a_{r,k}^{\sigma}\right].
    \label{eq:sm_exact_two_replica_source_history}
\end{equation}

\subsection{Auxiliary replicas for the quenched logarithm}
\label{subsec:sm_auxiliary_replicas_for_quenched_log}

We evaluate the quenched logarithm $\mathbb E_J[\ln Z^{(2)}_{J,p}(\lambda)]$ using the replica trick~\cite{MezardParisiVirasoro1987,CastellaniCavagna2005}
\begin{equation}
    \mathbb E_J\left[\ln Z^{(2)}_{J,p}(\lambda)\right]=\lim_{n\to0}\frac{1}{n}\ln\mathbb E_J\left[\left(Z^{(2)}_{J,p}(\lambda)\right)^n\right],
    \label{eq:sm_replica_trick}
\end{equation}
where the right-hand side is first computed for positive integer $n$ and then analytically continued to $n\to0$. For integer $n$, each auxiliary replica $\alpha=1,\ldots,n$ contains two physical replicas $\sigma=A,B$, and each physical replica carries a QAOA history indexed by $r\in\mathcal I_p$. We combine these labels as $I=(\alpha,\sigma,r)$, which labels a Keldysh spin inside a physical replica inside an auxiliary disorder replica. For a site $j$, the corresponding spin variable is denoted $a_{I,j}=a^{\alpha,\sigma}_{r,j}$. After replicating Eq.~\eqref{eq:sm_exact_two_replica_source_history}, the disorder-dependent phase takes the compact form
\begin{equation}
    \exp\left[\frac{i}{\sqrt N}\sum_{j<k}J_{jk}S_{jk}\right],\quad\textrm{with}~S_{jk}=\sum_{\alpha=1}^n\sum_{\sigma=A,B}\sum_{r\in\mathcal I_p}\Gamma_ra^{\alpha,\sigma}_{r,j}a^{\alpha,\sigma}_{r,k}.
    \label{eq:sm_replicated_phase}
\end{equation}
The $\lambda$-dependent overlap factor in Eq.~\eqref{eq:sm_exact_two_replica_source_history} is also replicated. In the original two-physical-replica problem it couples only the terminal spins $a^A_{0,j}$ and $a^B_{0,j}$. After introducing the auxiliary replicas, each $\alpha$ carries its own such $A$--$B$ source, while distinct auxiliary replicas remain uncoupled by $\lambda$. The replicated source factor is therefore
\begin{equation}
    \exp\left[\lambda\sum_{\alpha=1}^n\sum_{j=1}^Na^{\alpha,A}_{0,j}a^{\alpha,B}_{0,j}\right].
    \label{eq:sm_replicated_source}
\end{equation}

\subsection{Gaussian SK disorder average}
\label{subsec:sm_gaussian_disorder_average}

We now average Eq.~\eqref{eq:sm_replicated_phase} over the SK couplings. Since the $J_{jk}$ are independent standard Gaussian variables, one has
\begin{equation}
    \mathbb E_J\left[\exp\left(\frac{i}{\sqrt N}\sum_{j<k}J_{jk}S_{jk}\right)\right]=\exp\left(-\frac{1}{2N}\sum_{j<k}S_{jk}^2\right),
    \label{eq:sm_gaussian_average}
\end{equation}
where the square couples different sites. To make the dependence on the history overlaps explicit, we use $I=(\alpha,\sigma,r)$ and write $\Gamma_I\equiv\Gamma_r$
\begin{equation}
    S_{jk}=\sum_I\Gamma_Ia_{I,j}a_{I,k},\quad\mathrm{and}\quad\sum_{j<k}S_{jk}^2=\sum_{I,I'}\Gamma_I\Gamma_{I'}\sum_{j<k}a_{I,j}a_{I',j}a_{I,k}a_{I',k}.
    \label{eq:sm_sjk_square_expanded}
\end{equation}
The disorder-averaged expression depends on a replicated spin configuration only through site averages of products $a_{I,j}a_{I',j}$. These are overlaps between two Keldysh histories labeled by the composite indices $I$ and $I'$, in the same sense that $q_{AB}$ is an overlap between two sampled bitstrings. We denote the corresponding site average by
\begin{equation}
    \mathsf{Q}^{(n)}_{II'}\bigl[\{a\}\bigr]=\frac{1}{N}\sum_{j=1}^Na_{I,j}a_{I',j},\quad I=(\alpha,\sigma,r),\quad I'=(\alpha',\sigma',r'),
    \label{eq:sm_history_overlap_site_average}
\end{equation}
where the diagonal entries obey $\mathsf{Q}^{(n)}_{II}[\{a\}]=1$. For fixed $I,I'$, the site-pair sum can be written exactly as
\begin{equation}
    \sum_{j<k}a_{I,j}a_{I',j}a_{I,k}a_{I',k}=\frac{1}{2}\left[\left(\sum_{j=1}^Na_{I,j}a_{I',j}\right)^2-\sum_{j=1}^N\left(a_{I,j}a_{I',j}\right)^2\right]=\frac{N^2}{2}\Bigl(\mathsf{Q}^{(n)}_{II'}\bigl[\{a\}\bigr]\Bigr)^2-\frac{N}{2}.
    \label{eq:sm_pair_identity}
\end{equation}
Substituting this identity into Eq.~\eqref{eq:sm_gaussian_average} gives
\begin{equation}
    -\frac{1}{2N}\sum_{j<k}S_{jk}^2=-\frac{N}{4}\sum_{I,I'}\Gamma_I\Gamma_{I'}\Bigl(\mathsf{Q}^{(n)}_{II'}\bigl[\{a\}\bigr]\Bigr)^2+\frac{1}{4}\sum_{I,I'}\Gamma_I\Gamma_{I'}.
    \label{eq:sm_gaussian_exponent_in_Qsite}
\end{equation}
The final term vanishes because
\begin{equation}
    \sum_I \Gamma_I=\sum_{\alpha=1}^n\sum_{\sigma=A,B}\sum_{r\in\mathcal I_p}\Gamma_r=0,
    \label{eq:sm_gamma_sum_composite_zero}
\end{equation}
using the contour cancellation $\sum_{r\in\mathcal I_p}\Gamma_r=0$. Thus, for any fixed replicated spin configuration, the disorder average becomes
\begin{equation}
    \exp\left[-\frac{N}{4}\sum_{I,I'}\Gamma_I\Gamma_{I'}\Bigl(\mathsf{Q}^{(n)}_{II'}\bigl[\{a\}\bigr]\Bigr)^2\right].
    \label{eq:sm_disorder_in_terms_of_Qsite}
\end{equation}
Equation~\eqref{eq:sm_disorder_in_terms_of_Qsite} leaves an interaction only through overlaps between replicated histories. We call a history coordinate active in the disorder term when $\Gamma_r\neq0$. The terminal coordinate is therefore inactive in the untilted disorder term because $\Gamma_0=0$, although it remains in the one-site trace through the mixer weights and the explicit source $e^{\lambda a_0^Aa_0^B}$.

\subsection{Collective history action and large-\texorpdfstring{$N$}{N} saddle}
\label{subsec:sm_thermodynamic_saddle_action}

The disorder average has reduced the $N$-spin problem to empirical overlaps between Keldysh histories, but these overlaps still depend on the microscopic histories at all sites. The remaining mean-field step is to promote them to collective variables. Enforcing their definitions with conjugate fields then factorizes the $N$-site trace into identical one-site problems, so that the thermodynamic limit is controlled by a finite-dimensional saddle at fixed $p$. The Gaussian disorder average expresses the replicated moment $\mathbb E_J[(Z^{(2)}_{J,p}(\lambda))^n]$, introduced for the quenched logarithm in Eq.~\eqref{eq:sm_replica_trick}, through the site averages $\mathsf{Q}^{(n)}_{II'}[\{a\}]$ introduced above. These quantities are still functions of the microscopic histories at all sites. To separate the sites, we promote the site averages to independent collective variables $\mathsf{Q}^{(n)}_{II'}$ and enforce their definitions with conjugate Lagrange multiplier fields $\widehat{\mathsf{Q}}^{(n)}_{II'}$. At the exponential scale relevant for the large-$N$ limit, this constraint may be represented as
\begin{equation}
    1\sim\int d\mathsf{Q}^{(n)}d\widehat{\mathsf{Q}}^{(n)}
    \exp\left\{
    \frac{N}{2}\sum_{I,I'}\widehat{\mathsf{Q}}^{(n)}_{II'}
    \left[
    \mathsf{Q}^{(n)}_{II'}-\frac{1}{N}\sum_{j=1}^Na_{I,j}a_{I',j}
    \right]
    \right\},
    \label{eq:sm_collective_identity}
\end{equation}
where $\sim$ denotes equality at the exponential scale, meaning that normalization factors and contour-dependent constants that are subexponential in $N$ are omitted. Equivalently, Eq.~\eqref{eq:sm_collective_identity} imposes
\begin{equation}
    \mathsf{Q}^{(n)}_{II'}=\mathsf{Q}^{(n)}_{II'}\bigl[\{a\}\bigr]=\frac{1}{N}\sum_{j=1}^Na_{I,j}a_{I',j}.
    \label{eq:sm_collective_overlap_constraint}
\end{equation}
Here $\mathsf{Q}^{(n)}_{II'}[\{a\}]$ denotes the site average computed from a microscopic replicated history configuration, while $\mathsf{Q}^{(n)}_{II'}$ denotes the corresponding collective variable. After the constraint is inserted, the microscopic histories appear only through sums over individual sites. The $N$-site trace therefore factorizes into identical one-site traces, leaving an integral over the collective overlaps and their conjugate fields. This is the standard collective-variable reduction used in SK-QAOA calculations~\cite{FarhiGoldstoneGutmannZhou2022,BassoFarhiMarwahaVillalongaZhou2022,BassoGamarnikMeiZhou2022,BoulebnaneKhanLiuLarsonHermanShaydulinPistoia2026}, here applied to the two-sample overlap generating function $Z^{(2)}_{J,p}(\lambda)$ defined in Eq.~\eqref{eq:sm_two_replica_source}.

Diagonal entries $\mathsf{Q}^{(n)}_{II}=1$ are fixed. Combining the collective-variable identity with the disorder contribution in Eq.~\eqref{eq:sm_disorder_in_terms_of_Qsite}, the disorder-averaged replicated moment $\mathbb E_J[(Z^{(2)}_{J,p}(\lambda))^n]$ can be written (up to subexponential prefactors and with the conjugate fields integrated along steepest-descent contours) as
\begin{equation}
    \mathbb E_J\left[\left(Z^{(2)}_{J,p}(\lambda)\right)^n\right]\sim\int d\mathsf Q^{(n)}d\widehat{\mathsf Q}^{(n)}\exp\left[N\mathcal A_n\bigl(\mathsf Q^{(n)},\widehat{\mathsf Q}^{(n)},\lambda\bigr)\right],
    \label{eq:sm_collective_integral_with_conjugates}
\end{equation}
where
\begin{equation}
    \mathcal A_n\bigl(\mathsf{Q}^{(n)},\widehat{\mathsf{Q}}^{(n)},\lambda\bigr)=-\frac{1}{4}\sum_{I,I'}\Gamma_I\Gamma_{I'}\bigl(\mathsf{Q}^{(n)}_{II'}\bigr)^2+\frac{1}{2}\sum_{I,I'}\widehat{\mathsf{Q}}^{(n)}_{II'}\mathsf{Q}^{(n)}_{II'}+\ln \mathcal Z_n\bigl(\widehat{\mathsf{Q}}^{(n)},\lambda\bigr).
    \label{eq:sm_action_with_conjugates}
\end{equation}
The one-site trace is
\begin{equation}
    \mathcal Z_n\bigl(\widehat{\mathsf{Q}}^{(n)},\lambda\bigr)=\sum_{\{\boldsymbol a^{\alpha,A},\boldsymbol a^{\alpha,B}\}_{\alpha=1}^n}\prod_{\alpha=1}^nf_{\boldsymbol\beta}\bigl(\boldsymbol a^{\alpha,A}\bigr)f_{\boldsymbol\beta}\bigl(\boldsymbol a^{\alpha,B}\bigr)\exp\left[\lambda\sum_{\alpha=1}^na^{\alpha,A}_0a^{\alpha,B}_0-\frac{1}{2}\sum_{I,I'}\widehat{\mathsf{Q}}^{(n)}_{II'}a_Ia_{I'}\right].
    \label{eq:sm_single_site_trace_with_conjugates}
\end{equation}
Here $a_I$ denotes a one-site spin with label $I=(\alpha,\sigma,r)$. The collective-variable reduction has converted the original $N$-site problem into the $N$-th power of this one-site trace. The large-$N$ integral is therefore evaluated by steepest descent. At a stationary point of $\mathcal A_n(\mathsf{Q}^{(n)},\widehat{\mathsf{Q}}^{(n)},\lambda)$, the first derivative with respect to each collective variable that is varied must vanish. Taking the derivative of Eq.~\eqref{eq:sm_action_with_conjugates} with respect to $\mathsf{Q}^{(n)}_{II'}$ gives
\begin{equation}
    0=\frac{\partial\mathcal A_n}{\partial \mathsf{Q}^{(n)}_{II'}}=-\frac{1}{2}\Gamma_I\Gamma_{I'}\mathsf{Q}^{(n)}_{II'}+\frac{1}{2}\widehat{\mathsf{Q}}^{(n)}_{II'}.
    \label{eq:sm_conjugate_stationarity}
\end{equation}
Therefore, at stationarity
\begin{equation}
    \widehat{\mathsf{Q}}^{(n)}_{II'}=\Gamma_I\Gamma_{I'}\mathsf{Q}^{(n)}_{II'}.
\label{eq:sm_conjugate_elimination}
\end{equation}
Substituting this relation into Eq.~\eqref{eq:sm_action_with_conjugates} gives the reduced collective action
\begin{equation}
    \mathbb E_J\left[\left(Z^{(2)}_{J,p}(\lambda)\right)^n\right]\sim\int d\mathsf Q^{(n)}\exp\bigl[N\mathcal A_n(\mathsf Q^{(n)},\lambda)\bigr],
    \label{eq:sm_replicated_saddle_integral}
\end{equation}
with
\begin{equation}
    \mathcal A_n\bigl(\mathsf{Q}^{(n)},\lambda\bigr)=\frac{1}{4}\sum_{I,I'}\Gamma_I\Gamma_{I'}\bigl(\mathsf{Q}^{(n)}_{II'}\bigr)^2+\ln\mathcal Z_n(\mathsf{Q}^{(n)},\lambda),
    \label{eq:sm_replicated_action}
\end{equation}
and
\begin{equation}
    \mathcal Z_n\bigl(\mathsf{Q}^{(n)},\lambda\bigr)=\sum_{\{\boldsymbol a^{\alpha,A},\boldsymbol a^{\alpha,B}\}_{\alpha=1}^n}\prod_{\alpha=1}^nf_{\boldsymbol\beta}\bigl(\boldsymbol a^{\alpha,A}\bigr)f_{\boldsymbol\beta}\bigl(\boldsymbol a^{\alpha,B}\bigr)\exp\left[\lambda\sum_{\alpha=1}^na^{\alpha,A}_0a^{\alpha,B}_0-\frac{1}{2}\sum_{I,I'}\Gamma_I\Gamma_{I'}\mathsf{Q}^{(n)}_{II'}a_Ia_{I'}\right].
    \label{eq:sm_single_site_trace_replicated}
\end{equation}
Equation~\eqref{eq:sm_conjugate_elimination} eliminates the conjugate fields and gives the reduced action in Eq.~\eqref{eq:sm_replicated_action}. Their effect remains in the one-site trace through the substitution $\widehat{\mathsf Q}^{(n)}_{II'}=\Gamma_I\Gamma_{I'}\mathsf Q^{(n)}_{II'}$. The remaining large-$N$ integral over $\mathsf Q^{(n)}$ is evaluated by steepest descent. Stationarity with respect to the varied entries of $\mathsf Q^{(n)}$ gives
\begin{equation}
    0=\frac{\partial \mathcal A_n}{\partial \mathsf{Q}^{(n)}_{II'}}=\frac{1}{2}\Gamma_I\Gamma_{I'}\Bigl[\mathsf{Q}^{(n)}_{II'}-\bigl\langle a_Ia_{I'}\bigr\rangle_{\mathsf{Q}^{(n)},\lambda}\Bigr],
    \label{eq:sm_saddle_variation}
\end{equation}
where the expectation is taken with respect to the normalized one-site trace in Eq.~\eqref{eq:sm_single_site_trace_replicated}. Only pairs of active coordinates enter the disorder-generated quadratic term. Thus, for pairs of active coordinates
\begin{equation}
    \mathsf{Q}^{(n)}_{II'}=\bigl\langle a_Ia_{I'}\bigr\rangle_{\mathsf{Q}^{(n)},\lambda}.
    \label{eq:sm_replicated_saddle_equations}
\end{equation}
The interpretation is direct: the overlap imposed as a collective variable in the effective one-site problem must equal the overlap generated by that same one-site problem. Entries with $\Gamma_I\Gamma_{I'}=0$ are not fixed by this saddle. Terminal observables involving $r=0$ are computed directly from the one-site measure. Denoting by $\mathsf{Q}^{(n)}_\star(\lambda)$ a stationary solution of Eq.~\eqref{eq:sm_replicated_saddle_equations}, the quenched free-entropy density $\phi_p(\lambda)$ defined in Eq.~\eqref{eq:sm_quenched_source_density} follows from the replica identity as
\begin{equation}
    \phi_p(\lambda)=\lim_{n\to0}\frac{1}{n}\Bigl\{\mathcal A_n\bigl[\mathsf{Q}^{(n)}_\star(\lambda),\lambda\bigr]-\mathcal A_n\bigl[\mathsf{Q}^{(n)}_\star(0),0\bigr]\Bigr\}.
    \label{eq:sm_phi_from_replicated_action}
\end{equation}
The subtraction fixes the normalization $\phi_p(0)=0$, matching the exact finite-$N$ identity $Z^{(2)}_{J,p}(0)=1$. Finally, the physical overlap is obtained by differentiating the quenched free-entropy density $\phi_p(\lambda)$ with respect to the overlap source $\lambda$. Since $\mathsf{Q}^{(n)}_\star(\lambda)$ is stationary, the envelope theorem implies that the derivative of $\mathsf{Q}^{(n)}_\star(\lambda)$ does not contribute. Therefore
\begin{equation}
    q_p(\lambda)=\partial_\lambda\phi_p(\lambda)=\lim_{n\to0}\frac{1}{n}\left\langle\sum_{\alpha=1}^na^{\alpha,A}_0a^{\alpha,B}_0\right\rangle_{\mathsf{Q}^{(n)}_\star(\lambda),\lambda}.
    \label{eq:sm_source_derivative_replicated}
\end{equation}
This last equation separates the two overlap notions used in the construction. The collective variables $\mathsf{Q}^{(n)}_{II'}$ describe correlations between replicated Keldysh histories in the disorder-averaged one-site problem. The physical overlap $q_p(\lambda)$ is the overlap between the two QAOA samples, obtained as the response to the explicit source $\lambda$.

\subsection{Auxiliary-replica ansatz used in the main text}
\label{subsec:sm_auxiliary_replica_diagonal_branch}

The saddle equations~\eqref{eq:sm_replicated_saddle_equations} are defined before taking the $n\to0$ limit, and the collective matrix may in principle contain correlations between distinct auxiliary replicas $\alpha$ and $\alpha'$. We use the ansatz stated in the main text, in which overlaps between distinct auxiliary replicas are set to zero, while the two physical replicas retain their coupled structure
\begin{equation}
    \mathsf{Q}^{(n)}_{(\alpha,\sigma,r),(\alpha',\sigma',r')}=\delta_{\alpha,\alpha'}\mathsf{Q}_{\sigma r,\sigma' r'}.
    \label{eq:sm_auxiliary_replica_diagonal_branch}
\end{equation}
We refer to this below as the auxiliary-replica-diagonal ansatz. This restriction is stronger than replica symmetry alone, which would also allow a common nonzero overlap between distinct auxiliary replicas. Consequently, loss of stability of the ansatz used here signals the need to include correlations between auxiliary replicas, but does not by itself establish replica-symmetry breaking. Within each auxiliary replica, the reduced collective matrix $\mathsf Q$ contains the two physical QAOA samples and can be written as
\begin{equation}
    \mathsf{Q}=
    \begin{pmatrix}
        \mathsf{G}^A & \mathsf{R}\\
        \mathsf{R}^{\mathsf T} & \mathsf{G}^B
    \end{pmatrix}.
    \label{eq:sm_two_replica_block_matrix}
\end{equation}
The within-sample blocks $\mathsf{G}^A$ and $\mathsf{G}^B$ describe Keldysh-history correlations inside each physical sample, while the cross-sample block $\mathsf{R}$ describes correlations between them:
\begin{equation}
    \mathsf{G}^\sigma_{rr'}=\mathsf{Q}_{\sigma r,\sigma r'},\qquad\mathsf{R}_{rr'}=\mathsf{Q}_{A r,B r'}.
    \label{eq:sm_two_replica_block_entries}
\end{equation}
Since the source treats the two physical replicas symmetrically, the continued solution satisfies $\mathsf{G}^A=\mathsf{G}^B$. At $\lambda=0$, the two physical replicas factorize and the cross-sample block vanishes, $\mathsf{R}=0$. Under this ansatz, the replicated action factorizes into identical contributions from the auxiliary replicas. The $n\to0$ limit then reduces the quenched free-entropy density $\phi_p(\lambda)$ to a variational problem for the two physical replicas within one auxiliary replica
\begin{equation}
    \phi_p(\lambda)=\mathcal F_p\bigl[\mathsf{Q}_\star(\lambda),\lambda\bigr]-\mathcal F_p\bigl[\mathsf{Q}_\star(0),0\bigr],
    \label{eq:sm_phi_two_replica_branch}
\end{equation}
where $\mathsf{Q}_\star(\lambda)$ is the stationary solution within the ansatz above and $\mathcal F_p$ is the reduced two-sample action within one auxiliary replica. Explicitly, using the signed phases of Eq.~\eqref{eq:sm_gamma_signed} and the block coordinates introduced in Eqs.~\eqref{eq:sm_two_replica_block_matrix} and~\eqref{eq:sm_two_replica_block_entries}, the reduced action is
\begin{equation}
    \mathcal F_p(\mathsf{Q},\lambda)=\frac{1}{4}\sum_{\sigma,\sigma'=A,B}\sum_{r,r'\in\mathcal I_p}\Gamma_r\Gamma_{r'}\bigl(\mathsf{Q}_{\sigma r,\sigma' r'}\bigr)^2+\ln \mathcal Z_p(\mathsf{Q},\lambda),
    \label{eq:sm_two_replica_action}
\end{equation}
with
\begin{equation}
    \mathcal Z_p(\mathsf{Q},\lambda)=\sum_{\boldsymbol a^A,\boldsymbol a^B}f_{\boldsymbol\beta}\bigl(\boldsymbol a^A\bigr)f_{\boldsymbol\beta}\bigl(\boldsymbol a^B\bigr)\exp\left[\lambda a_0^A a_0^B-\frac{1}{2}\sum_{\sigma,\sigma'=A,B}\sum_{r,r'\in\mathcal I_p}\Gamma_r\Gamma_{r'}\mathsf{Q}_{\sigma r,\sigma' r'}a_r^\sigma a_{r'}^{\sigma'}\right].
    \label{eq:sm_two_replica_local_trace}
\end{equation}
The sums over $\boldsymbol a^A$ and $\boldsymbol a^B$ run over two one-site QAOA histories. The terminal variables $a_0^A$ and $a_0^B$ enter through the mixer weights and the explicit source $\lambda$. Since $\Gamma_0=0$, terms with $r=0$ or $r'=0$ do not contribute to the SK disorder quadratic form. The saddle equations following from Eq.~\eqref{eq:sm_two_replica_action} are
\begin{equation}
    \mathsf{Q}_{\sigma r,\sigma' r'}=\left\langle a_r^\sigma a_{r'}^{\sigma'}\right\rangle_{\mathsf{Q},\lambda},
    \label{eq:sm_two_replica_saddle_equations}
\end{equation}
for entries with $\Gamma_r\Gamma_{r'}\neq0$. The expectation is taken with respect to the normalized single-site measure defined by Eq.~\eqref{eq:sm_two_replica_local_trace}. The physical overlap selected by the source is obtained by differentiating the action with respect to the output-overlap source $\lambda$. In particular, by the envelope theorem
\begin{equation}
    q_p(\lambda)=\partial_\lambda\phi_p(\lambda)=\left\langle a_0^A a_0^B\right\rangle_{\mathsf{Q}_\star(\lambda),\lambda}.
    \label{eq:sm_terminal_overlap_from_saddle}
\end{equation}
For $\lambda>0$, the source induces a nonzero cross-sample block $\mathsf R_\star(\lambda)$. The continuation used to select this saddle and construct $V_p(q)$ is described next.

\subsection{Continuation in the overlap source}
\label{subsec:sm_source_connected_continuation}

The continuation used in this work is initialized at the uncoupled two-physical-replica saddle. At $\lambda=0$, the two physical replicas are statistically independent, and the two-physical-replica saddle has the product form
\begin{equation}
    \mathsf{Q}_\star(0)=
    \begin{pmatrix}
        \mathsf{G}_\star & 0\\
        0 & \mathsf{G}_\star
    \end{pmatrix}.
    \label{eq:sm_zero_source_two_replica_saddle}
\end{equation}
Here $\mathsf{G}_\star$ is the one-sample SK-QAOA history saddle at the chosen QAOA depth and angles. For $\lambda>0$, the continued solution $\mathsf{Q}_\star(\lambda)$ is obtained by continuing this product saddle: the stationarity equations in Eq.~\eqref{eq:sm_two_replica_saddle_equations} are solved at increasing values of $\lambda$, using the previously converged stationary point as the initial condition for the next source value. This prescription selects the solution continuously connected to the unbiased two-sample QAOA distribution at $\lambda=0$. For $\lambda>0$, the continued saddle has the block form of Eq.~\eqref{eq:sm_two_replica_block_matrix}, with the product solution at $\lambda=0$ in Eq.~\eqref{eq:sm_zero_source_two_replica_saddle} deformed by a source-induced cross-sample block $\mathsf{R}_\star(\lambda)\neq0$. At each point on this continued solution, the quenched free-entropy density $\phi_p(\lambda)$ is evaluated from Eq.~\eqref{eq:sm_phi_two_replica_branch}, and the physical overlap $q_p(\lambda)$ is computed from Eq.~\eqref{eq:sm_terminal_overlap_from_saddle}. The overlap potential $V_p(q)$ defined in Sec.~\ref{subsec:sm_field_selected_potential} is then obtained parametrically from Eq.~\eqref{eq:sm_parametric_potential}, with the overlap coordinate given by $q=q_p(\lambda)$, and is shown in Fig.~2a of the main text. The next subsection uses multistart solutions of the same fixed-source saddle equations to search for disconnected branches relevant to the fixed-overlap problem.

\subsection{Multistart search for additional saddle branches}
\label{subsec:sm_constrained_overlap_branch_search}

\begin{figure*}[!t]
\centering
    \includegraphics[width=1\textwidth]{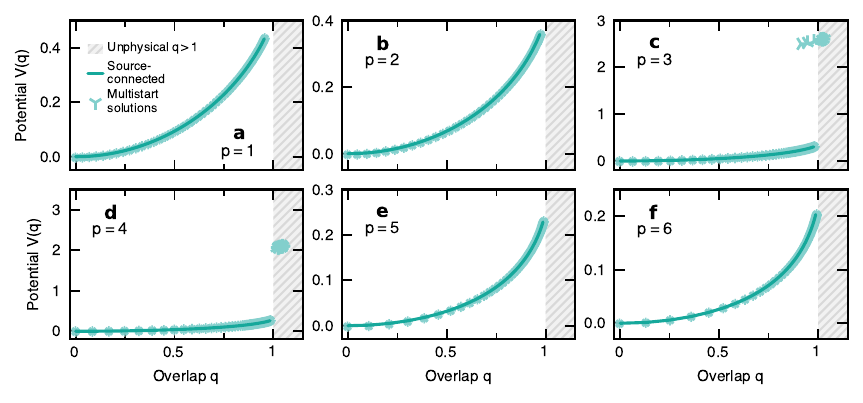}
    \caption{Potential $V_p(q)$ for (a) $p=1$, (b) $p=2$, (c) $p=3$, (d) $p=4$, (e) $p=5$, (f) $p=6$. Solid curves denote the continued solution, while markers denote converged multistart searches of the same saddle equations, initialized from random $A\leftrightarrow B$-symmetric starting points. The hatched region marks the unphysical range $q>1$.}
    \label{fig:constrained_saddle_potentials}
\end{figure*}

We test whether continuation in $\lambda$ misses stationary branches relevant to the fixed-overlap Franz--Parisi construction. In this construction, the output overlap $q_{AB}=q$ is imposed directly, with $\lambda$ acting as its Lagrange multiplier~\cite{MezardParisiVirasoro1987,CastellaniCavagna2005,FranzParisi1995,FranzParisi1997,FranzParisi1998EffectivePotential}. The search is restricted to real two-replica stationary points preserving the $A\leftrightarrow B$ symmetry of the physical samples and retains the same ansatz for the auxiliary replicas. The stability of that ansatz is examined separately in Sec.~\ref{sec:sm_auxiliary_replica_stability}.

Let $\boldsymbol Q$ denote the vector of independent variables in this reduced real, $A\leftrightarrow B$-symmetric two-replica problem, and let $\mathcal F_p^{\rm red}(\boldsymbol Q,\lambda)$ denote the corresponding restriction of the reduced two-sample action in Eq.~\eqref{eq:sm_two_replica_action}. At fixed source $\lambda$, the continued solution solves $\nabla_{\boldsymbol{Q}}\mathcal F_p^{\rm red}(\boldsymbol{Q},\lambda)=\boldsymbol{0}$ and is initialized at the uncoupled saddle when $\lambda=0$. The fixed-overlap stationary problem can instead be written with $\lambda$ as a Lagrange multiplier enforcing the output overlap
\begin{equation}
    \nabla_{\boldsymbol{Q}}\mathcal F_p^{\rm red}(\boldsymbol{Q},\lambda)=\boldsymbol{0},\qquad\partial_\lambda\mathcal F_p^{\rm red}(\boldsymbol{Q},\lambda)=q.
    \label{eq:sm_constrained_overlap_stationary_system}
\end{equation}
Any fixed-overlap stationary point is therefore also a fixed-source stationary point at its conjugate value of $\lambda$. We consequently search for additional branches by solving the fixed-source equations from many independent initial conditions and recording the induced overlap. The distinction lies in branch selection: continuation follows the solution connected to $\lambda=0$, whereas the multistart search can reveal disconnected stationary solutions.

In the numerical search, both the random initial conditions and the fixed-point iteration are restricted to the same $A\leftrightarrow B$-symmetric sector, matching the symmetry of the output-overlap factor $e^{\lambda\sum_j a^A_{0,j}a^B_{0,j}}$ in Eq.~\eqref{eq:sm_exact_two_replica_source_history}. For each depth $p$ and source value $\lambda$, we first compute the continued reference solution by continuation from $\lambda=0$. We then perform $\approx 100$ independent random starts at the same fixed value of $\lambda$. Each independent component of the reduced real overlap vector is drawn independently from $\operatorname{Unif}[-1,1]$, the natural range of an overlap variable. Starting from these initial conditions, we solve the fixed-source saddle equations and record, for each converged solution $\boldsymbol{Q}_{\rm sol}(\lambda)$, the induced overlap $q_{\rm sol}=\partial_\lambda \mathcal F_p^{\rm red}(\boldsymbol{Q}_{\rm sol},\lambda)$ and the potential value
\begin{equation}
    V_{\rm sol}=\lambda q_{\rm sol}-\mathcal F_p^{\rm red}(\boldsymbol{Q}_{\rm sol},\lambda)+\mathcal F_p^{\rm red}(\boldsymbol{Q}_0,0).
    \label{eq:sm_constrained_saddle_source_dual_value}
\end{equation}
Here $\boldsymbol{Q}_0$ denotes the $\lambda=0$-connected solution in the same reduced coordinate system. The final term is the same subtraction used in Eq.~\eqref{eq:sm_phi_two_replica_branch} which enforces $\phi_p(0)=0$ and places $V_{\rm sol}$ on the same additive scale as the parametric potential $V_p(q_p(\lambda))$ in Eq.~\eqref{eq:sm_parametric_potential}.

The result is shown in Fig.~\ref{fig:constrained_saddle_potentials}. The solid curve is the continued solution, while the markers show the converged solutions obtained from random $A\leftrightarrow B$-symmetric starts. In most cases the independent starts collapse onto the continued solution, so the markers are superimposed on the solid curve. Thus, within the restricted symmetry sector, the multistart search
reveals no additional lower-potential branch in the physical overlap range $|q|\le1$. At larger depths, especially in Figs.~\ref{fig:constrained_saddle_potentials}c and~\ref{fig:constrained_saddle_potentials}d, the search does find additional converged stationary solutions that are numerically distinct from the continued solution. However, these solutions are not competitive because they lie at substantially larger potential values than the continued solution at comparable overlap $q$, and several lie in the unphysical region $q>1$, outside the allowed range of Ising overlaps.

Within the sampled depths and source ranges, this supports the continuation construction used in the main text. At depth $p=1$, this numerical conclusion is consistent with the exact fixed-overlap analysis of Sec.~\ref{subsec:sm_depth_one_fixed_overlap}, where the continued real branch is shown to remain nonsingular throughout the physical overlap range. The separate question of whether correlations between auxiliary replicas destabilize the ansatz used here is addressed in Sec.~\ref{sec:sm_auxiliary_replica_stability}.

\section{Overlap response and correlation identity}
\label{sec:sm_overlap_response}

The previous section gives the source response $q_p(\lambda)$ and the overlap potential $V_p(q)$ in the thermodynamic limit. We now analyze the local response near $\lambda=0$ and connect the curvature of $V_p$ to two-spin correlation functions of the QAOA distribution.

\subsection{Local Taylor expansion}
\label{subsec:sm_local_taylor_expansion}

We begin with the small-$\lambda$ expansion of the continued solution. By the spin-flip symmetry established above, the two-replica overlap distribution is even at $\lambda=0$. Consequently, $\phi_p(\lambda)=\phi_p(-\lambda)$ and $q_p(\lambda)=-q_p(-\lambda)$. Assuming the continued solution is analytic at the origin, and using the normalization $\phi_p(0)=0$, we expand
\begin{equation}
    \phi_p(\lambda)=\frac{\chi_p}{2}\lambda^2+\frac{a_{4,p}}{4}\lambda^4+\frac{a_{6,p}}{6}\lambda^6+O(\lambda^8).
    \label{eq:sm_phi_local_expansion}
\end{equation}
The coefficient $\chi_p$ is the overlap susceptibility $\chi_p=\phi_p''(0)$. The normalization of the quartic and sixth-order terms is chosen so that differentiating Eq.~\eqref{eq:sm_phi_local_expansion} gives the simple response expansion
\begin{equation}
    q_p(\lambda)=\chi_p\lambda+a_{4,p}\lambda^3+a_{6,p}\lambda^5+O(\lambda^7).
    \label{eq:sm_q_local_expansion}
\end{equation}
Inverting this relation and substituting into the Legendre transform gives the local expansion of the overlap potential
\begin{equation}
    V_p(q)=\frac{q^2}{2\chi_p}-\frac{a_{4,p}}{4\chi_p^4}q^4+\left(\frac{a_{4,p}^2}{2\chi_p^7}-\frac{a_{6,p}}{6\chi_p^6}\right)q^6+O(q^8).
    \label{eq:sm_potential_local_expansion}
\end{equation}
In particular $V_p''(0)=\chi_p^{-1}$.

\subsection{Overlap susceptibility as squared two-spin correlation density}
\label{subsec:sm_susceptibility_correlation_identity}

The susceptibility $\chi_p$ can be written as a two-spin correlation observable of the QAOA distribution. For a fixed disorder instance and finite $N$, the extensive two-replica overlap is $Nq_{AB}=\sum_{j=1}^N z_j^A z_j^B$. Taking $\lambda=0$ in Eq.~\eqref{eq:sm_source_second_derivative} gives the finite-size variance identity
\begin{equation}
    \frac{1}{N^2}\left.\frac{\partial^2}{\partial\lambda^2}\ln Z^{(2)}_{J,p}(\lambda)\right|_{\lambda=0}=\left\langle q_{AB}^2\right\rangle^{\otimes 2}_{J,p}-\left(\left\langle q_{AB}\right\rangle^{\otimes 2}_{J,p}\right)^2.
    \label{eq:sm_finite_size_overlap_variance}
\end{equation}
The first moment vanishes by the spin-flip symmetry discussed above. Using the independence of the two physical replicas at $\lambda=0$ gives
\begin{equation}
    N^2\left\langle q_{AB}^2\right\rangle^{\otimes 2}_{J,p}=\sum_{j,k=1}^N\left\langle z_j^A z_k^A\right\rangle_{J,p}\left\langle z_j^B z_k^B\right\rangle_{J,p}=\sum_{j,k=1}^N\left\langle z_j z_k\right\rangle_{J,p}^2.
    \label{eq:sm_overlap_variance_correlation_square}
\end{equation}
Taking the disorder average, dividing by $N$, and sending $N\to\infty$ gives
\begin{equation}
    \chi_p=\lim_{N\to\infty}\frac{1}{N}\mathbb E_J\left[\sum_{j,k=1}^N\left\langle\hat Z_j\hat Z_k\right\rangle_{J,p}^2\right].
    \label{eq:sm_chi_correlation_power_identity}
\end{equation}
The local overlap susceptibility is therefore the disorder-averaged sum of squared two-spin correlations per spin.

\section{Energy-tilted overlap geometry}
\label{sec:sm_energy_tilted_overlap}

We now turn from typical QAOA outputs to atypical samples in the low-energy tail. We ask how their overlap geometry evolves as progressively lower energies are selected at fixed $p$.

\subsection{Energy conditioning, normalized tilt, and overlap source}
\label{subsec:sm_tilted_source}

For a fixed disorder realization $J$, we define the energy density of a configuration by $e_J(\boldsymbol z)=C_J(\boldsymbol z)/N$. The corresponding energy-density distribution induced by a QAOA sample is
\begin{equation}
    \rho_{J,p,N}(e)=\sum_{\boldsymbol z}P_{J,p}(\boldsymbol z)\delta\bigl[e-e_J(\boldsymbol z)\bigr],
    \label{eq:sm_qaoa_energy_distribution}
\end{equation}
where $\delta$ denotes a delta mass on the finite-instance energy spectrum, or equivalently the limit of an infinitesimally narrow energy shell around $e$. Integrating $\rho_{J,p,N}(e)$ over its low-energy tail gives the probability of obtaining better-than-typical energies when sampling from the QAOA distribution. To isolate QAOA outputs at a chosen energy density, we condition the distribution on the corresponding energy shell
\begin{equation}
    P_{J,p}(\boldsymbol z\mid e)=\frac{P_{J,p}(\boldsymbol z)\delta\bigl[e-e_J(\boldsymbol z)\bigr]}{\rho_{J,p,N}(e)}.
    \label{eq:sm_energy_conditioned_qaoa_distribution}
\end{equation}
Within this shell, configurations retain the relative probabilities assigned by the QAOA. The conditioned distribution therefore describes QAOA outputs at fixed energy density $e$ and is the natural ensemble for probing their overlap geometry at that energy. To access these fixed-energy ensembles without imposing a hard energy constraint, we introduce a field $\eta$ conjugate to the energy. It reweights the QAOA energy distribution by the exponential factor $e^{-N\eta e}$, with normalization given by its Laplace transform
\begin{equation}
    Z_{J,p}^{(1)}(\eta)=\int de\rho_{J,p,N}(e)e^{-N\eta e}=\sum_{\boldsymbol z}P_{J,p}(\boldsymbol z)e^{-\eta C_J(\boldsymbol z)}.
    \label{eq:sm_tilted_energy_laplace_transform}
\end{equation}
For $\eta\geq0$, the normalized energy-tilted distribution is
\begin{equation}
    P_{J,p}^{(\eta)}(\boldsymbol z)=\frac{P_{J,p}(\boldsymbol z)e^{-\eta C_J(\boldsymbol z)}}{Z_{J,p}^{(1)}(\eta)},
    \label{eq:sm_energy_tilted_distribution}
\end{equation}
where positive $\eta$ transfers probability toward lower-energy shells without changing the QAOA circuit. The energy distribution induced by the tilted ensemble is
\begin{equation}
    \rho_{J,p,N}^{(\eta)}(e)=\sum_{\boldsymbol z}P_{J,p}^{(\eta)}(\boldsymbol z)\delta\bigl[e-e_J(\boldsymbol z)\bigr]=\frac{\rho_{J,p,N}(e)e^{-N\eta e}}{Z_{J,p}^{(1)}(\eta)}.
    \label{eq:sm_tilted_energy_distribution}
\end{equation}
At a fixed energy density, the exponential factor is common to all configurations and cancels upon conditioning. Consequently
\begin{equation}
    P_{J,p}^{(\eta)}(\boldsymbol z\mid e)=P_{J,p}(\boldsymbol z\mid e).
    \label{eq:sm_tilted_conditioned_equivalence}
\end{equation}
Thus the tilt changes only the probability assigned to each energy shell and not the distribution of QAOA outputs within it. Since $C_J=Ne$ is extensive, an $O(1)$ displacement of the energy density from its typical value is expected to have a probability exponentially small in $N$. At fixed QAOA depth, we therefore use the standard large-deviation description for typical disorder realizations~\cite{Touchette2009LargeDeviations}
\begin{equation}
    \rho_{J,p,N}(e)\sim e^{-N I_p(e)},
    \label{eq:sm_qaoa_energy_large_deviation_form}
\end{equation}
where $I_p(e)$ is a self-averaging rate function. The rate vanishes at the typical QAOA energy and gives the exponential probability cost per spin of sampling energy density $e$ from the untilted distribution. The conjugate thermodynamic quantity is the one-sample energy-tilted free-entropy density
\begin{equation}
    \varphi_p^{(1)}(\eta)=\lim_{N\to\infty}\frac{1}{N}\mathbb E_J\Bigl[\ln Z_{J,p}^{(1)}(\eta)\Bigr].
    \label{eq:sm_single_replica_tilted_density}
\end{equation}
Using Eq.~\eqref{eq:sm_tilted_energy_laplace_transform}, the large-deviation form in Eq.~\eqref{eq:sm_qaoa_energy_large_deviation_form} gives by Laplace's method
\begin{equation}
    \varphi_p^{(1)}(\eta)=\max_e\bigl[-\eta e-I_p(e)\bigr].
    \label{eq:sm_tilted_energy_duality}
\end{equation}
When the maximizing energy is unique and $\varphi_p^{(1)}(\eta)$ is differentiable, the tilted energy distribution in Eq.~\eqref{eq:sm_tilted_energy_distribution} concentrates at
\begin{equation}
    E_p(\eta)=-\partial_\eta\varphi_p^{(1)}(\eta),\qquad E_p(0)=E_p.
    \label{eq:sm_tilted_selected_energy}
\end{equation}
Combining this concentration with Eq.~\eqref{eq:sm_tilted_conditioned_equivalence} shows that, at the level of thermodynamic observables, the tilted ensemble is equivalent to the original QAOA distribution conditioned on energy density $E_p(\eta)$. Increasing $\eta$ therefore selects progressively deeper energy shells of the untilted QAOA output. The corresponding lower-tail rate follows from the conjugate Legendre transform
\begin{equation}
    I_p(e)=\max_{\eta\geq0}\left[-\eta e-\varphi_p^{(1)}(\eta)\right],
    \label{eq:sm_tilted_energy_rate_function}
\end{equation}
where $\eta\geq0$ selects energies below the untilted mean. Along a differentiable tilted branch,
\begin{equation}
    I_p\bigl[E_p(\eta)\bigr]=-\eta E_p(\eta)-\varphi_p^{(1)}(\eta).
    \label{eq:sm_tilted_energy_rate_parametric}
\end{equation}
The normalization at $\eta=0$ gives $I_p[E_p(0)]=0$. Convexity of $\varphi_p^{(1)}$ implies $dE_p/d\eta\leq0$ and
\begin{equation}
    \frac{d}{d\eta}I_p\bigl[E_p(\eta)\bigr]=-\eta\frac{dE_p}{d\eta}\geq0.
    \label{eq:sm_tilted_energy_rate_monotonicity}
\end{equation}
The rate therefore increases as the tilt selects lower energies. Along this lower-tail branch, energies below $E_p(\eta)$ are exponentially more costly, so the cumulative probability is dominated at exponential order by the threshold energy itself. For a typical disorder instance
\begin{equation}
    -\lim_{N\to\infty}\frac{1}{N}\ln\Pr_{P_{J,p}}\left[\frac{C_J(\boldsymbol z)}{N}\leq E_p(\eta)\right]=I_p\bigl[E_p(\eta)\bigr].
    \label{eq:sm_tilted_energy_lower_tail_probability}
\end{equation}
The tilted construction therefore selects an energy shell of the original QAOA output and quantifies its exponential probability cost under ordinary QAOA sampling.

The depth-zero limit recovers the standard statistical-mechanics correspondence. At $p=0$, the QAOA distribution is uniform, i.e., $P_{J,0}(\boldsymbol z)=2^{-N}$, and
\begin{equation}
    \rho_{J,0,N}(e)=2^{-N}\Omega_{J,N}(e),\qquad\Omega_{J,N}(e)=\sum_{\boldsymbol z}\delta\bigl[e-e_J(\boldsymbol z)\bigr].
    \label{eq:sm_p0_density_of_states_relation}
\end{equation}
The conditioned distribution in Eq.~\eqref{eq:sm_energy_conditioned_qaoa_distribution} is then uniform within the selected energy shell and coincides with the microcanonical SK ensemble. The tilted distribution in Eq.~\eqref{eq:sm_energy_tilted_distribution} becomes the canonical SK Gibbs distribution at inverse temperature $\eta$. Thus the $p=0$ limit recovers the standard canonical--microcanonical correspondence. For $p>0$, the conditioning-and-tilting construction is instead defined relative to the generally nonuniform QAOA distribution, whose relative weights are preserved within each energy shell. In this case, $\eta$ is an energy-conjugate bias and does not define a physical inverse temperature of the QAOA output. With each sample drawn from $P_{J,p}^{(\eta)}$, the tilted two-replica source is
\begin{equation}
    Z_{J,p}^{(2)}(\lambda,\eta)=\left\langle e^{\lambda Nq_{AB}}\right\rangle_{J,p,\eta}^{\otimes2}.
    \label{eq:sm_tilted_two_replica_source}
\end{equation}
In terms of the untilted QAOA distribution,
\begin{equation}
    Z_{J,p}^{(2)}(\lambda,\eta)=\frac{\left\langle e^{\lambda Nq_{AB}-\eta C_J(\boldsymbol z^A)-\eta C_J(\boldsymbol z^B)}\right\rangle_{J,p}^{\otimes2}}{\left[Z_{J,p}^{(1)}(\eta)\right]^2},
    \label{eq:sm_tilted_two_replica_source_normalized}
\end{equation}
where the denominator ensures $Z_{J,p}^{(2)}(0,\eta)=1$. The corresponding quenched free-entropy density is
\begin{equation}
    \phi_p(\lambda,\eta)=\lim_{N\to\infty}\frac{1}{N}\mathbb E_J\Bigl[\ln Z_{J,p}^{(2)}(\lambda,\eta)\Bigr],
    \label{eq:sm_tilted_source_density}
\end{equation}
with $\phi_p(0,\eta)=0$ and $\phi_p(\lambda,0)=\phi_p(\lambda)$. For the saddle construction, it is useful to separate the unnormalized tilted numerator
\begin{equation}
    \Phi_p^{(2)}(\lambda,\eta)=\lim_{N\to\infty}\frac{1}{N}\mathbb E_J\left[\ln\left\langle e^{\lambda Nq_{AB}-\eta C_J(\boldsymbol z^A)-\eta C_J(\boldsymbol z^B)}\right\rangle_{J,p}^{\otimes2}\right],
    \label{eq:sm_tilted_unnormalized_two_replica_density}
\end{equation}
from the one-sample normalization. Thus
\begin{equation}
    \phi_p(\lambda,\eta)=\Phi_p^{(2)}(\lambda,\eta)-2\varphi_p^{(1)}(\eta).
    \label{eq:sm_tilted_subtracted_density}
\end{equation}
The tilted overlap response and susceptibility are
\begin{equation}
    q_{p,\eta}(\lambda)=\partial_\lambda\phi_p(\lambda,\eta),\qquad\chi_{p,\eta}=\left.\partial_\lambda^2\phi_p(\lambda,\eta)\right|_{\lambda=0}.
    \label{eq:sm_tilted_overlap_response}
\end{equation}
At $\lambda=0$, the two physical replicas are independent tilted samples and, along a concentrating thermodynamic branch, both select the energy density $E_p(\eta)$. The susceptibility $\chi_{p,\eta}$ therefore probes the local overlap response of QAOA outputs conditioned on that energy density. When the source-overlap map is smooth, the corresponding potential is defined by the same Legendre relation
\begin{equation}
    V_{p,\eta}\bigl[q_{p,\eta}(\lambda)\bigr]=\lambda q_{p,\eta}(\lambda)-\phi_p(\lambda,\eta).
    \label{eq:sm_tilted_potential_parametric}
\end{equation}
Whenever the local source-overlap map is invertible at the origin, its curvature is $V_{p,\eta}''(0)=\chi_{p,\eta}^{-1}$.

\subsection{Terminal-slice deformation of the Keldysh saddle}
\label{subsec:sm_tilted_terminal_slice}

In the Keldysh representation of the unnormalized tilted numerator, the structural change is that the terminal history coordinate becomes active in the disorder term. In the untilted overlap source, the terminal spin $a_0$ appears only through $e^{\lambda a_0^A a_0^B}$, and the signed phase vector satisfies $\Gamma_0=0$. With energy tilting, each measured configuration receives the additional factor $e^{-\eta C_J(\boldsymbol z)}$. Since $a_0$ is the measured spin, this is equivalent to assigning the terminal slice the complex Keldysh coefficient
\begin{equation}
    \Gamma_r^{(\eta)}=\Gamma_r+i\eta\delta_{r,0}.
    \label{eq:sm_tilted_gamma_vector}
\end{equation}
Indeed, the terminal contribution to the disorder exponent is
$i\Gamma_0^{(\eta)}C_J(\boldsymbol a_0)=-\eta C_J(\boldsymbol a_0)$. Thus the set of active history coordinates expands from the nonterminal slices to the full history, including $r=0$. Using the same history weight $f_{\boldsymbol\beta}(\boldsymbol a)$ as in Sec.~\ref{sec:sm_keldysh_formalism}, the unnormalized numerator appearing in Eq.~\eqref{eq:sm_tilted_two_replica_source_normalized} can be written as
\begin{equation}
    \widetilde Z_{J,p}^{(2)}(\lambda,\eta)=\sum_{\{\boldsymbol a_j^A,\boldsymbol a_j^B\}_{j=1}^N}\prod_{j=1}^Nf_{\boldsymbol\beta}\bigl(\boldsymbol a_j^A\bigr)f_{\boldsymbol\beta}\bigl(\boldsymbol a_j^B\bigr)e^{\lambda\sum_{j=1}^N a_{0,j}^A a_{0,j}^B}\exp\left[\frac{i}{\sqrt N}\sum_{j<k}J_{jk}\sum_{\sigma=A,B}\sum_{r\in\mathcal I_p}\Gamma_r^{(\eta)}a_{r,j}^{\sigma}a_{r,k}^{\sigma}\right].
    \label{eq:sm_tilted_history_source}
\end{equation}
The normalized two-replica generating function in Eq.~\eqref{eq:sm_tilted_two_replica_source_normalized} is obtained from this numerator by division by the two one-sample tilted normalizations. Equivalently, at the level of the quenched free-entropy density, this gives the subtraction in Eq.~\eqref{eq:sm_tilted_subtracted_density}. After introducing auxiliary replicas and averaging the Gaussian disorder, the reduction of Eqs.~\eqref{eq:sm_history_overlap_site_average}--\eqref{eq:sm_disorder_in_terms_of_Qsite} applies with the replacement $\Gamma_r\to\Gamma_r^{(\eta)}$ from Eq.~\eqref{eq:sm_tilted_gamma_vector}. We use again the label $I=(\alpha,\sigma,r)$ and write $\Gamma_I^{(\eta)}=\Gamma_r^{(\eta)}$. Since $\Gamma_0^{(\eta)}\neq0$ for $\eta>0$, terminal slices become active in the disorder term: collective variables involving terminal indices now enter the SK quadratic term. After promoting the site averages to collective overlaps $\mathsf{Q}^{(n)}_{II'}$, the disorder-generated contribution to the replicated action is
\begin{equation}
    \exp\left[-\frac{N}{4}\sum_{I,I'}\Gamma_I^{(\eta)}\Gamma_{I'}^{(\eta)}\bigl(\mathsf{Q}^{(n)}_{II'}\bigr)^2\right],
    \label{eq:sm_tilted_disorder_average_Q}
\end{equation}
up to field-independent factors that are subexponential in $N$ and do not contribute to the thermodynamic free-entropy density. The energy-tilted saddle equations are therefore obtained from the untilted saddle equations in Eq.~\eqref{eq:sm_two_replica_saddle_equations} by the replacement $\Gamma_r\to\Gamma_r^{(\eta)}$ from Eq.~\eqref{eq:sm_tilted_gamma_vector}, together with the explicit overlap source $\lambda a_0^A a_0^B$. The same replacement in the one-sample problem defines the tilted saddle $\mathsf G_\eta$, and the selected energy density is evaluated from Eq.~\eqref{eq:sm_tilted_selected_energy} along this stationary branch. At $\lambda=0$, the two physical replicas are independent tilted samples, so the two-replica saddle factorizes as
\begin{equation}
    \mathsf{Q}_\star(0,\eta)=
    \begin{pmatrix}
        \mathsf{G}_\eta & 0\\
        0 & \mathsf{G}_\eta
    \end{pmatrix}.
    \label{eq:sm_tilted_product_saddle}
\end{equation}
Let $\mathcal F_p^{(2)}(\mathsf{Q},\lambda,\eta)$ denote the tilted counterpart of the reduced two-sample action $\mathcal F_p$ in Eq.~\eqref{eq:sm_two_replica_action}, evaluated for the unnormalized two-replica tilted numerator. Let $\mathsf Q_\star(\lambda,\eta)$ be its stationary point on the branch connected, at fixed $\eta$, to the product saddle in Eq.~\eqref{eq:sm_tilted_product_saddle}. Then
\begin{equation}
    \Phi_p^{(2)}(\lambda,\eta)=\mathcal F_p^{(2)}\bigl[\mathsf{Q}_\star(\lambda,\eta),\lambda,\eta\bigr].
    \label{eq:sm_tilted_unnormalized_stationary_density}
\end{equation}
The normalized energy-tilted free-entropy density is
\begin{equation}
    \phi_p(\lambda,\eta)=\mathcal F_p^{(2)}\bigl[\mathsf{Q}_\star(\lambda,\eta),\lambda,\eta\bigr]-2\varphi_p^{(1)}(\eta).
    \label{eq:sm_tilted_stationary_density}
\end{equation}
The overlap response $q_{p,\eta}(\lambda)$ is obtained by differentiating with respect to the output-overlap source. By the envelope theorem, the implicit derivative of $\mathsf{Q}_\star(\lambda,\eta)$ does not contribute, so
\begin{equation}
    q_{p,\eta}(\lambda)=\partial_\lambda \phi_p(\lambda,\eta)=\left\langle a_0^A a_0^B\right\rangle_{\mathsf{Q}_\star(\lambda,\eta),\lambda,\eta}.
    \label{eq:sm_tilted_terminal_overlap}
\end{equation}
The resulting depth-one terminal-slice deformation is worked out explicitly in Sec.~\ref{subsec:sm_depth_one_energy_tilted_response}.

\subsection{Tilted susceptibility and the low-energy-tail threshold}
\label{subsec:sm_tilted_susceptibility_interpretation}

\begin{figure}[!t]
    \centering
    \includegraphics[width=0.5\columnwidth]{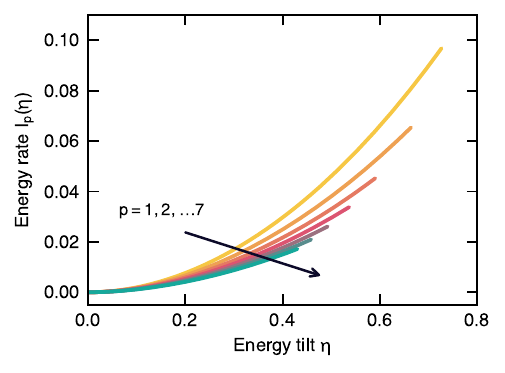}
    \caption{Energy rate $I_p\bigl(E_p(\eta)\bigr)$ from Eq.~\eqref{eq:sm_tilted_energy_rate_parametric} for $p=1,2,\ldots,7$, evaluated along the continued branch for $0\leq\eta\leq\eta_c(p)$. The endpoint at $\eta_c(p)$ gives the lower-tail rate $I_c(p)$ shown in Fig.~3c of the main text.}
    \label{fig:tilted_energy_rate}
\end{figure}

The endpoint $\eta_c(p)$ marks the loss of local stability of the continued auxiliary-replica-diagonal saddle. At $p=0$, this is the exact Gibbs transition into the Parisi spin-glass phase. The probability cost of reaching the corresponding energy from the untilted QAOA distribution follows directly from Eq.~\eqref{eq:sm_tilted_energy_lower_tail_probability}. We define
\begin{equation}
    I_c(p)=I_p\bigl[E_p(\eta_c(p))\bigr].
    \label{eq:sm_critical_lower_tail_rate}
\end{equation}
It gives the exponential cost per spin of sampling an energy density at or below $E_p(\eta_c(p))$ from the untilted QAOA distribution. Figure~\ref{fig:tilted_energy_rate} shows how the rate accumulates as the tilt selects progressively lower energies. Its endpoint $I_c(p)$ remains positive at every depth studied, so the corresponding tail remains exponentially suppressed at each fixed depth. As shown in Fig.~3c of the main text, this suppression rate decreases with depth and is compatible with $I_c(p)\sim p^{-1.2}$ over the largest three depths.

We now evaluate the local tilted susceptibility defined in Eq.~\eqref{eq:sm_tilted_overlap_response}. The finite-source tilted saddle is obtained by continuing the product solution in Eq.~\eqref{eq:sm_tilted_product_saddle} in $\lambda$, using the terminal-slice deformation described in Sec.~\ref{subsec:sm_tilted_terminal_slice}. The susceptibility is obtained by differentiating the saddle expression for $q_{p,\eta}(\lambda)$ in Eq.~\eqref{eq:sm_tilted_terminal_overlap} at $\lambda=0$. The energy tilt preserves the global spin-flip symmetry, so repeating the argument of Sec.~\ref{subsec:sm_susceptibility_correlation_identity} gives
\begin{equation} 
    \chi_{p,\eta}=\lim_{N\to\infty}\frac{1}{N}\mathbb E_J\sum_{j,k=1}^N\left\langle \hat Z_j\hat Z_k\right\rangle_{J,p,\eta}^2.
    \label{eq:sm_tilted_chi_correlation_identity}
\end{equation}

\begin{figure}[!t]
    \centering
    \includegraphics[width=0.85\columnwidth]{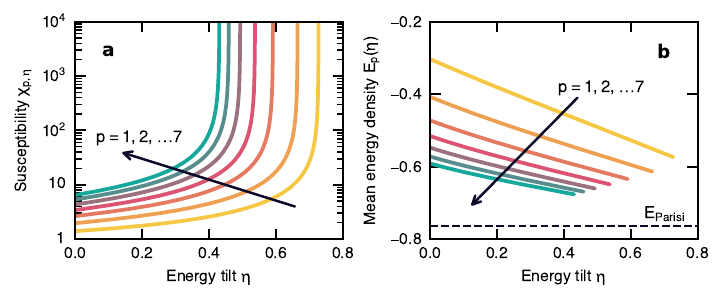}
    \caption{(a) Overlap susceptibility $\chi_{p,\eta}$ versus the energy tilt $\eta$ for $p=1,2,\ldots,7$. (b) Thermodynamic mean energy density $E_p(\eta)$ versus $\eta$ for $p=1,2,\ldots,7$. The dashed line marks $E_{\mathrm{Parisi}}=-0.7631667265\ldots$. The same data are replotted against the relative energy distance in Fig.~3 of the main text.}
    \label{fig:energy_vs_eta}
\end{figure}

Figure~\ref{fig:energy_vs_eta} shows $\chi_{p,\eta}$ and the corresponding mean energy density $E_p(\eta)$ as functions of the tilt. The energy density is evaluated on the corresponding one-sample tilted saddle and is shown over the range $\eta<\eta_c(p)$ for which the ansatz remains locally stable. For $p=1$, the energy curve is independently reproduced by the exact depth-one saddle calculation of Sec.~\ref{subsec:sm_depth_one_energy_tilted_response}. Eliminating $\eta$ between these quantities gives the relative energy coordinate $1-|E_p(\eta)/E_{\mathrm{Parisi}}|$ used in Fig.~3 of the main text. At $\eta=0$, the tilted distribution reduces to the untilted QAOA distribution, and $\chi_{p,0}=\chi_p$.

\subsection{Classical Gibbs benchmark at zero QAOA depth}
\label{subsec:sm_p0_gibbs_benchmark}

We now evaluate explicitly the depth-zero Gibbs benchmark introduced in Sec.~\ref{subsec:sm_tilted_source}. The tilted distribution is
\begin{equation}
    P_{J,0}^{(\eta)}(\boldsymbol z)=\frac{e^{-\eta C_J(\boldsymbol z)}}{\sum_{\boldsymbol z'}e^{-\eta C_J(\boldsymbol z')}},
    \label{eq:sm_p0_gibbs_distribution}
\end{equation}
which is the classical SK Gibbs distribution at inverse temperature $\eta$~\cite{SherringtonKirkpatrick1975,BinderYoung1986,MezardParisiVirasoro1987}. This limit provides an exact benchmark of the energy-tilt normalization, the overlap susceptibility, and the auxiliary-replica stability diagnostic. In the Keldysh construction, no forward or backward circuit slices remain at $p=0$. The terminal spin is the only active coordinate once the energy tilt is introduced, with $\Gamma_0^{(\eta)}=i\eta$ from Eq.~\eqref{eq:sm_tilted_gamma_vector}. The auxiliary-replica-diagonal ansatz in Eq.~\eqref{eq:sm_auxiliary_replica_diagonal_branch} then reduces to
the paramagnetic equilibrium saddle
\begin{equation}
    \mathsf Q_{\alpha\alpha'}=\delta_{\alpha,\alpha'}.
    \label{eq:sm_p0_paramagnetic_auxiliary_saddle}
\end{equation}
The diagonal collective variables satisfy $\mathsf Q_{\alpha\alpha}=1$, while the overlaps between distinct auxiliary replicas vanish. For $n$ auxiliary replicas, inserting this saddle into the one-sample replicated action gives a quadratic contribution $-n\eta^2/4$ and a one-site contribution $n\eta^2/2$. Taking the replica limit $n\to0$ therefore gives the normalized one-sample free-entropy density
\begin{equation}
    \varphi_0^{(1)}(\eta)=\frac{\eta^2}{4},\qquad E_0(\eta)=-\partial_\eta\varphi_0^{(1)}(\eta)=-\frac{\eta}{2},
    \label{eq:sm_p0_free_entropy_energy}
\end{equation}
on the paramagnetic branch for $0\leq\eta\leq1$. The normalization $P_{J,0}(\boldsymbol z)=2^{-N}$ removes the additive $\ln 2$ from the conventional SK Gibbs free entropy. At $\eta=1$, the selected energy is $E_0(1)=-1/2$, equal to the known SK transition energy density~\cite{SherringtonKirkpatrick1975,BinderYoung1986,MezardParisiVirasoro1987}. To obtain the overlap response, we next consider two physical replicas. Within one auxiliary replica, the reduced collective matrix contains only their terminal spins and takes the form
\begin{equation}
    \mathsf Q=
    \begin{pmatrix}
        1 & q\\
        q & 1
    \end{pmatrix}.
    \label{eq:sm_p0_two_replica_matrix}
\end{equation}
Substituting this matrix and $\Gamma_0^{(\eta)}=i\eta$ into the tilted counterpart of Eq.~\eqref{eq:sm_two_replica_action}, followed by the normalization subtraction in Eq.~\eqref{eq:sm_tilted_stationary_density}, gives
\begin{equation}
    \phi_0\bigl(\lambda,\eta\bigr)=-\frac{\eta^2}{2}q^2_{0,\eta}(\lambda)+\ln\cosh\Bigl[\lambda+\eta^2q_{0,\eta}(\lambda)\Bigr].
    \label{eq:sm_p0_two_replica_free_entropy}
\end{equation}
Stationarity with respect to the physical overlap gives
\begin{equation}
    q_{0,\eta}(\lambda)=\tanh\bigl[\lambda+\eta^2q_{0,\eta}(\lambda)\bigr].
    \label{eq:sm_p0_overlap_saddle}
\end{equation}
For $\eta<1$, the zero-source solution is $q_{0,\eta}(0)=0$. Differentiating Eq.~\eqref{eq:sm_p0_overlap_saddle} with respect to $\lambda$ at the origin yields
\begin{equation}
    \chi_{0,\eta}=\left.\partial_\lambda q_{0,\eta}(\lambda)\right|_{\lambda=0}=\frac{1}{1-\eta^2},\qquad V_{0,\eta}''(0)=1-\eta^2.
    \label{eq:sm_p0_tilted_susceptibility}
\end{equation}
Through the correlation identity in Eq.~\eqref{eq:sm_tilted_chi_correlation_identity}, this is the standard equilibrium spin-glass susceptibility of the paramagnetic SK model. It diverges at the known transition point $\eta_c(0)=1$. The corresponding auxiliary-replica stability spectrum is derived independently in Sec.~\ref{subsec:sm_p0_replicon_spectrum}, where the stability gap closes at the same point. The depth-zero construction therefore reproduces the classical SK transition in both the physical overlap response and the stability of the auxiliary-replica-diagonal saddle. The large-deviation construction of Sec.~\ref{subsec:sm_tilted_source} further gives, along the paramagnetic branch
\begin{equation}
    I_0\bigl(E_0(\eta)\bigr)=-\eta E_0(\eta)-\varphi_0^{(1)}(\eta)=\frac{\eta^2}{4}, \qquad 0\leq\eta\leq1.
    \label{eq:sm_p0_energy_rate}
\end{equation}
At the SK transition, $\eta_c(0)=1$ and $E_0(\eta_c(0))=-1/2$, so the critical energy lies below the untilted mean $E_0(0)=0$, i.e., $\Delta E_c(0)=1/2$. Its lower-tail rate is $I_c(0)=I_0(E_0(\eta_c(0)))=1/4$, showing that this energy is exponentially rare under the untilted uniform distribution but typical in the critical Gibbs distribution. For $\eta>1$, the equilibrium Gibbs distribution enters the Parisi replica-symmetry-breaking phase, described by a nontrivial overlap structure between replicas~\cite{Parisi1979,MezardParisiVirasoro1987,BinderYoung1986,CastellaniCavagna2005}. The depth-zero benchmark therefore shows that the susceptibility divergence marks the boundary of the auxiliary-replica-diagonal description and the onset of the equilibrium spin-glass overlap structure. It provides the reference for the depth-one nonlinear continuation constructed in Sec.~\ref{sec:sm_p1_tilted_parisi_transition}.

\section{Depth-one analytic solution}
\label{sec:sm_depth_one_solution}

At depth one, both the untilted and energy-tilted thermodynamic problems reduce to finite-dimensional saddles. For the untilted distribution, the two-sample saddle contains four fields, yielding an exact overlap susceptibility, an independent correlation-function check, and a fixed-overlap branch analysis. With energy tilting, the one-sample saddle contains three fields and the overlap response reduces to a finite $9\times9$ operator, allowing the critical tilt and its energetic and large-deviation benchmarks to be determined explicitly.

At $p=1$, each one-site Keldysh history contains one forward spin, one terminal measured spin, and one backward spin. We denote the forward and backward history spins by $u$ and $v$, respectively, so that in the notation of Sec.~\ref{sec:sm_keldysh_formalism} one has $u=a_{-1}$, $z=a_0$, and $v=a_1$. The depth-one state is
\begin{equation}
    |\Psi_{J,1}\rangle=e^{-i\beta \hat B}e^{-i\gamma \hat C_J}|+\rangle^{\otimes N},
    \label{eq:sm_p1_state}
\end{equation}
where the cost and mixer operators, $\hat C_J$ and $\hat B$, are defined in Eqs.~\eqref{eq:sm_quantum_cost} and~\eqref{eq:sm_mixer}, respectively. Throughout the $p=1$ section, we use $\gamma\equiv\gamma_1$ and $\beta\equiv\beta_1$.

\subsection{Terminal-spin reduction at depth one}
\label{subsec:sm_p1_terminal_source}

The mixer matrix element is
\begin{equation}
    M_\beta(z,u)=\langle z|e^{-i\beta \hat X}|u\rangle=\cos\beta\delta_{z,u}-i\sin\beta\delta_{z,-u}.
    \label{eq:sm_p1_mixer_element}
\end{equation}
Inserting a computational-basis resolution of the identity $\hat{\mathbb{1}}_{2^N}=\sum_{\boldsymbol z}|\boldsymbol z\rangle\langle\boldsymbol z|$ before the mixer gives
\begin{equation}
    \langle \boldsymbol z|\Psi_{J,1}\rangle=2^{-N/2}\sum_{\boldsymbol u}e^{-i\gamma C_J(\boldsymbol u)}\prod_{j=1}^N M_\beta(z_j,u_j),
    \label{eq:sm_p1_backprop_amplitude}
\end{equation}
and therefore
\begin{equation}
    P_{J,1}(\boldsymbol z)=2^{-N}\sum_{\boldsymbol u,\boldsymbol v}e^{-i\gamma C_J(\boldsymbol u)+i\gamma C_J(\boldsymbol v)}\prod_{j=1}^NM_\beta(z_j,u_j)M^*_\beta(z_j,v_j).
    \label{eq:sm_p1_probability_uv}
\end{equation}
We now insert Eq.~\eqref{eq:sm_p1_probability_uv} into the two-replica source of Eq.~\eqref{eq:sm_two_replica_source}. For each physical replica $\sigma=A,B$, we introduce forward and backward histories $\boldsymbol u^\sigma$ and $\boldsymbol v^\sigma$. The measured spins $\boldsymbol z^A$ and $\boldsymbol z^B$ only appear locally, so they can be summed out site by site. This gives
\begin{equation}
    Z^{(2)}_{J,1}(\lambda)=2^{-2N}\sum_{\boldsymbol u^A,\boldsymbol v^A,\boldsymbol u^B,\boldsymbol v^B}e^{-i\gamma C_J(\boldsymbol u^A)+i\gamma C_J(\boldsymbol v^A)-i\gamma C_J(\boldsymbol u^B)+i\gamma C_J(\boldsymbol v^B)}\prod_{j=1}^NK_\lambda\bigl(u_j^A,v_j^A,u_j^B,v_j^B\bigr),
    \label{eq:sm_p1_source_after_terminal_sum}
\end{equation}
where the one-site terminal kernel is
\begin{equation}
    K_\lambda\bigl(u^A,v^A,u^B,v^B\bigr)=\cosh\lambda A_0\bigl(u^A,v^A\bigr)A_0\bigl(u^B,v^B\bigr)+\sinh\lambda A_1\bigl(u^A,v^A\bigr)A_1\bigl(u^B,v^B\bigr),
    \label{eq:sm_p1_terminal_kernel}
\end{equation}
with
\begin{equation}
    A_0(u,v)=\sum_{z=\pm1}M_\beta(z,u)M^*_\beta(z,v)=\delta_{u,v},
    \label{eq:sm_p1_A0}
\end{equation}
and
\begin{equation}
    A_1(u,v)=\sum_{z=\pm1}zM_\beta(z,u)M^*_\beta(z,v)=\frac{u+v}{2}\cos(2\beta)+i\frac{u-v}{2}\sin(2\beta).
    \label{eq:sm_p1_A1}
\end{equation}
Thus the terminal sum separates into two contributions: the $A_0$ contribution is present at $\lambda=0$ and enforces forward--backward consistency, while the $A_1$ contribution carries the response to the overlap source.

\subsection{Four-field depth-one saddle}
\label{subsec:sm_p1_four_field_saddle}

We now apply the same SK disorder average and collective-overlap reduction as in Sec.~\ref{sec:sm_keldysh_formalism}. On the solution connected to $\lambda=0$, the depth-one saddle reduces to four fields
\begin{equation}
    \mathsf{Q}_{u^A v^A}=\mathsf{Q}_{u^B v^B}\equiv \rho,\quad \mathsf{Q}_{u^A v^B}=\mathsf{Q}_{v^A u^B}\equiv \sigma,\quad \mathsf{Q}_{u^A u^B}\equiv m+i\nu,\quad \mathsf{Q}_{v^A v^B}\equiv m-i\nu.
    \label{eq:sm_p1_four_fields}
\end{equation}
Here $\rho$ is the forward--backward history overlap within either physical sample. The fields $\sigma$, $m$, and $\nu$ parameterize source-induced correlations between the two physical samples. At $\lambda=0$, the two physical replicas factorize and the continued saddle is
\begin{equation}
    \rho=1,\quad\sigma=0,\quad m=0,\quad\nu=0.
    \label{eq:sm_p1_zero_source_saddle}
\end{equation}
At optimal angles $\gamma=1/2$ and $\beta=\pi/8$~\cite{FarhiGoldstoneGutmannZhou2022,BassoFarhiMarwahaVillalongaZhou2022}, the local trace entering the saddle can be written in terms of two factors
\begin{equation}
    \mathcal Z_0(\rho,\sigma,m,\nu)=4e^{\rho/2}\cosh\left(\frac{\sigma-m}{2}\right),
    \label{eq:sm_p1_Z0}
\end{equation}
and
\begin{equation}
    \mathcal Z_1(\rho,\sigma,m,\nu)=4\left[\frac{1}{2}e^{\rho/2}\sinh\left(\frac{\sigma-m}{2}\right)+\frac{1}{2}e^{-\rho/2}\sinh\left(\frac{\sigma+m}{2}\right)+\sin\left(\frac{\nu}{2}\right)\right].
    \label{eq:sm_p1_Z1}
\end{equation}
The normalized free-entropy density is obtained by evaluating
\begin{equation}
    F_1(\rho,\sigma,m,\nu,\lambda)=-2\ln 2-\frac{1}{4}-\frac{1}{4}(\rho^2+\sigma^2)+\frac{1}{4}(m^2-\nu^2)+\ln\Bigl[\cosh(\lambda)\mathcal Z_0(\rho,\sigma,m,\nu)+\sinh(\lambda)\mathcal Z_1(\rho,\sigma,m,\nu)\Bigr],
    \label{eq:sm_p1_four_field_functional}
\end{equation}
at the stationary point connected to Eq.~\eqref{eq:sm_p1_zero_source_saddle}. Here, $F_1$ is the reduced scalar functional of the four depth-one fields and the source with the constants chosen so that $F_1(1,0,0,0,0)=0$. If the continued stationary point is denoted by $(\rho_\star(\lambda),\sigma_\star(\lambda),m_\star(\lambda),\nu_\star(\lambda))$, then
\begin{equation}
    \phi_1(\lambda)=F_1\bigl(\rho_\star(\lambda),\sigma_\star(\lambda),m_\star(\lambda),\nu_\star(\lambda),\lambda\bigr).
    \label{eq:sm_p1_phi_stationary}
\end{equation}
The stationary fields satisfy $\partial_\rho F_1=\partial_\sigma F_1=\partial_m F_1=\partial_\nu F_1=0$. The physical overlap follows by differentiating with respect to the overlap source. By the envelope theorem, the implicit derivative of the stationary fields does not contribute, and
\begin{equation}
    q_1(\lambda)=\partial_\lambda\phi_1(\lambda)=\left.\frac{\sinh(\lambda)\mathcal Z_0+\cosh(\lambda)\mathcal Z_1}{\cosh(\lambda)\mathcal Z_0+\sinh(\lambda)\mathcal Z_1}\right|_\star.
    \label{eq:sm_p1_overlap_response}
\end{equation}
The subscript $\star$ means that the fields are evaluated at the continued stationary point.

\subsection{Small-source susceptibility from the saddle}
\label{subsec:sm_p1_saddle_susceptibility}

We now expand the saddle explicitly around $\lambda=0$. We write
\begin{equation}
    \rho_\star(\lambda)=1+\rho_1\lambda+O(\lambda^2),\quad\sigma_\star(\lambda)=\sigma_1\lambda+O(\lambda^2),\quad m_\star(\lambda)=m_1\lambda+O(\lambda^2),\quad\nu_\star(\lambda)=\nu_1\lambda+O(\lambda^2).
    \label{eq:sm_p1_field_linear_expansion}
\end{equation}
Substituting this ansatz into the four stationarity equations and keeping only the terms linear in $\lambda$ gives
\begin{equation}
    -\frac{1}{2}\rho_1=0,\quad-\frac{1}{4}\sigma_1-\frac{1}{4}m_1+\frac{1+e^{-1}}{4}=0,\quad-\frac{1}{4}\sigma_1+\frac{3}{4}m_1+\frac{e^{-1}-1}{4}=0,\quad-\frac{1}{2}\nu_1+\frac{e^{-1/2}}{2}=0.
    \label{eq:sm_p1_linearized_saddle_equations}
\end{equation}
Solving these four linear equations gives
\begin{equation}
    \rho_1=0,\quad\sigma_1=\frac{1}{2}+e^{-1},\quad m_1=\frac{1}{2},\quad\nu_1=e^{-1/2}.
    \label{eq:sm_p1_linearized_saddle_solution}
\end{equation}
To extract the susceptibility, we define the local bias $w=\mathcal Z_1/\mathcal Z_0$. Equation~\eqref{eq:sm_p1_overlap_response} can then be written as
\begin{equation}
    q_1(\lambda)=\left.\frac{\tanh(\lambda)+w}{1+w\tanh(\lambda)}\right|_\star.
    \label{eq:sm_p1_q_from_bias}
\end{equation}
At the $\lambda=0$ saddle, $w=0$. Its first derivatives at $(\rho,\sigma,m,\nu)=(1,0,0,0)$ are
\begin{equation}
    \partial_\rho w=0,\quad\partial_\sigma w=\frac{1+e^{-1}}{4},\quad\partial_m w=\frac{e^{-1}-1}{4},\quad\partial_\nu w=\frac{e^{-1/2}}{2}.
    \label{eq:sm_p1_bias_derivatives}
\end{equation}
Combining Eqs.~\eqref{eq:sm_p1_linearized_saddle_solution} and~\eqref{eq:sm_p1_bias_derivatives}, the linear response is
\begin{equation}
    \chi_1=\left.\frac{dq_1}{d\lambda}\right|_{\lambda=0}=1+\frac{1+e^{-1}}{4}\left(\frac{1}{2}+e^{-1}\right)+\frac{e^{-1}-1}{4}\frac{1}{2}+\frac{e^{-1/2}}{2}e^{-1/2}=1+\frac{1}{e}+\frac{1}{4e^2}=\left(1+\frac{1}{2e}\right)^2.
    \label{eq:sm_p1_chi_from_saddle}
\end{equation}
Therefore
\begin{equation}
    V_1''(0)=\chi_1^{-1}=\left(1+\frac{1}{2e}\right)^{-2}.
    \label{eq:sm_p1_stiffness_from_saddle}
\end{equation}

\subsection{Correlation-function check}
\label{subsec:sm_p1_correlation_check}

The same result can be obtained without introducing the overlap source by using the correlation identity in Eq.~\eqref{eq:sm_chi_correlation_power_identity}. This is a useful check because it gives a direct interpretation of the three terms in Eq.~\eqref{eq:sm_p1_chi_from_saddle}. Following Ref.~\cite{PhysRevA.109.012429}, the depth-one back-propagation calculation starts from
\begin{equation}
    e^{i\beta\hat B}\hat Z_j e^{-i\beta\hat B}=\cos(2\beta)\hat Z_j+\sin(2\beta)\hat Y_j.
    \label{eq:sm_p1_mixer_backprop_Z}
\end{equation}
Conjugating $\hat Y_j$ through the phase separator gives a trigonometric function of the neighboring $\hat Z$ operators. Expanding the resulting expression in the SK scaling $C_J=N^{-1/2}\sum_{j<k}J_{jk}z_j z_k$, and then setting $\gamma=1/2$ and $\beta=\pi/8$, gives for $j\neq k$ in the large-$N$ limit
\begin{equation}
    \bigl\langle \hat Z_j\hat Z_k\bigr\rangle_J\simeq-\frac{J_{jk}}{\sqrt{eN}}+\frac{1}{2eN}\sum_{l\neq j,k}J_{jl}J_{lk}
    \label{eq:sm_p1_two_point_large_N}
\end{equation}
The first term is the direct edge contribution. The second term is the contribution of two-step paths $j\to l\to k$. The two terms have zero disorder mean and are uncorrelated at leading order, so
\begin{equation}
    \mathbb E_J\left[\left\langle \hat Z_j\hat Z_k\right\rangle_{J}^2\right]=\frac{1}{eN}+\frac{1}{4Ne^2},\quad j\neq k.
    \label{eq:sm_p1_two_point_square}
\end{equation}
Using Eq.~\eqref{eq:sm_chi_correlation_power_identity}, the diagonal terms $j=k$ contribute $1$ because $\hat Z_j^2=\hat{\mathbb{1}}_2$, while the off-diagonal terms contribute $1/e+1/(4e^2)$. Thus
\begin{equation}
    \chi_1=1+\frac{1}{e}+\frac{1}{4e^2}=\left(1+\frac{1}{2e}\right)^2,
    \label{eq:sm_p1_chi_from_correlations}
\end{equation}
in agreement with the saddle result in Eq.~\eqref{eq:sm_p1_chi_from_saddle}.

\subsection{Source-eliminated form and local overlap potential}
\label{subsec:sm_p1_fixed_overlap_local}

The source can also be eliminated at fixed values of the four internal fields, giving a compact representation of the same source-defined overlap potential. Define
\begin{equation}
    d_-=\frac{\sigma-m}{2},\quad d_+=\frac{\sigma+m}{2}.
    \label{eq:sm_p1_dpm}
\end{equation}
The part of $F_1$ independent of the explicit source is
\begin{equation}
    B(\rho,\sigma,m,\nu)=-\frac{1}{4}-\frac{1}{4}(\rho^2+\sigma^2)+\frac{1}{4}(m^2-\nu^2)+\frac{\rho}{2}+\ln\cosh(d_-),
    \label{eq:sm_p1_B_function}
\end{equation}
and the local bias is
\begin{equation}
    w(\rho,\sigma,m,\nu)=\frac{\mathcal Z_1}{\mathcal Z_0}=\frac{1}{2}\tanh (d_-)+\frac{1}{2}e^{-\rho}\frac{\sinh (d_+)}{\cosh (d_-)}+e^{-\rho/2}\frac{\sin(\nu/2)}{\cosh (d_-)}.
    \label{eq:sm_p1_w_bias}
\end{equation}
For fixed $\rho$, $\sigma$, $m$, and $\nu$, Eq.~\eqref{eq:sm_p1_q_from_bias} gives
\begin{equation}
    \lambda(q,w)=\operatorname{arctanh}(q)-\operatorname{arctanh}(w).
    \label{eq:sm_p1_lambda_of_q}
\end{equation}
The corresponding binary relative-entropy cost is
\begin{equation}
    I_w(q)=\frac{1+q}{2}\ln\frac{1+q}{1+w}+\frac{1-q}{2}\ln\frac{1-q}{1-w}.
    \label{eq:sm_p1_binary_cost}
\end{equation}
Therefore the source-eliminated overlap potential is obtained by evaluating
\begin{equation}
    V_1(q)=I_{w_q}(q)-B(\rho_q,\sigma_q,m_q,\nu_q),
    \label{eq:sm_p1_fixed_overlap_potential}
\end{equation}
where $w_q=w(\rho_q,\sigma_q,m_q,\nu_q)$, and $(\rho_q,\sigma_q,m_q,\nu_q)$ is the stationary solution of $I_w(q)-B$ connected to $(\rho,\sigma,m,\nu,q)=(1,0,0,0,0)$. This branch specification is essential since Eq.~\eqref{eq:sm_p1_fixed_overlap_potential} describes the source-defined potential and not an unrestricted search over all fixed-overlap saddles. To extend the local response beyond linear order, we substitute power-series ans\"atze for the four stationary fields into the $p=1$ saddle equations and match successive powers of $\lambda$. At each order, the new coefficients enter linearly, allowing the recursion to be solved symbolically in closed form. Writing $h=e^{-1/2}$ for compactness, this gives
\begin{align}
    q_1(\lambda)
    &=\frac{(h^2+2)^2}{4}\lambda+\frac{h^8-16h^6-62h^4-62h^2-16}{48}\lambda^3\nonumber\\
    &\quad+\frac{512+4035h^2+7788h^4+3436h^6-2440h^8-1872h^{10}-244h^{12}}{3840}\lambda^5+O(\lambda^7).
    \label{eq:sm_p1_q_series_closed}
\end{align}
Numerically,
\begin{equation}
    q_1(\lambda)=1.401713262\lambda-0.999533133\lambda^3+0.823839762\lambda^5+O(\lambda^7).
    \label{eq:sm_p1_q_series}
\end{equation}
The linear coefficient is the exact susceptibility derived above, $\chi_1=[1+1/(2e)]^2$. The corresponding overlap potential has the local expansion
\begin{align}
    V_1(q)
    &=
    \frac{2}{(h^2+2)^2}q^2+\frac{4\left(16+62h^2+62h^4+16h^6-h^8\right)}{3(h^2+2)^8}q^4
    \nonumber\\
    &\quad
    +\frac{8\left(264h^{16}+2208h^{14}+13544h^{12}+51012h^{10}+104148h^8+115069h^6+68756h^4+21492h^2+3072\right)}{45(h^2+2)^{14}}q^6\nonumber\\
    &\quad+O(q^8).
    \label{eq:sm_p1_V_series_closed}
\end{align}
Numerically
\begin{equation}
    V_1(q)=0.356706335q^2+0.064729235q^4+0.028881604q^6+O(q^8).
    \label{eq:sm_p1_V_series}
\end{equation}
The coefficient of $q^2$ is $1/(2\chi_1)$, so $V_1''(0)=\chi_1^{-1}=[1+1/(2e)]^{-2}$. The positive quartic coefficient gives an additional positive correction to the local potential away from $q=0$.

\subsection{Fixed-overlap stationary problem}
\label{subsec:sm_depth_one_fixed_overlap}

The source construction follows the overlap branch continuously connected to the ordinary QAOA distribution at $\lambda=0$. A complementary question is whether fixing the overlap directly reveals a disconnected stationary branch that cannot be reached by tuning its conjugate source. At depth one, this can be tested analytically using the same source-eliminated functional. We impose the output overlap $q$ and search for stationary points of the fixed-overlap functional. This is the depth-one analogue of the fixed-overlap Franz--Parisi construction, where one asks whether fixing the overlap reveals branches not reached by varying a conjugate source~\cite{FranzParisi1995,FranzParisi1997,FranzParisi1998EffectivePotential,MezardParisiVirasoro1987,CastellaniCavagna2005}. The constrained functional is
\begin{equation}
    \mathcal C_1(q,\rho,\sigma,m,\nu)=I_w(q)-B(\rho,\sigma,m,\nu),
    \label{eq:sm_p1_constrained_functional}
\end{equation}
where $B$, $w$, and $I_w(q)$ are the functions defined in Eqs.~\eqref{eq:sm_p1_B_function}, \eqref{eq:sm_p1_w_bias}, and \eqref{eq:sm_p1_binary_cost}, respectively. The difference from Eq.~\eqref{eq:sm_p1_fixed_overlap_potential} is that the stationary point is now not assumed to be the continuation of the zero-overlap saddle. Instead, we ask whether other stationary points of $\mathcal C_1$ in Eq.~\eqref{eq:sm_p1_constrained_functional} exist. For the derivation it is convenient to use the combinations $d_-$ and $d_+$ already introduced in Eq.~\eqref{eq:sm_p1_dpm}. Since the change of variables $(\sigma,m)\leftrightarrow(d_-,d_+)$ is invertible, stationarity with respect to $(\rho,\sigma,m,\nu)$ is equivalent to stationarity with respect to $(\rho,d_-,d_+,\nu)$. In this subsection, when no ambiguity is possible, we write $B$ and $w$ as functions of $(\rho,d_-,d_+,\nu)$ by substituting $\sigma=d_-+d_+$ and $m=d_+-d_-$. Thus Eqs.~\eqref{eq:sm_p1_B_function} and~\eqref{eq:sm_p1_w_bias} become
\begin{equation}
    \begin{aligned}
        B\bigl(\rho,d_-,d_+,\nu\bigr)
        &=-\frac14-\frac{\rho^2}{4}+\frac{\rho}{2}-d_-d_+-\frac{\nu^2}{4}+\ln\cosh d_-,\\
        w\bigl(\rho,d_-,d_+,\nu\bigr)
        &=\frac12\tanh d_-+\frac12e^{-\rho}\frac{\sinh d_+}{\cosh d_-}+e^{-\rho/2}\frac{\sin(\nu/2)}{\cosh d_-}.
    \end{aligned}
    \label{eq:sm_p1_fixed_B_w_d_variables}
\end{equation}
We collect these four fixed-overlap fields into the vector
\begin{equation}
    \boldsymbol{\theta}=(\rho,d_-,d_+,\nu).
    \label{eq:sm_p1_fixed_theta_vector}
\end{equation}
All gradients and Hessians in this subsection are taken with respect to $\boldsymbol{\theta}$ unless stated otherwise. The fixed-overlap stationary equations follow by differentiating Eq.~\eqref{eq:sm_p1_constrained_functional}. Since
\begin{equation}
    \frac{\partial I_w(q)}{\partial w}=\frac{w-q}{1-w^2},
    \label{eq:sm_p1_fixed_Iw_derivative}
\end{equation}
one obtains
\begin{equation}
    \nabla_{\boldsymbol{\theta}} B=\frac{w-q}{1-w^2}\nabla_{\boldsymbol{\theta}}w.
    \label{eq:sm_p1_fixed_stationarity}
\end{equation}
This equation is the fixed-overlap version of the depth-one saddle equations. The fixed-overlap and fixed-source stationary problems are related exactly. Given a stationary point of Eq.~\eqref{eq:sm_p1_fixed_stationarity} with $|w|<1$, define the reconstructed source
\begin{equation}
    \lambda_q=\operatorname{arctanh}q-\operatorname{arctanh}w,\quad \tau=\tanh\lambda_q,\quad s=\frac{\tau}{1+\tau w}.
    \label{eq:sm_p1_fixed_reconstructed_source}
\end{equation}
Then
\begin{equation}
    q=\frac{\tau+w}{1+\tau w},\quad s=\frac{q-w}{1-w^2}.
    \label{eq:sm_p1_fixed_q_source_relation}
\end{equation}
Substituting Eq.~\eqref{eq:sm_p1_fixed_q_source_relation} into Eq.~\eqref{eq:sm_p1_fixed_stationarity} gives
\begin{equation}
    \nabla_{\boldsymbol{\theta}} B+s\nabla_{\boldsymbol{\theta}} w=\boldsymbol{0}.
    \label{eq:sm_p1_fixed_s_stationarity}
\end{equation}
The corresponding scalar source function is
\begin{equation}
    F_\tau\bigl(\rho,d_-,d_+,\nu\bigr)=B\bigl(\rho,d_-,d_+,\nu\bigr)+\ln\Bigl[1+\tau w\bigl(\rho,d_-,d_+,\nu\bigr)\Bigr],
    \label{eq:sm_p1_fixed_source_action}
\end{equation}
where the additive constant $\ln\cosh\lambda_q$ has been omitted because it does not affect stationarity. Thus every fixed-overlap stationary point reconstructs a fixed-source stationary point, and conversely every fixed-source stationary point gives a fixed-overlap stationary point with overlap $q=(\tau+w)/(1+\tau w)$. A constrained branch not reached by source continuation would therefore appear as an additional real stationary branch of $F_\tau$. For $q\ge0$, the solution continued from $q=0$ remains in the real sector
\begin{equation}
    d_-\ge0, \qquad d_+\ge0,\qquad 0\le\nu<\pi.
    \label{eq:sm_p1_fixed_principal_sector}
\end{equation}
We refer to the connected solution in this sector as the continued real branch. The negative-overlap sector follows by the symmetry
\begin{equation}
    \bigl(d_-,d_+,\nu,q\bigr)\mapsto\bigl(-d_-,-d_+,-\nu,-q\bigr),\qquad\rho\mapsto\rho.
    \label{eq:sm_p1_fixed_negative_q_symmetry}
\end{equation}
In the sector of Eq.~\eqref{eq:sm_p1_fixed_principal_sector}, Eq.~\eqref{eq:sm_p1_fixed_s_stationarity} is equivalent to
\begin{equation}
    \begin{aligned}
        1-\rho
        &=\frac{s}{\cosh d_-}\left[e^{-\rho}\sinh d_++e^{-\rho/2}\sin(\nu/2)\right],\\
        d_+
        &=\tanh d_-+s\left(\frac12-\tanh d_-w\right),\\
        d_-
        &=\frac{s e^{-\rho}\cosh d_+}{2\cosh d_-},\\
        \nu
        &=\frac{s e^{-\rho/2}\cos(\nu/2)}{\cosh d_-}.
    \end{aligned}
    \label{eq:sm_p1_fixed_principal_equations}
\end{equation}
These equations impose a useful identity. Define
\begin{equation}
    A=d_-\tanh(d_+),\qquad D=\nu\tan(\nu/2).
    \label{eq:sm_p1_fixed_A_D_def}
\end{equation}
Combining the first, third, and fourth equations in Eq.~\eqref{eq:sm_p1_fixed_principal_equations} gives
\begin{equation}
    \rho=1-2A-D.
    \label{eq:sm_p1_fixed_rho_identity}
\end{equation}
This removes the explicit source parameter from one direction of the stationary problem and will be used below to simplify the stability analysis. At $\tau=s=q=0$, Eq.~\eqref{eq:sm_p1_fixed_principal_equations} gives the $\lambda=0$ saddle
\begin{equation}
    \rho=1,\qquad d_-=0,\qquad d_+=0,\qquad\nu=0.
    \label{eq:sm_p1_fixed_origin}
\end{equation}
This is the same saddle as Eq.~\eqref{eq:sm_p1_zero_source_saddle}. The Hessian of $F_\tau$ at $\tau=0$, in the variable order $(\rho,d_-,d_+,\nu)$, is
\begin{equation}
    \mathsf{H}_0=
    \begin{pmatrix}
        -1/2 & 0 & 0 & 0\\
        0 & 1 & -1 & 0\\
        0 & -1 & 0 & 0\\
        0 & 0 & 0 & -1/2
    \end{pmatrix}.
    \label{eq:sm_p1_fixed_zero_hessian}
\end{equation}
Since $\det \mathsf{H}_0=-1/4$, the implicit-function theorem gives a unique analytic real branch through Eq.~\eqref{eq:sm_p1_fixed_origin}. This is the continued solution studied above. Its local source-overlap slope is the susceptibility $\chi_1=[1+1/(2e)]^2$ from Eq.~\eqref{eq:sm_p1_chi_from_saddle}, so the map between $\lambda$ and $q$ is locally invertible around the origin. It remains to check whether an additional real stationary branch can appear at finite source. Such a branch can be born from the connected component only through a spinodal, i.e., a fold of the stationary branch where the Hessian of $F_\tau$ from Eq.~\eqref{eq:sm_p1_fixed_source_action} becomes singular. The source Hessian is
\begin{equation}
    \mathsf{H}_\tau=\nabla_{\boldsymbol{\theta}}^2 F_\tau=\nabla_{\boldsymbol{\theta}}^2 B+s\nabla_{\boldsymbol{\theta}}^2 w-s^2\bigl(\nabla_{\boldsymbol{\theta}}w\bigr)\bigl(\nabla_{\boldsymbol{\theta}}w\bigr)^{\mathsf T}.
    \label{eq:sm_p1_fixed_source_hessian}
\end{equation}
Using the stationary equations of Eq.~\eqref{eq:sm_p1_fixed_principal_equations}, this Hessian simplifies substantially. In the variable order $(\rho,d_-,d_+,\nu)$, it becomes
\begin{equation}
    \mathsf{H}_\tau=
    \begin{pmatrix}
        A+\frac D4-\frac{(2A+D)^2}{4}-\frac12
        &
        d_+\left(A+\frac D2\right)
        &
        \frac{d_-(2A+D-2)}{2}
        &
        \frac{\nu(2A+D-1)}{4}
        \\[1mm]
        d_+\left(A+\frac D2\right)
        &
        1-A-D-d_+^2
        &
        -1-d_-d_+
        &
        -\frac{\nu d_+}{2}
        \\[1mm]
        \frac{d_-(2A+D-2)}{2}
        &
        -1-d_-d_+
        &
        A-d_-^2
        &
        -\frac{\nu d_-}{2}
        \\[1mm]
        \frac{\nu(2A+D-1)}{4}
        &
        -\frac{\nu d_+}{2}
        &
        -\frac{\nu d_-}{2}
        &
        -\frac12-\frac D4-\frac{\nu^2}{4}
    \end{pmatrix}.
    \label{eq:sm_p1_fixed_onshell_hessian}
\end{equation}
All explicit factors of $s$, $\tau$, $e^{-\rho}$, and $\cosh d_-$ have cancelled. Thus the possible spinodal condition is reduced to an algebraic sign test on the stationary manifold. To track the eigenvalue signature, reorder the variables as $(d_-,\rho,d_+,\nu)$, and let $\Delta_k$ be the leading $k\times k$ principal minor of the reordered Hessian. At $\tau=0$, one has
\begin{equation}
    \Delta_1=1,\qquad\Delta_2=-\frac12,\qquad\Delta_3=\frac12,\qquad\Delta_4=-\frac14.
    \label{eq:sm_p1_fixed_zero_minors}
\end{equation}
Thus the Hessian has one positive and three negative directions. If the sign pattern
\begin{equation}
    \Delta_1>0,\qquad\Delta_2<0,\qquad\Delta_3>0,\qquad\Delta_4<0
    \label{eq:sm_p1_fixed_minor_sign_pattern}
\end{equation}
persists, the Hessian remains nonsingular along the continued real branch and keeps the same number of positive and negative eigenvalues. The first two signs are controlled directly. The first minor is
\begin{equation}
    \Delta_1=1-A-D-d_+^2.
    \label{eq:sm_p1_fixed_delta1}
\end{equation}
The $\rho\rho$ entry of Eq.~\eqref{eq:sm_p1_fixed_onshell_hessian} can be rewritten, using Eq.~\eqref{eq:sm_p1_fixed_rho_identity}, as
\begin{equation}
    \mathsf{H}_{\rho\rho}=A+\frac D4-\frac{(2A+D)^2}{4}-\frac12=-\frac{(1-2A-D)^2+D+1}{4}<0.
    \label{eq:sm_p1_fixed_Hrho_negative}
\end{equation}
Therefore
\begin{equation}
    \Delta_2=\Delta_1 \mathsf{H}_{\rho\rho}-d_+^2\left(A+\frac D2\right)^2.
    \label{eq:sm_p1_fixed_delta2}
\end{equation}
Hence $\Delta_1>0$ implies $\Delta_2<0$. The remaining nontrivial sign checks are $\Delta_3>0$ and $\Delta_4<0$. We evaluate these signs along the branch specified by Eq.~\eqref{eq:sm_p1_fixed_principal_sector}. The continuation from $\tau=0$ to the endpoint $\tau\to1$ is regular in the internal fields. Along this branch
\begin{equation}
    0.6980279\lesssim\rho\le 1,\qquad0\le d_-\lesssim 0.1978974,\qquad0\le d_+\lesssim 0.5001306,\qquad0\le\nu\lesssim 0.4831227.
    \label{eq:sm_p1_fixed_field_bounds}
\end{equation}
Equivalently
\begin{equation}
    0\le A \lesssim 0.0914721,\qquad0\le D \lesssim 0.1190280,\qquad0\le 2A+D \lesssim 0.3019722.
    \label{eq:sm_p1_fixed_AD_bounds}
\end{equation}
Using the exact Hessian \eqref{eq:sm_p1_fixed_onshell_hessian}, the leading minors remain uniformly separated from zero
\begin{equation}
    \Delta_1 \ge 0.539369,\qquad-\Delta_2 \ge 0.222295,\qquad\Delta_3 \ge 0.474934,\qquad-\Delta_4 \ge 0.249064.
    \label{eq:sm_p1_fixed_minor_margins}
\end{equation}
\begin{figure}[!t]
    \centering
    \includegraphics[width=0.5\columnwidth]{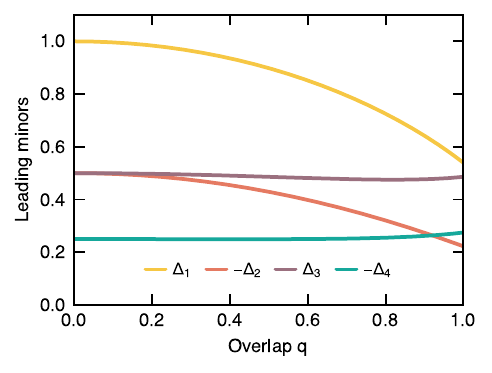}
    \caption{Sign-corrected leading principal minors $\Delta_1$, $-\Delta_2$, $\Delta_3$, and $-\Delta_4$ of the Hessian $\mathsf{H}_\tau$ in Eq.~\eqref{eq:sm_p1_fixed_onshell_hessian}, evaluated along the continued real branch of Eq.~\eqref{eq:sm_p1_fixed_principal_equations} and plotted against the induced overlap $q$.}
    \label{fig:p1_fixed_overlap_minors}
\end{figure}
Thus, for $0\le\tau<1$, the finite-source Hessian $\mathsf{H}_\tau$ has a fixed eigenvalue signature: one positive eigenvalue and three negative eigenvalues. In particular, $\det \mathsf{H}_\tau=\Delta_4$ (see Fig.~\ref{fig:p1_fixed_overlap_minors}) never vanishes on the continued real branch, so this branch does not undergo a spinodal.

The same conclusion can be expressed independently of the source parametrization. Let $\mathsf K_s=\nabla_{\boldsymbol\theta}^2(B+sw)$ denote the Hessian of the fixed-$s$ stationary equations. Combining the fixed-$s$ stationarity condition, the matrix determinant lemma, and $\tau=s/(1-sw)$ gives
\begin{equation}
    \det\mathsf H_\tau=\det\mathsf K_s\left(1+s^2\frac{dw}{ds}\right)=\det\mathsf K_s(1-sw)^2\frac{d\tau}{ds}.
    \label{eq:sm_p1_fixed_det_factorization}
\end{equation}
Thus a zero of $\det\mathsf H_\tau$ would signal either a singular fixed-$s$ stationary problem or a fold of the source parametrization. The uniform bound in Eq.~\eqref{eq:sm_p1_fixed_minor_margins} rules out both mechanisms on the continued real branch. We therefore find that the continued real branch is locally unique at $q=0$, remains nonsingular for all $0\le\tau<1$, and reaches $q\to1$ as $\tau\to1$. Within the sector of Eq.~\eqref{eq:sm_p1_fixed_principal_sector}, no additional real fixed-overlap stationary branch is found beyond the one obtained by source continuation. By the symmetry in Eq.~\eqref{eq:sm_p1_fixed_negative_q_symmetry}, the same conclusion applies for $q\le0$.

At $p=1$, the Franz--Parisi potential computed from the source is not missing a separate real fixed-overlap branch within this sector since imposing the overlap directly gives the same saddle selected by a conjugate source. Thus the weak overlap response and positive local stiffness found above reflect the absence of an additional real fixed-overlap branch analogous to the secondary Franz--Parisi branches that appear in glassy mean-field problems~\cite{FranzParisi1995,FranzParisi1997,FranzParisi1998EffectivePotential,MezardParisiVirasoro1987,CastellaniCavagna2005}. This statement is restricted to the real solution family satisfying Eq.~\eqref{eq:sm_p1_fixed_principal_sector} since possible complex stationary points of the analytically continued Keldysh saddle equations are not excluded. However, their contribution would depend on the steepest-descent contour of the original collective-variable integral and would not constitute an additional real fixed-overlap branch by itself.

\subsection{Energy-tilted depth-one response}
\label{subsec:sm_depth_one_energy_tilted_response}

The energy-tilted construction of Sec.~\ref{sec:sm_energy_tilted_overlap} becomes especially explicit at depth one. The key difference from the untilted calculation above is that the terminal measured spin is no longer a purely local variable. As explained in Sec.~\ref{subsec:sm_tilted_terminal_slice}, the energy tilt assigns the terminal slice a nonzero Keldysh coefficient through Eq.~\eqref{eq:sm_tilted_gamma_vector}. The terminal history coordinate must therefore be included among the coordinates active in the disorder term.

We use the depth-one notation of Eq.~\eqref{eq:sm_p1_state}, with forward spin $u$, terminal spin $z$, and backward spin $v$. For compactness, we order the active history coordinates as $(v,z,u)$. In this ordering, the coefficients $\Gamma^{(\eta)}_r$ entering the disorder-averaged saddle are collected in the diagonal matrix
\begin{equation}
    \mathsf{D}_\eta=\operatorname{diag}\bigl(\gamma,i\eta,-\gamma\bigr).
    \label{eq:sm_p1_tilted_D_matrix}
\end{equation}
The one-sample tilted saddle $\mathsf{G}_\eta$ appearing in Eq.~\eqref{eq:sm_tilted_product_saddle} can then be parametrized by three real variables
\begin{equation}
    \mathsf{G}_\eta=
    \begin{pmatrix}
        1 & a+ib & \rho \\
        a+ib & 1 & a-ib \\
        \rho & a-ib & 1
    \end{pmatrix}.
    \label{eq:sm_p1_tilted_single_replica_saddle}
\end{equation}
Here $\rho$ is the overlap between the backward and forward slices, while $a\pm i b$ describe the correlations between the terminal slice and the two Keldysh branches. The one-site trace contains only the eight histories $(v,z,u)\in\{\pm1\}^3$, so the saddle equations can be written in closed form. Define the scalar combinations
\begin{equation}
    k=\gamma^2\rho,\qquad X=2\gamma\eta a,\qquad Y=2\gamma\eta b,
    \label{eq:sm_p1_tilted_kxy}
\end{equation}
and
\begin{align}
    U(Y) &= \cos^2\beta e^{Y}+\sin^2\beta e^{-Y},\nonumber\\
    V(Y) &= \cos^2\beta e^{Y}-\sin^2\beta e^{-Y},\nonumber\\
    T(k,X) &= \sin(2\beta)e^{-k}\sin X,\nonumber\\
    \mathcal Z_{\rm loc}(k,X,Y) &=e^k U(Y)+T(k,X).
    \label{eq:sm_p1_tilted_auxiliary_functions}
\end{align}
Then the tilted depth-one saddle is determined by
\begin{equation}
    \rho = \frac{e^k U(Y)-T(k,X)}{\mathcal Z_{\rm loc}(k,X,Y)},\quad a = \frac{e^k V(Y)}{\mathcal Z_{\rm loc}(k,X,Y)},\quad b = \frac{\sin(2\beta)e^{-k}\cos X}{\mathcal Z_{\rm loc}(k,X,Y)}.
    \label{eq:sm_p1_tilted_saddle_equations}
\end{equation}
At $\eta=0$, these equations reduce to
\begin{equation}
    \rho=1,\quad a=\cos(2\beta),\quad b=\sin(2\beta)e^{-2\gamma^2}.
    \label{eq:sm_p1_tilted_zero_eta_saddle}
\end{equation}
For the optimized $p=1$ angles $\gamma=1/2$ and $\beta=\pi/8$~\cite{FarhiGoldstoneGutmannZhou2022}, this gives $a=1/\sqrt 2$ and $b=e^{-1/2}/\sqrt 2$. As established generally in Sec.~\ref{subsec:sm_tilted_source}, increasing $\eta$ selects monotonically lower energies, with $E_1(\eta)=-\partial_\eta\varphi_1^{(1)}(\eta)$ along a differentiable thermodynamic branch. For the depth-one tilted saddle in Eq.~\eqref{eq:sm_p1_tilted_saddle_equations}, the normalized one-sample tilted free-entropy density is
\begin{equation}
    \varphi_1^{(1)}(\eta)=-\frac{\gamma^2}{2}+\frac{\eta^2}{4}-\frac{\gamma^2\rho^2}{2}-2\gamma\eta ab+\ln\mathcal Z_{\rm loc}(k,X,Y),
    \label{eq:sm_p1_tilted_single_replica_free_entropy}
\end{equation}
evaluated at the solution of Eq.~\eqref{eq:sm_p1_tilted_saddle_equations}. The normalization is chosen so that $\varphi_1^{(1)}(0)=0$, consistent with $Z_{J,1}^{(1)}(0)=1$. The saddle equations imply
\begin{equation}
    \partial_k\ln \mathcal Z_{\rm loc}=\rho,\qquad\partial_X\ln \mathcal Z_{\rm loc}=b,\qquad\partial_Y\ln \mathcal Z_{\rm loc}=a.
    \label{eq:sm_p1_tilted_local_trace_derivatives}
\end{equation}
Therefore, by the envelope theorem, the implicit $\eta$-dependence of $\rho$, $a$, and $b$ does not contribute to $\partial_\eta\varphi_1^{(1)}(\eta)$. Using Eq.~\eqref{eq:sm_p1_tilted_kxy}, one obtains
\begin{equation}
    E_1(\eta)=-\partial_\eta\varphi_1^{(1)}(\eta)=-\frac{\eta}{2}-2\gamma a b.
    \label{eq:sm_p1_tilted_energy_saddle}
\end{equation}
For the optimized depth-one angles used here, $\gamma=1/2$, this reduces to $E_1(\eta)=-\eta/2-ab$. At $\eta=0$, Eq.~\eqref{eq:sm_p1_tilted_zero_eta_saddle} gives $E_1(0)=-1/(2\sqrt e)$, in agreement with the $p=1$ QAOA energy. Using $h=e^{-1/2}$ and solving Eq.~\eqref{eq:sm_p1_tilted_saddle_equations} order by order in $\eta$ and substituting into Eq.~\eqref{eq:sm_p1_tilted_energy_saddle} gives a local expansion for the selected energy
\begin{equation}
    \begin{aligned}
        E_1(\eta)
        &=-\frac{h}{2}+\frac{h^2-2}{4}\eta+\frac{h(3h^2+2)}{16}\eta^2-\frac{h^2(8h^2+1)}{12}\eta^3+\frac{h(643h^4-220h^2-4)}{768}\eta^4 \\
        &\quad-\frac{h^2(181h^4-440h^2+6)}{480}\eta^5-\frac{h(55257h^6+104370h^4-13636h^2-8)}{92160}\eta^6+O(\eta^7).
    \end{aligned}
    \label{eq:sm_p1_tilted_energy_series}
\end{equation}
Numerically
\begin{equation}
    \begin{aligned}
        E_1(\eta)
        &=-0.303265330-0.408030140\eta+0.117653238\eta^2-0.120880142\eta^3 \\
        &\quad+0.001648302\eta^4+0.100684976\eta^5-0.077998839\eta^6+O(\eta^7).
    \end{aligned}
    \label{eq:sm_p1_tilted_energy_series_numeric}
\end{equation}
The same one-sample saddle also gives the depth-one energy rate introduced in Sec.~\ref{subsec:sm_tilted_source}. Substituting Eqs.~\eqref{eq:sm_p1_tilted_single_replica_free_entropy} and~\eqref{eq:sm_p1_tilted_energy_saddle} into Eq.~\eqref{eq:sm_tilted_energy_rate_parametric}, and using $\gamma=1/2$, gives
\begin{equation}
    I_1\bigl[E_1(\eta)\bigr]=\frac{\eta^2}{4}+2\eta a b+\frac{1+\rho^2}{8}-\ln\mathcal Z_{\rm loc}(k,X,Y),
    \label{eq:sm_p1_tilted_energy_rate}
\end{equation}
evaluated on the solution of Eq.~\eqref{eq:sm_p1_tilted_saddle_equations}. This expression vanishes at $\eta=0$, as required by $I_1[E_1(0)]=0$. Defining the energy displacement from the typical depth-one value by
\begin{equation}
    \Delta E_1(\eta)=E_1(0)-E_1(\eta),
    \label{eq:sm_p1_tilted_energy_displacement}
\end{equation}
the small-$\eta$ energy series gives
\begin{equation}
    I_1\bigl[E_1(\eta)\bigr] =\frac{2-e^{-1}}{8}\eta^2+O(\eta^3)=\frac{2e}{2e-1}\bigl[\Delta E_1(\eta)\bigr]^2+O\Bigl(\bigl[\Delta E_1(\eta)\bigr]^3\bigr).
    \label{eq:sm_p1_local_energy_rate}
\end{equation}
Thus the lower-tail probability cost is locally quadratic in the energy improvement away from the typical depth-one QAOA energy. We now use this tilted one-sample saddle to compute the local response to the overlap source. At $\lambda=0$, the two physical replicas are independent tilted samples, so the two-replica saddle factorizes as in Eq.~\eqref{eq:sm_tilted_product_saddle}. Let $\mathsf{R}_\eta$ denote the linear response of the cross-sample history block to the source
\begin{equation}
    \mathsf{R}_\eta=\left.\frac{\partial \mathsf{R}(\lambda,\eta)}{\partial\lambda}\right|_{\lambda=0}.
    \label{eq:sm_p1_tilted_response_matrix_definition}
\end{equation}
Linearizing the saddle equation around the product tilted solution gives
\begin{equation}
    \mathsf{R}_\eta+\mathsf{G}_\eta \mathsf{D}_\eta \mathsf{R}_\eta \mathsf{D}_\eta \mathsf{G}_\eta= \boldsymbol{c}_\eta \boldsymbol{c}_\eta^{\mathsf T},
    \label{eq:sm_p1_tilted_linear_response_equation}
\end{equation}
where $\boldsymbol{c}_\eta=\mathsf{G}_\eta \boldsymbol{e}_z$ with $\boldsymbol{e}_z=(0,1,0)^{\mathsf T}$ selecting the terminal slice in the $(v,z,u)$ ordering. The tilted overlap susceptibility is the output-output component of this response
\begin{equation}
    \chi_{1,\eta}=\boldsymbol{e}_z^{\mathsf T} \mathsf{R}_\eta \boldsymbol{e}_z.
    \label{eq:sm_p1_tilted_chi_from_response}
\end{equation}
Equivalently
\begin{equation}
    \chi_{1,\eta}=\bigl(\boldsymbol{e}_z\otimes \boldsymbol{e}_z\bigr)^{\mathsf T}\Bigl[\mathsf{I}_9+(\mathsf{G}_\eta \mathsf{D}_\eta)\otimes(\mathsf{G}_\eta \mathsf{D}_\eta)\Bigr]^{-1}\bigl(\boldsymbol{c}_\eta\otimes \boldsymbol{c}_\eta\bigr).
    \label{eq:sm_p1_tilted_chi_closed_form}
\end{equation}
This is the depth-one analogue of the general tilted susceptibility calculation in Eq.~\eqref{eq:sm_tilted_overlap_response}. It reduces the energy-tilted response to a three-variable saddle and a finite $9\times9$ response operator. As a check, Eq.~\eqref{eq:sm_p1_tilted_chi_closed_form} at $\eta=0$ gives $\chi_{1,0}=(1+1/(2e))^2$, in agreement with Eq.~\eqref{eq:sm_p1_chi_from_saddle}. The same formula gives an analytic small-$\eta$ expansion. Since the Jacobian of Eq.~\eqref{eq:sm_p1_tilted_saddle_equations} is nonsingular at $\eta=0$, the three saddle variables have regular power series in $\eta$. As in the small-$\lambda$ expansion above, the coefficients are obtained by expanding the finite set of saddle equations and the response equation order by order. Solving order by order and substituting into Eq.~\eqref{eq:sm_p1_tilted_chi_closed_form} gives
\begin{align}
    \chi_{1,\eta}
    &=
    \left(1+\frac{1}{2e}\right)^2+\frac{8e^2+3e-3}{4e^{5/2}}\eta+\frac{16e^4+41e^3-32e^2+28e+1}{16e^4}\eta^2\nonumber\\
    &\quad
    +\frac{288e^4+136e^3+621e^2-173e-36}{96e^{9/2}}\eta^3\nonumber\\
    &\quad
    +\frac{192e^6+1102e^5+1441e^4-1094e^3-219e^2+270e+3}{192e^6}\eta^4+O(\eta^5).
    \label{eq:sm_p1_tilted_chi_series}
\end{align}
Numerically
\begin{equation}
    \chi_{1,\eta}=1.401713262+1.318845191\eta+1.760292599\eta^2+2.608096299\eta^3+3.832134331\eta^4+O(\eta^5).
    \label{eq:sm_p1_tilted_chi_series_numeric}
\end{equation}
Combining the linear terms of Eqs.~\eqref{eq:sm_p1_tilted_energy_series} and~\eqref{eq:sm_p1_tilted_chi_series} eliminates the tilt and gives the local energy--geometry relation
\begin{equation}
    \chi_{1,\eta}=\chi_1+\frac{8e^2+3e-3}{e^{3/2}(2e-1)}\bigl[E_1(0)-E_1(\eta)\bigr]+O\Bigl(\bigl[E_1(0)-E_1(\eta)\bigr]^2\Bigr).
    \label{eq:sm_p1_local_energy_geometry_relation}
\end{equation}
The linear coefficient is positive and equals $3.232224932\ldots$. Thus, already at depth one, moving into the lower-energy tail produces a linear enhancement of the overlap susceptibility. Together with Eq.~\eqref{eq:sm_p1_local_energy_rate}, this shows that the susceptibility changes linearly with the energy displacement into the lower tail, while the associated sampling cost begins quadratically. We next identify the finite-$\eta$ endpoint of this response.

While the series in Eq.~\eqref{eq:sm_p1_tilted_chi_series} captures the initial growth of $\chi_{1,\eta}$, the finite-$\eta$ divergence observed in Fig.~3 of the main text is controlled directly by the linear response operator in Eq.~\eqref{eq:sm_p1_tilted_chi_closed_form}. Let
\begin{equation}
    \mathsf{M}_\eta=\mathsf{G}_\eta \mathsf{D}_\eta.
    \label{eq:sm_p1_tilted_response_matrix}
\end{equation}
The response can diverge only when
\begin{equation}
    \det\Bigl[\mathsf{I}_9+\mathsf{M}_\eta\otimes \mathsf{M}_\eta\Bigr]=0.
    \label{eq:sm_p1_tilted_instability_determinant}
\end{equation}
Equivalently, if $\mu_j$ are the eigenvalues of $\mathsf{M}_\eta$, a singularity occurs when $1+\mu_j\mu_k=0$ for some pair $(j,k)$. Along the tilted continuation followed here, the first zero of this determinant is equivalently located by the crossing of one eigenvalue of $\mathsf{M}_\eta$ through the imaginary unit $i$. This gives a scalar condition for $\eta_c(1)$. Writing
\begin{equation}
    t_2=\frac{\rho^2-1}{4}+2\eta a b,\qquad d_\eta=\det\bigl(\mathsf{G}_\eta\bigr),
    \label{eq:sm_p1_tilted_t2_det}
\end{equation}
the condition $\det(\mathsf{M}_\eta-i\mathsf{I}_3)=0$ becomes
\begin{equation}
    \eta+t_2+\frac{\eta}{4}d_\eta=1.
    \label{eq:sm_p1_tilted_instability_scalar}
\end{equation}
Solving Eqs.~\eqref{eq:sm_p1_tilted_saddle_equations} and~\eqref{eq:sm_p1_tilted_instability_scalar} numerically gives
\begin{equation}
    \eta_c(1)=0.7266931845\ldots.
    \label{eq:sm_p1_tilted_eta_c}
\end{equation}
This value marks the singularity of the depth-one continued tilted response. As shown in Sec.~\ref{subsec:sm_p1_replicon_spectrum}, the auxiliary-replica stability gap closes at the same point. Using Eq.~\eqref{eq:sm_p1_tilted_energy_saddle}, the corresponding selected energy density is
\begin{equation}
    E_1\bigl(\eta_c(1)\bigr)=-0.573725333\ldots.
    \label{eq:sm_p1_tilted_eta_c_energy}
\end{equation}
Together with $E_1(0)=-1/(2\sqrt e)$, this gives
\begin{equation}
    \Delta E_c(1)=E_1(0)-E_1\bigl(\eta_c(1)\bigr)=0.270460004\ldots.
    \label{eq:sm_p1_critical_energy_displacement}
\end{equation}
Evaluating Eq.~\eqref{eq:sm_p1_tilted_energy_rate} on the same saddle gives
\begin{equation}
    I_c(1)=I_1\left[E_1\bigl(\eta_c(1)\bigr)\right]=0.096830650\ldots.
    \label{eq:sm_p1_critical_energy_rate}
\end{equation}
These values provide the depth-one analytic benchmarks for Figs.~3b and~3c of the main text. The critical energy lies substantially below the untilted depth-one value $E_1(0)=-0.303265\ldots$, but remains above the Parisi value $E_{\mathrm{Parisi}}=-0.763166\ldots$.

At finite $N$, the normalized tilted distribution in Eq.~\eqref{eq:sm_energy_tilted_distribution} is well defined for all $\eta$ and, if the limit $\eta\to\infty$ is taken before the thermodynamic limit, concentrates on the lowest-energy configurations in the support of the QAOA distribution. The thermodynamic calculation above is more limited: it follows the same auxiliary-replica ansatz and the saddle continued from $\eta=0$. It should not be extrapolated to $\eta\to\infty$. Indeed, Eq.~\eqref{eq:sm_p1_tilted_energy_saddle} already contains an explicit contribution $-\eta/2$, so a formal continuation of the same saddle would eventually drive the energy below any finite thermodynamic ground-state density. The singularity therefore marks both the limit of the response calculation on the continued tilted saddle and the loss of local stability of the auxiliary-replica-diagonal ansatz. The nonlinear auxiliary-replica continuation for $0<\eta-\eta_c(1)\ll1$ is constructed in Sec.~\ref{sec:sm_p1_tilted_parisi_transition}, where a controlled near-critical expansion identifies the continuous Parisi saddle that replaces the unstable auxiliary-replica-diagonal branch.

\section{Stability to correlations between auxiliary replicas}
\label{sec:sm_auxiliary_replica_stability}

The main-text calculation sets the overlaps between distinct auxiliary replicas to zero while allowing the two physical QAOA samples to develop mutual overlap. We test whether this ansatz is locally stable by introducing infinitesimal correlations between two auxiliary replicas and linearizing the saddle equations, in the spirit of the de Almeida--Thouless stability analysis~\cite{deAlmeidaThouless1978,MezardParisiVirasoro1987,CastellaniCavagna2005}. The resulting stability gap measures the distance from an instability: a nonzero gap means that the ansatz remains locally stable, whereas gap closure signals that a more general auxiliary-replica overlap structure must be considered.

\subsection{Linear stability operator}
\label{subsec:sm_replicon_stability_operator}

A composite coordinate within one auxiliary replica is denoted by $x=(\sigma,r)$, where $\sigma\in\{A,B\}$ labels the two physical replicas and $r\in\mathcal I_p$ labels the Keldysh-history slice. In the untilted problem, only slices with $\Gamma_r\neq0$ are active in the disorder term. The terminal coordinate $r=0$ is therefore excluded from the disorder-generated collective variables because $\Gamma_0=0$, although it still enters the one-site measure through the explicit source $\lambda a_0^A a_0^B$. With energy tilting, $\Gamma_0^{(\eta)}=i\eta$, so the terminal coordinates $(A,0)$ and $(B,0)$ also become active in the disorder term. For any active coordinate $x=(\sigma,r)$, we write $\Gamma_x^{(\eta)}=\Gamma_r^{(\eta)}$. The corresponding quadratic disorder kernel is
\begin{equation}
    \mathsf{B}_{xy}^{(\eta)}=-\Gamma_x^{(\eta)}\Gamma_y^{(\eta)}.
    \label{eq:sm_replicon_disorder_kernel}
\end{equation}
Consider two distinct auxiliary replicas $\alpha\neq\alpha'$ and introduce an infinitesimal mutual-overlap block
\begin{equation}
    \delta\mathsf{Q}_{xy}=\mathsf{Q}^{(n)}_{(\alpha,x),(\alpha',y)}.
    \label{eq:sm_replicon_offdiagonal_block}
\end{equation}
The ansatz used here sets this block to zero. The overlap source couples the two physical replicas inside each auxiliary replica, but distinct auxiliary replicas remain uncoupled. Hence the one-site measure factorizes between $\alpha$ and $\alpha'$, and the self-consistent mutual overlap between the two auxiliary replicas factorizes as
\begin{equation}
    \mathsf{Q}^{(n)}_{(\alpha,x),(\alpha',y)}=\left\langle a_x^\alpha a_y^{\alpha'}\right\rangle_\star=\left\langle a_x^\alpha\right\rangle_\star\left\langle a_y^{\alpha'}\right\rangle_\star=0.
    \label{eq:sm_replicon_offdiagonal_vanishes}
\end{equation}
The last equality follows from the spin-flip symmetry of the continued saddle. We next linearize the corresponding saddle equation around this zero-overlap solution. To first order, the perturbation $\delta\mathsf{Q}_{x'y'}$ inserts the spin product $a_{x'}^\alpha a_{y'}^{\alpha'}$ into the one-site measure. Because the unperturbed measure factorizes between $\alpha$ and $\alpha'$, the response factorizes into one two-point function within each auxiliary replica. The resulting linearized map is
\begin{equation}
    \bigl(\mathsf{L}^{\rm rep}\delta\mathsf{Q}\bigr)_{xy}=\sum_{x',y'}\mathsf{B}_{x'y'}^{(\eta)}\bigl\langle a_xa_{x'}\bigr\rangle_\star\bigl\langle a_ya_{y'}\bigr\rangle_\star\delta\mathsf{Q}_{x'y'}.
    \label{eq:sm_replicon_linearized_map}
\end{equation}
Here $a_x\equiv a_r^\sigma$ for $x=(\sigma,r)$, all sums run over the active coordinates, and $\langle\cdot\rangle_\star$ denotes expectation in the normalized one-site measure at $\mathsf{Q}_\star(\lambda,\eta)$. The operator $\mathsf L^{\rm rep}$ is the Jacobian of the self-consistency map for correlations between distinct auxiliary replicas. The linearized saddle equation is $\delta\mathsf{Q}=\mathsf{L}^{\rm rep}\delta\mathsf{Q}$. Thus an infinitesimal correlated solution can appear only when $\mathsf{L}^{\rm rep}$ has an eigenvalue equal to one. We define the stability gap $\Delta_{\rm rep}(p,\lambda,\eta)$ as the smallest value of $|1-\epsilon|$ among the eigenvalues $\epsilon$ of $\mathsf{L}^{\rm rep}$ at the saddle. The corresponding instability condition is
\begin{equation}
    \Delta_{\rm rep}(p,\lambda,\eta)\to0.
    \label{eq:sm_replicon_instability_condition}
\end{equation}
In the classical SK stability problem, the analogous criterion is a vanishing de Almeida--Thouless stability eigenvalue. Here the Keldysh one-site weights can be complex, so the linearized map need not be Hermitian and the stability gap is defined using the complex distance to $\epsilon=1$. If $\Delta_{\rm rep}$ remains $O(1)$ at $\lambda=\eta=0$, the untilted QAOA saddle is locally stable against correlations between auxiliary replicas. If the gap vanishes at finite $\lambda$ or finite $\eta$, the ansatz with vanishing mutual overlap between auxiliary replicas loses local stability and a nonlinear solution with nonzero mutual overlaps between auxiliary replicas must be sought.

\subsection{Depth-zero analytic stability spectrum}
\label{subsec:sm_p0_replicon_spectrum}

The depth-zero Gibbs benchmark of Sec.~\ref{subsec:sm_p0_gibbs_benchmark} provides a closed-form test of the auxiliary-replica stability operator. At $p=0$, the only active coordinate is the terminal spin because $\Gamma_0^{(\eta)}=i\eta$. For the two physical replicas, the active coordinates are therefore $(A,0)$ and $(B,0)$. At $\lambda=0$ and on the paramagnetic branch $\eta<1$, the two physical replicas are independent and their one-site correlation matrix is
\begin{equation}
    \mathsf C_0=
    \begin{pmatrix}
        1 & 0\\
        0 & 1
    \end{pmatrix}.
    \label{eq:sm_p0_replicon_correlation_matrix}
\end{equation}
The disorder kernel in Eq.~\eqref{eq:sm_replicon_disorder_kernel} becomes
\begin{equation}
    \mathsf B_0^{(\eta)}=\eta^2
    \begin{pmatrix}
        1 & 1\\
        1 & 1
    \end{pmatrix}.
    \label{eq:sm_p0_replicon_disorder_kernel}
\end{equation}
Substituting these matrices into Eq.~\eqref{eq:sm_replicon_linearized_map} gives
\begin{equation}
    \bigl(\mathsf L^{\rm rep}_0\delta\mathsf Q\bigr)_{xy}=\eta^2\delta\mathsf Q_{xy}.
    \label{eq:sm_p0_replicon_map}
\end{equation}
All four components of the mutual-overlap block between two distinct auxiliary replicas are therefore eigenmodes with eigenvalue $\eta^2$. The stability gap defined in Sec.~\ref{subsec:sm_replicon_stability_operator} is
\begin{equation}
    \Delta_{\rm rep}(0,0,\eta)=\left|1-\eta^2\right|.
    \label{eq:sm_p0_replicon_gap}
\end{equation}
On the paramagnetic branch
\begin{equation}
    \Delta_{\rm rep}(0,0,\eta)=1-\eta^2,\qquad 0\leq\eta\leq1.
    \label{eq:sm_p0_paramagnetic_replicon_gap}
\end{equation}
The gap closes at $\eta=1$, exactly where the physical overlap susceptibility in Eq.~\eqref{eq:sm_p0_tilted_susceptibility} diverges. This reproduces the zero-field de Almeida--Thouless stability condition for the classical SK model and independently confirms that the gap in Eq.~\eqref{eq:sm_replicon_instability_condition} detects the known stability boundary of the paramagnetic solution~\cite{deAlmeidaThouless1978,MezardParisiVirasoro1987,CastellaniCavagna2005}.

\subsection{Depth-one analytic stability spectrum}
\label{subsec:sm_p1_replicon_spectrum}

The depth-zero calculation above contains only the terminal Gibbs spin, whereas the first nontrivial QAOA circuit produces a richer analytic stability problem at depth one. This stability problem is distinct from the fixed-overlap branch analysis in Sec.~\ref{subsec:sm_depth_one_fixed_overlap} where the Hessian tests whether the real four-field saddle equations for the two physical replicas develop a fold as the output overlap is imposed. Here, the two-physical-replica saddle is held fixed while perturbations coupling distinct auxiliary replicas are introduced, testing whether their assumed zero mutual overlap remains locally stable.

We first consider the energy-tilted problem at $\lambda=0$ where the two physical replicas are independent. The one-site correlation matrix and Keldysh-coefficient matrix factorize as
\begin{equation}
    \mathsf C_\star(0,\eta)=\mathsf G_\eta\oplus\mathsf G_\eta,\qquad\mathsf D_\star(\eta)=\mathsf D_\eta\oplus\mathsf D_\eta,
    \label{eq:sm_p1_tilted_replicon_product_matrices}
\end{equation}
where $\mathsf G_\eta$ and $\mathsf D_\eta$ are defined in Eqs.~\eqref{eq:sm_p1_tilted_single_replica_saddle} and~\eqref{eq:sm_p1_tilted_D_matrix}. Using the response matrix $\mathsf M_\eta$ defined in Eq.~\eqref{eq:sm_p1_tilted_response_matrix}, Eq.~\eqref{eq:sm_replicon_linearized_map} acts independently on each physical-replica block as
\begin{equation}
    \mathsf L^{\rm rep}_{1,\eta}\bigl[\delta\mathsf Q\bigr]=-\mathsf M_\eta\delta\mathsf Q\mathsf M_\eta^{\mathsf T}.
    \label{eq:sm_p1_tilted_replicon_map}
\end{equation}
If $\mu_j$ are the three eigenvalues of $\mathsf M_\eta$, the corresponding eigenvalues of the linearized stability map are
\begin{equation}
    \epsilon_{jk}=-\mu_j\mu_k,\qquad\Delta_{\rm rep}(1,0,\eta)=\min_{j,k}\left|1+\mu_j\mu_k\right|.
    \label{eq:sm_p1_tilted_replicon_gap}
\end{equation}
Equation~\eqref{eq:sm_p1_tilted_replicon_gap} shows that gap closure is identical to the determinant condition in Eq.~\eqref{eq:sm_p1_tilted_instability_determinant}, which makes the physical response operator singular. The susceptibility pole at $\eta_c(1)$ and the loss of local stability of the auxiliary-replica-diagonal saddle thus coincide exactly.

For the $\eta=0$ case at finite overlap source, the terminal coordinate is inactive and the stability map can be diagonalized explicitly. The active coordinates are the four nonterminal Keldysh spins $(u^A,v^A,u^B,v^B)$, where $u$ and $v$ denote the two nonterminal slices of the depth-one history. We use the same four fields as in Sec.~\ref{sec:sm_depth_one_solution}
\begin{equation}
    \mathsf{Q}_{u^A v^A}=\mathsf{Q}_{u^B v^B}=\rho,\quad\mathsf{Q}_{u^A v^B}=\mathsf{Q}_{v^A u^B}=\sigma,\quad\mathsf{Q}_{u^A u^B}=m+i\nu,\quad\mathsf{Q}_{v^A v^B}=m-i\nu.
    \label{eq:sm_p1_replicon_fields}
\end{equation}
The corresponding one-site two-point matrix on the active coordinates is
\begin{equation}
    \mathsf{C} =
    \begin{pmatrix}
        1 & \rho & m+i\nu & \sigma\\
        \rho & 1 & \sigma & m-i\nu\\
        m+i\nu & \sigma & 1 & \rho\\
        \sigma & m-i\nu & \rho & 1
    \end{pmatrix}.
\label{eq:sm_p1_correlation_matrix}
\end{equation}
The diagonal matrix $\mathsf E=\operatorname{diag}(1,-1,1,-1)$ records the relative Keldysh signs of the active coordinates in this ordering; an overall sign convention does not affect the products below. With $\gamma$ the $p=1$ phase angle, Eq.~\eqref{eq:sm_replicon_linearized_map} reduces to
\begin{equation}
    \mathsf{L}^{\rm rep}_1=-\gamma^2\left(\mathsf{C}\mathsf{E}\right)\otimes\left(\mathsf{C}\mathsf{E}\right).
    \label{eq:sm_p1_replicon_map}
\end{equation}
Thus the $16\times16$ stability problem reduces to the $4\times4$ spectrum of $\mathsf{C}\mathsf{E}$. If $\zeta_j$ are the four eigenvalues of $\mathsf{C}\mathsf{E}$, then the sixteen eigenvalues of the linearized map are
\begin{equation}
    \epsilon_{jk}=-\gamma^2\zeta_j\zeta_k,\qquad j,k=1,\ldots,4.
    \label{eq:sm_p1_replicon_map_eigenvalues}
\end{equation}
The distance from instability for each mode is therefore $|1-\epsilon_{jk}|=|1+\gamma^2\zeta_j\zeta_k|$. The four eigenvalues $\zeta_j$ can be written explicitly
\begin{equation}
    \zeta_{1,2}=-i\nu\pm\sqrt{(m-1)^2-(\rho-\sigma)^2},\qquad\zeta_{3,4}=+i\nu\pm\sqrt{(m+1)^2-(\rho+\sigma)^2}.
    \label{eq:sm_p1_zeta_eigenvalues}
\end{equation}
Equations~\eqref{eq:sm_p1_replicon_map_eigenvalues} and~\eqref{eq:sm_p1_zeta_eigenvalues} give the closed-form stability spectrum. At $\lambda=0$, the uncoupled product saddle has $\rho=1$ and $\sigma=m=\nu=0$. Then all four $\zeta_j$ vanish, so all sixteen eigenvalues $\epsilon_{jk}$ of $\mathsf{L}^{\rm rep}_1$ are zero. Hence the stability gap is one, and $\Delta_{\rm rep}(1,0,0)=1$. The depth-one continued solution is therefore strictly stable at $\lambda=\eta=0$. This analytic result is the stability counterpart of the local susceptibility calculation: the $p=1$ QAOA distribution has a finite overlap susceptibility, but the saddle remains locally stable against correlations between auxiliary replicas at $\lambda=0$.

\begin{figure*}[!t]
\centering
    \includegraphics[width=1\textwidth]{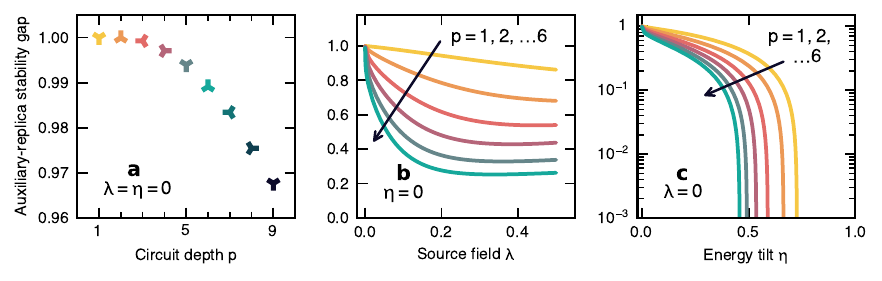}
    \caption{Auxiliary-replica stability gap $\Delta_{\rm rep}$ defined in Sec.~\ref{subsec:sm_replicon_stability_operator}: (a) as a function of circuit depth $p$ at $\lambda=\eta=0$, for $p=1,2,\ldots,9$; (b) as a function of the overlap source $\lambda$ at $\eta=0$, for $p=1,2,\ldots,6$; (c) as a function of the energy tilt $\eta$ at $\lambda=0$, for $p=1,2,\ldots,6$.}
    \label{fig:stability}
\end{figure*}

\subsection{Stability across depth, source, and energy tilt}
\label{subsec:sm_replicon_results}

For general fixed $p$, we evaluate the linearized map in Eq.~\eqref{eq:sm_replicon_linearized_map} on the converged continued saddle $\mathsf{Q}_\star(\lambda,\eta)$. In the untilted problem, each physical replica contributes $2p$ active nonterminal history coordinates, giving $4p$ active coordinates in total. With energy tilting, each replica also contributes its terminal coordinate, giving $4p+2$ active coordinates. The resulting finite-dimensional map is diagonalized directly, and the reported diagnostic is the stability gap $\Delta_{\rm rep}(p,\lambda,\eta)$ defined above. We report this gap only when the continued saddle equations are solved within numerical tolerance.

Figure~\ref{fig:stability}a gives the $\lambda=\eta=0$ stability gap as a function of QAOA depth $p$. The gap decreases with $p$, but remains separated from zero over the depths shown. Thus the untilted zero-source QAOA saddle used in the main text is locally stable against correlations between auxiliary replicas at $\lambda=\eta=0$. This distinguishes the growth of the physical overlap susceptibility $\chi_p$ from an auxiliary-replica instability. Indeed, at finite depth, the typical QAOA distribution becomes more susceptible to an overlap bias, while the auxiliary-replica ansatz remains locally stable.

Figure~\ref{fig:stability}b shows the same gap as a function of the overlap source $\lambda$ at $\eta=0$. The source biases the two physical replicas toward nonzero mutual overlap, and the gap decreases along the continued solution. The decrease is stronger at larger $p$, consistent with the fixed-source softening of the overlap potential $V_p(q)$ shown in the main text. Over the plotted source range, however, the gap remains nonzero, so the finite-source curves used to construct the potential remain locally stable against correlations between auxiliary replicas.

Figure~\ref{fig:stability}c shows the gap as a function of the energy tilt $\eta$ at $\lambda=0$. At $p=1$, Eqs.~\eqref{eq:sm_p1_tilted_replicon_gap} and~\eqref{eq:sm_p1_tilted_instability_determinant} show analytically that gap closure coincides exactly with the singularity of the energy-tilted susceptibility. At the larger depths evaluated numerically, the gap likewise approaches zero near the tilt at which $\chi_{p,\eta}$ grows rapidly. Thus the enhancement of the low-energy overlap response is accompanied by a weakening, and eventually a loss, of stability against correlations between auxiliary replicas. This occurs at progressively smaller $\eta$ as $p$ increases.

\section{Continuous Parisi phase beyond the tilted depth-one instability}
\label{sec:sm_p1_tilted_parisi_transition}

The divergence of the tilted susceptibility at $\eta_c(1)$ found in Sec.~\ref{subsec:sm_depth_one_energy_tilted_response} coincides with the loss of stability of the auxiliary-replica-diagonal saddle found in Sec.~\ref{subsec:sm_p1_replicon_spectrum}. This identifies a transition in the overlap geometry of the tilted depth-one QAOA distribution. We now determine the structure that emerges on its low-energy side for
\begin{equation}
    \delta\eta=\eta-\eta_c(1)>0,\qquad \delta\eta\ll1.
    \label{eq:sm_p1_parisi_delta_eta}
\end{equation}
The equilibrium SK transition at $p=0$ provides the natural reference. There, the paramagnetic solution loses stability at $\eta_c(0)=1$. A uniform nonzero replica overlap appears continuously but is itself unstable, and the equilibrium spin-glass phase is described by the continuous Parisi hierarchy~\cite{Parisi1979,deAlmeidaThouless1978}. The near-critical replica construction used below follows the standard SK treatment of this instability and its continuous Parisi resolution, reviewed pedagogically in Refs.~\cite{MezardParisiVirasoro1987,BinderYoung1986,CastellaniCavagna2005}. We therefore ask whether the depth-one QAOA transition follows the same sequence once the Keldysh-history degrees of freedom are included.

Near $\eta_c(1)$ the problem simplifies because a single auxiliary-replica overlap mode becomes soft while the remaining Keldysh-history fluctuations stay gapped. This permits the same near-critical Landau strategy as in the SK problem, now projected onto the QAOA critical mode and with coefficients dressed by the Keldysh-history structure. We first identify the soft mode and show that it directly couples to the output overlap. We then derive its nonlinear replicated action after allowing all gapped collective fields to relax. Finally, we determine the overlap structure selected beyond the transition and translate it back to the QAOA distribution. Throughout, we use the active history ordering $(v,z,u)$ and the depth-one matrices $\mathsf G_\eta$ and $\mathsf D_\eta$ defined in Eqs.~\eqref{eq:sm_p1_tilted_single_replica_saddle} and~\eqref{eq:sm_p1_tilted_D_matrix}.

\subsection{Soft overlap mode at the transition}
\label{subsec:sm_p1_tilted_critical_mode}

The susceptibility divergence signals that the quadratic stiffness of one collective overlap fluctuation vanishes at $\eta_c(1)$. The physical content of this mode depends on its projection onto the terminal history coordinate $z$, which corresponds to the output spin at a given site. A finite terminal component means that the instability directly reorganizes the overlap statistics of sampled bitstrings. At zero overlap source, $\lambda=0$, the two physical replicas factorize as in Eq.~\eqref{eq:sm_p1_tilted_replicon_product_matrices}. It is therefore sufficient to consider one physical replica and the mutual-overlap block between two distinct auxiliary replicas. On the active Keldysh-history coordinates $r,r'\in\{v,z,u\}$, we write
\begin{equation}
    \bigl[\mathsf P_{\alpha\alpha'}\bigr]_{rr'}=\mathsf Q^{(n)}_{(\alpha,r),(\alpha',r')},\qquad \alpha\neq\alpha'.
    \label{eq:sm_p1_auxiliary_overlap_block}
\end{equation}
The terminal entry $[\mathsf P_{\alpha\alpha'}]_{zz}$ is the replica representation of the overlap between independent outputs drawn from the same tilted QAOA distribution. The notation $\mathsf P_{\alpha\alpha'}$ distinguishes this auxiliary-replica overlap block from the physical cross-sample block $\mathsf R$ introduced in Eq.~\eqref{eq:sm_two_replica_block_matrix}. At the critical point we define
\begin{equation}
    \mathsf G_c=\mathsf G_{\eta_c(1)},\qquad \mathsf D_c=\mathsf D_{\eta_c(1)},\qquad \mathsf M_c=\mathsf G_c\mathsf D_c.
    \label{eq:sm_p1_critical_M_matrix}
\end{equation}
The critical one-sample saddle follows from Eq.~\eqref{eq:sm_p1_tilted_saddle_equations} together with the instability condition in Eq.~\eqref{eq:sm_p1_tilted_instability_scalar}. Numerically
\begin{equation}
    \rho_c=0.6675221370\ldots,\qquad a_c=0.6702402648\ldots,\qquad b_c=0.3138855605\ldots,
    \label{eq:sm_p1_critical_single_replica_fields}
\end{equation}
with $\eta_c(1)$ given in Eq.~\eqref{eq:sm_p1_tilted_eta_c}. Equation~\eqref{eq:sm_p1_tilted_instability_scalar} implies that $\mathsf M_c$ has an eigenvalue $i$. To write the corresponding eigenmode compactly, we define
\begin{equation}
    \Delta_c=5-\rho_c^2,\qquad \xi_c=2a_c+(1+\rho_c)b_c,\qquad \zeta_c=(\rho_c-1)a_c+2b_c,
    \label{eq:sm_p1_critical_mode_combinations}
\end{equation}
and
\begin{equation}
    \mathcal N_c=1+\frac{8\eta_c(1)\xi_c\zeta_c}{\Delta_c^2}.
    \label{eq:sm_p1_critical_mode_normalization}
\end{equation}
For the optimized depth-one angles $\gamma=1/2$ and $\beta=\pi/8$, the right eigenvector can be normalized so that its terminal component is unity
\begin{equation}
    \mathsf M_c\boldsymbol u_c=i\boldsymbol u_c,\qquad \boldsymbol e_z^{\mathsf T}\boldsymbol u_c=1,\qquad \boldsymbol u_c=\begin{pmatrix}\dfrac{2\eta_c(1)}{\Delta_c}(\xi_c+i\zeta_c)\\[2mm]1\\[1mm]\dfrac{2\eta_c(1)}{\Delta_c}(\xi_c-i\zeta_c)\end{pmatrix}=\begin{pmatrix}0.594797619+0.129219977i\\1\\0.594797619-0.129219977i\end{pmatrix}.
    \label{eq:sm_p1_critical_right_mode}
\end{equation}
Because the stability map in Eq.~\eqref{eq:sm_p1_tilted_replicon_map} contains an algebraic transpose, the corresponding critical overlap block is
\begin{equation}
    \mathsf P_c=\boldsymbol u_c\boldsymbol u_c^{\mathsf T},\qquad -\mathsf M_c\mathsf P_c\mathsf M_c^{\mathsf T}=\mathsf P_c,\qquad (\mathsf P_c)_{zz}=1.
    \label{eq:sm_p1_critical_history_block}
\end{equation}
Because $(\mathsf P_c)_{zz}=1$, the soft mode has finite terminal-spin weight and therefore directly controls the output-overlap response. The corresponding algebraic left eigenvector projects the response onto this soft direction. With $\boldsymbol\ell_c^{\mathsf T}\boldsymbol u_c=1$
\begin{equation}
    \boldsymbol\ell_c^{\mathsf T}\mathsf M_c=i\boldsymbol\ell_c^{\mathsf T},\qquad \boldsymbol\ell_c=\frac{1}{\mathcal N_c}\begin{pmatrix}\dfrac{\zeta_c-i\xi_c}{\Delta_c}\\[2mm]1\\[1mm]\dfrac{\zeta_c+i\xi_c}{\Delta_c}\end{pmatrix}=\begin{pmatrix}0.073386029-0.337794792i\\0.825400659\\0.073386029+0.337794792i\end{pmatrix}.
    \label{eq:sm_p1_critical_left_mode}
\end{equation}
The transition is controlled by this single overlap mode. Factoring the known eigenvalue $i$ from the characteristic polynomial of $\mathsf M_c$ gives the remaining two eigenvalues
\begin{equation}
    \mu_\pm=\frac{i[\eta_c(1)-1]\pm\sqrt{\eta_c(1)d_c-[\eta_c(1)-1]^2}}{2}=\pm0.295761079-0.136653408i,
    \label{eq:sm_p1_critical_nonsingular_M_eigenvalues}
\end{equation}
where $d_c=\det(\mathsf G_c)$. The nearest noncritical eigenvalue of the stability map remains a distance $0.893851\ldots$ from unity. The soft block $\mathsf P_c$ is therefore spectrally isolated from the remaining Keldysh-history directions, which can be treated perturbatively near the transition. The rate at which this mode softens fixes the quadratic coefficient of the effective theory. Let $\boldsymbol s=(\rho,a,b)^{\mathsf T}$ and write the one-sample saddle equations in Eq.~\eqref{eq:sm_p1_tilted_saddle_equations} as $\boldsymbol F(\boldsymbol s,\eta)=\boldsymbol0$. Along the continued one-sample saddle
\begin{equation}
    \dot{\boldsymbol s}_c\equiv\left.\frac{d\boldsymbol s}{d\eta}\right|_{\eta_c(1)}=-\left[\partial_{\boldsymbol s}\boldsymbol F\right]_c^{-1}\left[\partial_\eta\boldsymbol F\right]_c.
    \label{eq:sm_p1_critical_saddle_derivative}
\end{equation}
The one-sample saddle remains regular at $\eta_c(1)$, while the singularity develops in the overlap sector. We therefore obtain
\begin{equation}
    \dot{\mathsf M}_c=\left(\sum_j\dot s_{c,j}\frac{\partial\mathsf G}{\partial s_j}\right)_c\mathsf D_c+\mathsf G_c\operatorname{diag}(0,i,0),\qquad \dot\mu_c=\boldsymbol\ell_c^{\mathsf T}\dot{\mathsf M}_c\boldsymbol u_c.
    \label{eq:sm_p1_critical_eigenvalue_derivative}
\end{equation}
Defining
\begin{equation}
    \kappa_c=-2i\dot\mu_c=1.785504316\ldots,
    \label{eq:sm_p1_critical_eigenvalue_slope_coefficient}
\end{equation}
the soft eigenvalue behaves as
\begin{equation}
    1+\mu_c(\eta)^2=\kappa_c\bigl[\eta_c(1)-\eta\bigr]+O\left(\bigl[\eta_c(1)-\eta\bigr]^2\right).
    \label{eq:sm_p1_critical_eigenvalue_slope}
\end{equation}
Projecting the response in Eq.~\eqref{eq:sm_p1_tilted_chi_closed_form} onto this mode gives
\begin{equation}
    \chi_{1,\eta}=\frac{\bigl(\boldsymbol\ell_c^{\mathsf T}\boldsymbol c_c\bigr)^2}{\kappa_c[\eta_c(1)-\eta]}+O(1)=\frac{0.722547735\ldots}{\eta_c(1)-\eta}+O(1),
    \label{eq:sm_p1_critical_susceptibility_pole}
\end{equation}
where $\boldsymbol c_c=\mathsf G_c\boldsymbol e_z$ and $(\boldsymbol\ell_c^{\mathsf T}\boldsymbol c_c)^2=1.290112100\ldots$. Equivalently
\begin{equation}
    \chi_{1,\eta}^{-1}=r_{1,c}'\bigl[\eta_c(1)-\eta\bigr]+O\left(\bigl[\eta_c(1)-\eta\bigr]^2\right),\qquad r_{1,c}'=\frac{\kappa_c}{(\boldsymbol\ell_c^{\mathsf T}\boldsymbol c_c)^2}=1.383991605\ldots.
    \label{eq:sm_p1_critical_inverse_susceptibility}
\end{equation}
The overlap stiffness therefore vanishes linearly on approaching the transition. This is the quadratic (or mass) term of the Landau theory developed next and gives the thermodynamic susceptibility exponent $\gamma_{\rm crit}=1$.

\subsection{Landau theory for the soft overlap mode}
\label{subsec:sm_p1_tilted_projected_landau_action}

A continuous transition controlled by a single soft mode can be described by retaining that mode explicitly and allowing the remaining massive fields to adjust to it. We therefore decompose the overlap block between distinct auxiliary replicas as
\begin{equation}
    \mathsf P_{\alpha\alpha'}=q_{\alpha\alpha'}\mathsf P_c+\mathsf Y_{\alpha\alpha'},
    \label{eq:sm_p1_auxiliary_mode_decomposition}
\end{equation}
where $q_{\alpha\alpha'}$ is the scalar amplitude of the critical mode and $\mathsf Y_{\alpha\alpha'}$ contains the gapped Keldysh-history components. We fix the decomposition through
\begin{equation}
    \boldsymbol\ell_c^{\mathsf T}\mathsf Y_{\alpha\alpha'}\boldsymbol\ell_c=0.
    \label{eq:sm_p1_massive_mode_orthogonality}
\end{equation}
The diagonal collective blocks are allowed to relax at the same time. Eliminating these gapped degrees of freedom order by order leaves an effective action for the overlap amplitudes $q_{\alpha\alpha'}$, whose replica-index structure describes the mutual organization of independent samples drawn from the tilted distribution. The form of this effective action is constrained by the global spin-flip symmetry of the SK problem. In particular, each auxiliary replica can be flipped independently according to $\boldsymbol a^\alpha\mapsto s_\alpha\boldsymbol a^\alpha$, with $s_\alpha=\pm1$, which sends
\begin{equation}
    q_{\alpha\alpha'}\longmapsto s_\alpha s_{\alpha'}q_{\alpha\alpha'}.
    \label{eq:sm_p1_auxiliary_replica_spin_flip_symmetry}
\end{equation}
A cubic power involving a single replica pair changes sign under this symmetry. The lowest allowed cubic invariant instead couples three distinct replicas through the triangular product $q_{\alpha\alpha'}q_{\alpha\alpha''}q_{\alpha'\alpha''}$. This is the same cubic replica invariant that controls the zero-field SK transition~\cite{Parisi1979,MezardParisiVirasoro1987,BinderYoung1986,CastellaniCavagna2005}. We denote by $\mathcal A_{n,1}^{(1)}$ the one-physical-replica contribution to the replicated tilted action at $\lambda=0$, and by $\mathcal A_{n,1}^{(1,0)}$ its value on the auxiliary-replica-diagonal saddle. After the gapped fields are eliminated, the Landau expansion through quartic order is
\begin{equation}
    \mathcal A_{n,1}^{(1),\rm eff}-\mathcal A_{n,1}^{(1,0)}=\frac{r_1(\eta)}{2}\sum_{\alpha<\alpha'}q_{\alpha\alpha'}^2+w_1\sum_{\alpha<\alpha'<\alpha''}q_{\alpha\alpha'}q_{\alpha\alpha''}q_{\alpha'\alpha''}+\mathcal U_1\mathcal I_1+\mathcal U_2\mathcal I_2+\mathcal U_3\mathcal I_3+O\Bigl(q^5,\delta\eta\,q^3,\delta\eta^2q^2\Bigr),
    \label{eq:sm_p1_projected_landau_action}
\end{equation}
where the analytic continuation of the quadratic coefficient across the transition satisfies $r_1(\eta)=-\chi_{1,\eta}^{-1}+O(\delta\eta^2)=r_{1,c}'\delta\eta+O(\delta\eta^2)$. The independent quartic replica contractions are
\begin{equation}
    \mathcal I_1=\sum_{\alpha<\alpha'}q_{\alpha\alpha'}^4,
    \label{eq:sm_p1_quartic_invariant_I1}
\end{equation}
and
\begin{equation}
    \mathcal I_2=\sum_\alpha\sum_{\substack{\alpha'<\alpha''\\\alpha',\alpha''\neq\alpha}}q_{\alpha\alpha'}^2q_{\alpha\alpha''}^2,
    \label{eq:sm_p1_quartic_invariant_I2}
\end{equation}
and
\begin{equation}
    \mathcal I_3=\sum_{\alpha<\alpha'<\alpha''<\alpha'''}\Bigl(q_{\alpha\alpha'}q_{\alpha\alpha''}q_{\alpha'\alpha'''}q_{\alpha''\alpha'''}+q_{\alpha\alpha'}q_{\alpha\alpha'''}q_{\alpha'\alpha''}q_{\alpha''\alpha'''}+q_{\alpha\alpha''}q_{\alpha\alpha'''}q_{\alpha'\alpha''}q_{\alpha'\alpha'''}\Bigr).
    \label{eq:sm_p1_quartic_invariant_I3}
\end{equation}
The three invariants distinguish the possible ways in which four overlap amplitudes can share replica indices. Their coefficients determine whether the ordered phase can be described by a single mutual overlap or instead develops a hierarchy of overlaps. At depth one, these coefficients can be evaluated directly from the finite one-site history space. Using the critical mode defined in Eqs.~\eqref{eq:sm_p1_critical_M_matrix} and~\eqref{eq:sm_p1_critical_right_mode}, we define
\begin{equation}
    \boldsymbol h_c=\mathsf D_c\boldsymbol u_c,\qquad \vartheta_c(\boldsymbol a)=\boldsymbol h_c^{\mathsf T}\boldsymbol a,\qquad \boldsymbol a=(v,z,u)^{\mathsf T}.
    \label{eq:sm_p1_critical_projected_variable}
\end{equation}
The depth-one one-site trace contains only the eight histories $(v,z,u)\in\{\pm1\}^3$. Global spin-flip symmetry fixes the normalized critical Keldysh weight in terms of the pair correlations of $\mathsf G_c$
\begin{equation}
    \pi_c(v,z,u)=\frac{1}{8}\left[1+(a_c+ib_c)vz+\rho_cvu+(a_c-ib_c)zu\right],\qquad \langle f\rangle_c=\sum_{v,z,u=\pm1}\pi_c(v,z,u)f(v,z,u).
    \label{eq:sm_p1_critical_one_site_weight}
\end{equation}
Although this Keldysh weight is generally complex, the finite depth-one history space allows its projected moments to be evaluated exactly as finite sums. In particular,
\begin{equation}
    m_2\equiv\langle\vartheta_c^2\rangle_c=\boldsymbol h_c^{\mathsf T}\mathsf G_c\boldsymbol h_c=i\boldsymbol u_c^{\mathsf T}\mathsf D_c\boldsymbol u_c=-\eta_c(1)\mathcal N_c=-0.880412654\ldots,
    \label{eq:sm_p1_critical_projected_second_moment}
\end{equation}
\begin{equation}
    m_4\equiv\langle\vartheta_c^4\rangle_c=\sum_{v,z,u=\pm1}\pi_c(v,z,u)\vartheta_c(v,z,u)^4=0.389414212\ldots.
    \label{eq:sm_p1_critical_projected_fourth_moment}
\end{equation}
A critical-mode coupling between auxiliary replicas $\alpha$ and $\alpha'$ inserts $-q_{\alpha\alpha'}\vartheta_c(\boldsymbol a^\alpha)\vartheta_c(\boldsymbol a^{\alpha'})$. The connected triangular contraction therefore gives
\begin{equation}
    w_1=-m_2^3=\eta_c(1)^3\mathcal N_c^3=0.682431127\ldots\neq0.
    \label{eq:sm_p1_cubic_triangle_coefficient}
\end{equation}
The nonzero cubic coefficient shows that the depth-one transition inherits the leading replica invariant of the zero-field SK transition and allows the overlap order parameter to emerge linearly with the distance from criticality~\cite{Parisi1979,MezardParisiVirasoro1987,CastellaniCavagna2005}. How this emerging overlap is organized is then determined by the quartic terms, whose stationary values include the response of the gapped Keldysh fields. To isolate this contribution, we first hold those fields at their critical values, for which the one-site cumulants give the bare coefficients
\begin{equation}
    \mathcal U_1^{\rm bare}=\frac{m_4^2-3m_2^4}{24},\qquad \mathcal U_2^{\rm bare}=\frac{m_2^2(m_4-m_2^2)}{4},\qquad \mathcal U_3^{\rm bare}=m_2^4.
    \label{eq:sm_p1_bare_quartic_coefficients}
\end{equation}
Once the soft overlap develops, the massive diagonal and off-diagonal fields respond at order $q^2$, so their relaxation feeds back at order $q^4$ and renormalizes the physical quartic coefficients, as standard when massive modes are eliminated from a Landau theory~\cite{MezardParisiVirasoro1987}. We extract these relaxed coefficients using three finite auxiliary-replica configurations that isolate the independent quartic contractions: one pair for $k=2$, with $[\mathsf P_{12}]_{zz}=s$; two pairs sharing one replica for $k=3$, with $[\mathsf P_{12}]_{zz}=[\mathsf P_{13}]_{zz}=s$; and a four-pair cycle for $k=4$, with $[\mathsf P_{12}]_{zz}=[\mathsf P_{13}]_{zz}=[\mathsf P_{24}]_{zz}=[\mathsf P_{34}]_{zz}=s$. For each $k\in\{2,3,4\}$, let $\Phi^{[k]}(s,\boldsymbol y)$ denote the corresponding one-physical-replica contribution to the exact replicated action in Eq.~\eqref{eq:sm_replicated_action} before stationarity is imposed, where $\boldsymbol y$ collects the gapped diagonal and off-diagonal fields in shifted coordinates with $\boldsymbol y=\boldsymbol0$ at $s=0$. We define
\begin{equation}
    \mathsf H^{[k]}=\left[\nabla_{\boldsymbol y}\nabla_{\boldsymbol y}^{\mathsf T}\Phi^{[k]}\right]_{s=0,\boldsymbol y=\boldsymbol0},\qquad \boldsymbol B^{[k]}=\left[\partial_s^2\nabla_{\boldsymbol y}\Phi^{[k]}\right]_{s=0,\boldsymbol y=\boldsymbol0}.
    \label{eq:sm_p1_quartic_gapped_derivatives}
\end{equation}
The spectral separation established above makes $\mathsf H^{[k]}$ invertible on the gapped subspace. The induced relaxation is
\begin{equation}
    \boldsymbol y_\star^{[k]}(s)=-\frac{s^2}{2}\bigl(\mathsf H^{[k]}\bigr)^{-1}\boldsymbol B^{[k]}+O(s^3).
    \label{eq:sm_p1_quartic_gapped_relaxation}
\end{equation}
Substituting this response back into the action gives the relaxed fourth-order coefficient
\begin{equation}
    C^{[k]}=\frac{1}{4!}\left[\partial_s^4\Phi^{[k]}-3\bigl(\boldsymbol B^{[k]}\bigr)^{\mathsf T}\bigl(\mathsf H^{[k]}\bigr)^{-1}\boldsymbol B^{[k]}\right]_{s=0,\boldsymbol y=\boldsymbol0}.
    \label{eq:sm_p1_quartic_relaxed_schur_complement}
\end{equation}
The first contribution is the fixed-field fourth-order cumulant and the second is the correction generated by relaxation of the massive modes. The three configurations isolate the invariant coefficients according to
\begin{equation}
    C^{[2]}=\mathcal U_1,\qquad C^{[3]}=2\mathcal U_1+\mathcal U_2,\qquad C^{[4]}=4\mathcal U_1+4\mathcal U_2+\mathcal U_3.
    \label{eq:sm_p1_quartic_coefficient_relations}
\end{equation}
All quantities in Eq.~\eqref{eq:sm_p1_quartic_relaxed_schur_complement} are finite sums over $8^k$ one-site histories followed by finite matrix operations. Evaluating them at the critical saddle gives
\begin{equation}
    \mathcal U_1=-0.05411\ldots,\qquad \mathcal U_2=0.02297\ldots,\qquad \mathcal U_3=0.93294\ldots.
    \label{eq:sm_p1_relaxed_quartic_coefficients}
\end{equation}
Two combinations control the ordered phase
\begin{equation}
    g_1=4\mathcal U_1-4\mathcal U_2+\mathcal U_3=0.62466\ldots>0,
    \label{eq:sm_p1_parisi_g_coefficient}
\end{equation}
and
\begin{equation}
    8(\mathcal U_1-\mathcal U_2)=-0.6166\ldots<0.
    \label{eq:sm_p1_rs_replicon_coefficient}
\end{equation}
The first coefficient enters the continuous hierarchical solution and the second controls the stability of the uniform nonzero-overlap branch.

\subsection{Instability of the auxiliary-replica-symmetric continuation}
\label{subsec:sm_p1_tilted_rs_ordered_instability}

The simplest ordered state assigns the same nonzero mutual overlap to every pair of auxiliary replicas. Physically, this corresponds to a replica-symmetric phase characterized by a single overlap scale. We therefore set
\begin{equation}
    q_{\alpha\alpha'}=q_{\rm RS},\qquad \alpha\neq\alpha'.
    \label{eq:sm_p1_rs_ordered_ansatz}
\end{equation}
Taking $n\to0$ in Eq.~\eqref{eq:sm_p1_projected_landau_action} gives the stationarity condition
\begin{equation}
    0=r_1q_{\rm RS}-2w_1q_{\rm RS}^2+\bigl(4\mathcal U_1-8\mathcal U_2+6\mathcal U_3\bigr)q_{\rm RS}^3+O(q_{\rm RS}^4).
    \label{eq:sm_p1_rs_ordered_stationarity}
\end{equation}
The nonzero branch turns on continuously and linearly
\begin{equation}
    q_{\rm RS}=\frac{r_1}{2w_1}+O(r_1^2)=1.014016\ldots\delta\eta+O(\delta\eta^2).
    \label{eq:sm_p1_rs_ordered_onset}
\end{equation}
This linear onset follows directly from the cubic replica invariant and already matches the leading order-parameter scaling of the equilibrium SK transition~\cite{Parisi1979,MezardParisiVirasoro1987}. The stability of this branch is determined by fluctuations that redistribute the pair overlaps while leaving the uniform replica-symmetric component unchanged. These are the standard replicon fluctuations of spin-glass theory~\cite{deAlmeidaThouless1978,MezardParisiVirasoro1987,CastellaniCavagna2005}. We denote their signed Landau curvature by $\Lambda_{\rm rep}^{\rm RS}$. Marginality occurs at $\Lambda_{\rm rep}^{\rm RS}=0$, while a negative curvature means that the action is lowered by developing a nonuniform pattern of overlaps. The curvature is
\begin{equation}
    \Lambda_{\rm rep}^{\rm RS}=r_1-2w_1q_{\rm RS}+\bigl(12\mathcal U_1-16\mathcal U_2+6\mathcal U_3\bigr)q_{\rm RS}^2+O(q_{\rm RS}^3).
    \label{eq:sm_p1_rs_ordered_replicon_before_stationarity}
\end{equation}
Using Eq.~\eqref{eq:sm_p1_rs_ordered_stationarity} to eliminate $r_1$ gives
\begin{equation}
    \Lambda_{\rm rep}^{\rm RS}=8(\mathcal U_1-\mathcal U_2)q_{\rm RS}^2+O(q_{\rm RS}^3)<0.
    \label{eq:sm_p1_rs_ordered_replicon}
\end{equation}
The uniform nonzero-overlap branch is therefore unstable immediately above $\eta_c(1)$. The ordered side of the transition must allow different replica pairs to organize at different overlap scales. This is the same mechanism that selects replica-symmetry breaking at the equilibrium SK transition. Using the depth-zero coefficients collected in Eq.~\eqref{eq:sm_p0_parisi_landau_coefficients}, Eq.~\eqref{eq:sm_p1_rs_ordered_replicon} gives $\Lambda_{\rm rep}^{\rm RS}=-(2/3)q_{\rm RS}^2+\cdots$ at $p=0$, reproducing the standard instability of the replica-symmetric SK spin-glass solution~\cite{deAlmeidaThouless1978,Parisi1979,MezardParisiVirasoro1987,CastellaniCavagna2005}. The depth-one transition therefore follows the same route from a soft overlap mode to an unstable uniform-overlap state.

\subsection{Continuous Parisi hierarchy}
\label{subsec:sm_p1_tilted_parisi_solution}

The instability of the uniform-overlap branch motivates the continuous hierarchical organization familiar from the SK model. In the $n\to0$ limit, the mutual overlaps are described by the Parisi function $q(x)$, with $x\in[0,1]$ labeling levels of the replica hierarchy~\cite{Parisi1979,MezardParisiVirasoro1987,BinderYoung1986,CastellaniCavagna2005}. Here $q(x)$ is the scalar amplitude of the soft Keldysh-history block $\mathsf P_c$. Standard Parisi matrix algebra gives
\begin{equation}
    d_2=(q^2)_{\alpha\alpha}=-\int_0^1q(y)^2\,dy,
    \label{eq:sm_p1_parisi_d2}
\end{equation}
together with the Parisi transform
\begin{equation}
    \widehat q(x)=-xq(x)-\int_x^1q(y)\,dy.
    \label{eq:sm_p1_parisi_fourier_transform}
\end{equation}
We use $[q^k](x)$ for the off-diagonal Parisi function associated with the $k$th matrix product of $q_{\alpha\alpha'}$. Varying the Landau action in Eq.~\eqref{eq:sm_p1_projected_landau_action} gives
\begin{equation}
    0=r_1q(x)+w_1[q^2](x)+\mathcal U_3[q^3](x)+\bigl(4\mathcal U_2-2\mathcal U_3\bigr)d_2q(x)+g_1q(x)^3.
    \label{eq:sm_p1_parisi_stationarity}
\end{equation}
This is the depth-one analogue of the near-critical SK Parisi equation. The QAOA circuit enters through the coefficients of the effective theory, while the replica-index structure is the same as in the equilibrium SK problem. On an interval where $q(x)$ varies continuously, differentiating Eq.~\eqref{eq:sm_p1_parisi_stationarity} using the standard product rules for Parisi matrices gives
\begin{equation}
    0=r_1+2w_1\widehat q(x)+3\mathcal U_3\widehat q(x)^2+\bigl(4\mathcal U_2-2\mathcal U_3\bigr)d_2+3g_1q(x)^2.
    \label{eq:sm_p1_parisi_marginality_equation}
\end{equation}
The left-hand side is also the replicon curvature in the critical sector. Its vanishing throughout the continuous part of the solution expresses the marginal stability characteristic of the Parisi phase~\cite{Parisi1979,MezardParisiVirasoro1987,CastellaniCavagna2005}. Differentiating once more and using $d\widehat q/dx=-x\dot q(x)$ gives
\begin{equation}
    3g_1q(x)=x\left[w_1+3\mathcal U_3\widehat q(x)\right].
    \label{eq:sm_p1_parisi_differential_relation}
\end{equation}
The solution has a continuous part extending up to a breakpoint $x_c$, followed by a plateau at the Edwards--Anderson endpoint overlap $q_{\rm EA}$~\cite{EdwardsAnderson1975,Parisi1979}
\begin{equation}
    q(x)=\begin{cases}\displaystyle q_{\rm EA}\frac{x}{x_c}\sqrt{\frac{g_1+\mathcal U_3x_c^2}{g_1+\mathcal U_3x^2}}, & 0\leq x\leq x_c,\\[4mm]q_{\rm EA}, & x_c\leq x\leq1.\end{cases}
    \label{eq:sm_p1_parisi_order_parameter}
\end{equation}
The endpoint and breakpoint satisfy
\begin{equation}
    3g_1q_{\rm EA}=x_c\left(w_1-3\mathcal U_3q_{\rm EA}\right),
    \label{eq:sm_p1_parisi_endpoint_relation}
\end{equation}
and
\begin{equation}
    0=r_1-2w_1q_{\rm EA}+3(\mathcal U_3+g_1)q_{\rm EA}^2+\bigl(4\mathcal U_2-2\mathcal U_3\bigr)d_2.
    \label{eq:sm_p1_parisi_plateau_equation}
\end{equation}
For the profile in Eq.~\eqref{eq:sm_p1_parisi_order_parameter}
\begin{equation}
    d_2=-q_{\rm EA}^2\Bigg\{1-x_c+\frac{g_1+\mathcal U_3x_c^2}{x_c^2}\Bigg[\frac{x_c}{\mathcal U_3}-\frac{\sqrt{g_1}}{\mathcal U_3^{3/2}}\arctan\left(x_c\sqrt{\frac{\mathcal U_3}{g_1}}\right)\Bigg]\Bigg\}.
    \label{eq:sm_p1_parisi_d2_explicit}
\end{equation}
Equations~\eqref{eq:sm_p1_parisi_endpoint_relation}--\eqref{eq:sm_p1_parisi_d2_explicit} determine the near-critical solution for $0<\delta\eta\ll1$. The physics of this solution differs qualitatively from the uniform branch: instead of organizing around a single replica-symmetric overlap, independent tilted QAOA outputs span a continuous range of mutual-overlap scales. This continuum is marginal, with Eq.~\eqref{eq:sm_p1_parisi_marginality_equation} vanishing throughout the interval, precisely as in the near-critical continuous Parisi phase of the SK model~\cite{Parisi1979,MezardParisiVirasoro1987,BinderYoung1986,CastellaniCavagna2005}.

\subsection{Critical scaling and output-overlap geometry}
\label{subsec:sm_p1_tilted_parisi_critical_scaling}

We now translate the hierarchical saddle into the observable overlap geometry of the tilted QAOA outputs. Expanding Eqs.~\eqref{eq:sm_p1_parisi_endpoint_relation} and~\eqref{eq:sm_p1_parisi_plateau_equation} close to the transition gives
\begin{equation}
    q_{\rm EA}=\frac{r_{1,c}'}{2w_1}\delta\eta+O(\delta\eta^2)=1.014016\ldots\delta\eta+O(\delta\eta^2),
    \label{eq:sm_p1_parisi_qea_scaling}
\end{equation}
\begin{equation}
    x_c=\frac{3g_1r_{1,c}'}{2w_1^2}\delta\eta+O(\delta\eta^2)=2.7845\ldots\delta\eta+O(\delta\eta^2),
    \label{eq:sm_p1_parisi_breakpoint_scaling}
\end{equation}
and, on the continuous interval
\begin{equation}
    q(x)=\frac{w_1}{3g_1}x+O(\delta\eta^2)=0.36416\ldots x+O(\delta\eta^2),\qquad 0\leq x\leq x_c.
    \label{eq:sm_p1_parisi_ramp_scaling}
\end{equation}
The Edwards--Anderson endpoint overlap and the width of the continuous hierarchy both grow linearly with the distance from the transition. The full auxiliary-replica overlap block is, to leading order
\begin{equation}
    \mathsf P(x)=q(x)\mathsf P_c+O(\delta\eta^2).
    \label{eq:sm_p1_matrix_valued_parisi_order_parameter}
\end{equation}
Because the critical mode was normalized by $(\mathsf P_c)_{zz}=1$, its terminal component is
\begin{equation}
    \bigl[\mathsf P(x)\bigr]_{zz}=q(x)+O(\delta\eta^2).
    \label{eq:sm_p1_parisi_output_overlap_function}
\end{equation}
This equality therefore identifies $q(x)$, to leading order, with the hierarchy of overlaps between sampled QAOA bitstrings. To leading order, the Parisi function $q(x)$ describes the hierarchy of overlaps between measured bitstrings, while the Keldysh-history structure determines the soft mode and renormalizes its coefficients. The finite terminal projection carries this structure into the sampled configurations, so that in the positive-overlap sector selected by $\lambda\to0^+$, their overlap density is determined by $dx/dq$~\cite{Parisi1979,MezardParisiVirasoro1987,BinderYoung1986}. Near the transition
\begin{equation}
    \frac{dx}{dq}=\frac{3g_1}{w_1}+O(\delta\eta)=2.7460\ldots+O(\delta\eta),\qquad 0<q<q_{\rm EA}.
    \label{eq:sm_p1_parisi_continuous_overlap_density}
\end{equation}
The continuous part carries total weight $x_c=O(\delta\eta)$, while the plateau gives a delta contribution of weight $1-x_c$ at $q=q_{\rm EA}+O(\delta\eta^2)$. At strictly zero source, global spin-flip symmetry produces the corresponding structure at both positive and negative overlap. The transition therefore introduces a continuous hierarchy of output overlaps whose support grows linearly from the origin. The singular contribution to the stationary replicated action begins at cubic order in the distance from the transition
\begin{equation}
    \lim_{n\to0}\frac{1}{n}\left(\mathcal A_{n,1}^{(1),\rm Parisi}-\mathcal A_{n,1}^{(1,0)}\right)=-\frac{(r_{1,c}')^3}{48w_1^2}\delta\eta^3+O(\delta\eta^4)=-0.11859\ldots\delta\eta^3+O(\delta\eta^4).
    \label{eq:sm_p1_parisi_singular_action}
\end{equation}
Together with the linear onset of $q_{\rm EA}$ and the susceptibility pole obtained in Eq.~\eqref{eq:sm_p1_critical_susceptibility_pole}, this gives
\begin{equation}
    \beta_{\rm crit}=1,\qquad \gamma_{\rm crit}=1,\qquad \alpha_{\rm crit}=-1.
    \label{eq:sm_p1_parisi_critical_exponents}
\end{equation}
Here the singular free entropy is defined conventionally as $|\delta\eta|^{2-\alpha_{\rm crit}}$. The order-parameter exponent follows from Eq.~\eqref{eq:sm_p1_parisi_qea_scaling}, the susceptibility exponent from Eq.~\eqref{eq:sm_p1_critical_susceptibility_pole}, and the free-entropy singularity from Eq.~\eqref{eq:sm_p1_parisi_singular_action}. These are the mean-field critical exponents of the equilibrium SK transition~\cite{Parisi1979,BinderYoung1986,MezardParisiVirasoro1987,CastellaniCavagna2005}. We test these predictions with finite-size scaling in Sec.~\ref{subsec:sm_finite_size_tilted_critical_scaling}.

\subsection{Comparison with the Gibbs SK transition and scope}
\label{subsec:sm_p1_tilted_parisi_sk_comparison}

The connection to the equilibrium SK transition can be made quantitative by expanding the standard $p=0$ replicated action around the paramagnetic saddle in the same invariant basis~\cite{Parisi1979,MezardParisiVirasoro1987,BinderYoung1986,CastellaniCavagna2005}. This gives
\begin{equation}
    r_0(\eta)=2(\eta-1)+O\left((\eta-1)^2\right),\qquad w_0=1,\qquad \mathcal U_1^{(0)}=-\frac{1}{12},\qquad \mathcal U_2^{(0)}=0,\qquad \mathcal U_3^{(0)}=1.
    \label{eq:sm_p0_parisi_landau_coefficients}
\end{equation}
Consequently, $g_0=2/3$, and the near-critical solution reduces to
\begin{equation}
    q_{\rm EA}^{(0)}=\eta-1+O\left((\eta-1)^2\right),\qquad x_c^{(0)}=2(\eta-1)+O\left((\eta-1)^2\right),\qquad q_0(x)=\frac{x}{2}+O\left((\eta-1)^2\right).
    \label{eq:sm_p0_parisi_near_critical_solution}
\end{equation}
This is the standard continuous Parisi solution on the spin-glass side of the zero-field SK transition~\cite{Parisi1979,MezardParisiVirasoro1987,BinderYoung1986}. The comparison separates the universal critical structure from the depth-one QAOA dressing. The Keldysh histories modify the critical eigenvector and renormalize the coefficients $r_{1,c}'$, $w_1$, and $\mathcal U_i$, and therefore the numerical amplitudes of $q_{\rm EA}$ and $x_c$. The leading replica invariants remain those of the SK theory. The uniform nonzero-overlap branch is unstable in the same replicon sector, the ordered solution is a marginal continuous hierarchy, and the critical exponents agree with those of the equilibrium SK transition.

This gives a direct physical interpretation of the critical energy in Eq.~\eqref{eq:sm_p1_tilted_eta_c_energy}. For $\eta<\eta_c(1)$, the tilted depth-one distribution is described by the auxiliary-replica-diagonal saddle and has vanishing thermodynamic overlap at zero source. At $\eta_c(1)$ its overlap stiffness vanishes. For $0<\delta\eta\ll1$, the resulting ordered phase is described by the continuous Parisi hierarchy derived above. Because the soft mode has a finite terminal component, this hierarchy appears directly in the overlap distribution of the QAOA. The energy $E_1\bigl(\eta_c(1)\bigr)$ therefore marks the onset of Parisi overlap organization in the low-energy tail within the thermodynamic depth-one calculation.

The controlled result applies in the near-critical regime $0<\delta\eta\ll1$ at $p=1$. It establishes the structure immediately beyond the instability and provides the local critical behavior used throughout the main text. Extending the same construction deeper into the tilted tail requires following the nonlinear saddle beyond the Landau regime. At larger QAOA depths, the calculations of Sec.~\ref{sec:sm_auxiliary_replica_stability} establish the corresponding linear instabilities, while the decreasing $\Delta E_c(p)$ and $I_c(p)$ show that these thresholds move toward the typical distribution and become less exponentially costly. If both separations vanish as $p\to\infty$, the resulting scenario is that the Parisi-organized tail identified at finite depth merges asymptotically with the typical QAOA distribution.

\section{Finite-size crossover and critical scaling}
\label{sec:sm_finite_size}

The thermodynamic analysis predicts two distinct finite-size phenomena: a depth-dependent crossover in the ordinary QAOA distribution and critical rounding near the energy-tilted transitions. Exact state-vector simulations allow us to track both. We first characterize the crossover underlying Fig.~4 of the main text and then test the predicted critical scaling at $p=0$ and $p=1$. Throughout, $\chi_{p,N}=N\mathbb E_J\langle q_{AB}^2\rangle_{J,p,N}^{\otimes2}$ denotes the finite-size overlap susceptibility, and $\chi_{p,\infty}\equiv\chi_p$ its thermodynamic limit.

\subsection{Finite-size collapse diagnostic}
\label{subsec:sm_finite_size_collapse_diagnostic}

\begin{figure}[!t]
    \centering
    \includegraphics[width=0.85\columnwidth]{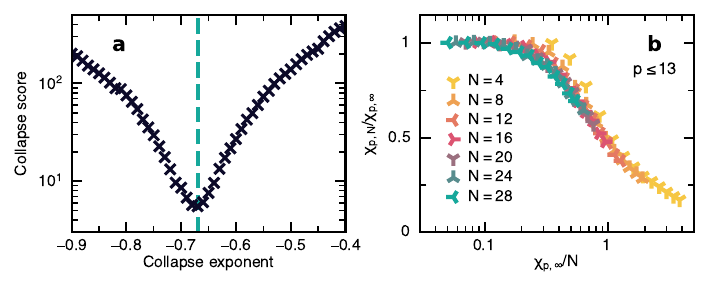}
    \caption{(a) Finite-size collapse score from Eq.~\eqref{eq:sm_finite_size_collapse_score} for the scaling variable $pN^\kappa$. The vertical dashed line marks the minimum of the score at $\kappa=-0.67$, the value used in Fig.~4 of the main text. (b) Parameter-free scaling analysis of the finite-size susceptibility data shown in Fig.~4 of the main text against $\chi_{p,\infty}/N$.}
    \label{fig:finite_size_collapse_sm}
\end{figure}

The collapse in Fig.~4 of the main text uses the empirical scaling variable $pN^{-0.67}$. We estimate this exponent using the data-collapse quality score of Ref.~\cite{PhysRevB.70.014418}, which quantifies how well rescaled data fall onto a scaling function inferred from the data themselves. For each trial exponent $\kappa$, we define $x_{N,p}(\kappa)=pN^\kappa$, $y_{N,p}=\chi_{p,N}/\chi_{p,\infty}$, and denote by $dy_{N,p}$ the statistical uncertainty of $y_{N,p}$. If $\kappa$ is well chosen, the points $y_{N,p}$ should lie close to a single function of $x_{N,p}(\kappa)$, independent of $N$. For each data point $(N,p)$, the method estimates this function locally from neighboring rescaled data at other system sizes. We denote by $Y_{N,p}(\kappa)$ the resulting estimate at $x_{N,p}(\kappa)$ and by $dY_{N,p}(\kappa)$ its uncertainty. The collapse score is
\begin{equation}
    \mathrm{Score}(\kappa)\propto\sum_{\{(N,p)\}}\frac{\Bigl[y_{N,p}-Y_{N,p}(\kappa)\Bigr]^2}{dy_{N,p}^2+dY_{N,p}^2},
    \label{eq:sm_finite_size_collapse_score}
\end{equation}
where smaller values indicate a better collapse. Figure~\ref{fig:finite_size_collapse_sm}a shows the score as a function of $\kappa$. Its clear minimum near $\kappa\simeq-0.67$ identifies the best collapse within this empirical one-parameter scaling analysis. Larger system sizes, greater depths, and an explicit treatment of subleading corrections would be required for a more precise estimate.

The corresponding crossover size $N^\star(p)\sim p^{1/0.67}\simeq p^{1.5}$ is numerically consistent with the thermodynamic growth $\chi_{p,\infty}\sim p^{1.5}$ shown in Fig.~4 of the main text. This agreement suggests that the susceptibility itself sets an overlap-response volume, $N^\star(p)\sim\chi_{p,\infty}$, and motivates the parameter-free scaling variable $\chi_{p,\infty}/N$ used in Fig.~\ref{fig:finite_size_collapse_sm}b. The resulting collapse is slightly less tight than that obtained with the optimized empirical variable $pN^{-0.67}$, with the largest deviations occurring at the smallest system sizes, but it captures the same crossover and supports the response-volume interpretation.

\subsection{Self-averaging of the finite-size overlap susceptibility}
\label{subsec:sm_finite_size_chi_self_averaging}

\begin{figure}[!t]
    \centering
    \includegraphics[width=0.85\columnwidth]{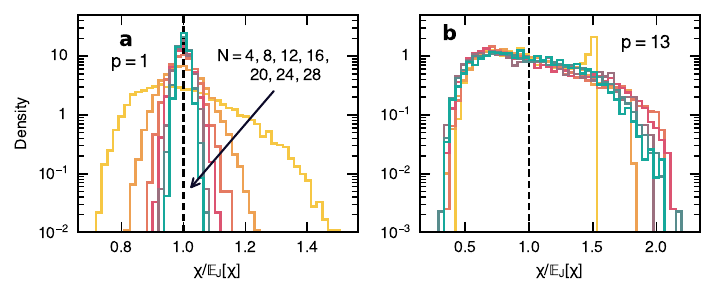}
    \caption{Distribution over disorder realizations of the normalized finite-size susceptibility $\chi_{J,p,N}/\mathbb E_J[\chi_{J,p,N}]$, with $\chi_{J,p,N}$ defined in Eq.~\eqref{eq:sm_finite_size_instance_chi}, for (a) $p=1$ and (b) $p=13$. System sizes $N=4,8,\ldots,28$ are averaged over $10^3$--$10^4$ random problem instances.}
    \label{fig:finite_size_chi_distribution}
\end{figure}

The finite-$N$ simulations also allow us to test whether the local overlap response is self-averaging across disorder instances. For a fixed instance $J$, the instance-resolved susceptibility of two independent physical replicas, with overlap $q_{AB}$ defined in Eq.~\eqref{eq:sm_two_replica_overlap}, is
\begin{equation}
    \chi_{J,p,N}=N\sum_{\boldsymbol z^A,\boldsymbol z^B}P_{J,p}\bigl(\boldsymbol z^A\bigr)P_{J,p}\bigl(\boldsymbol z^B\bigr)q_{AB}^2.
\label{eq:sm_finite_size_instance_chi}
\end{equation}
The corresponding disorder-averaged quantity is $\chi_{p,N}=\mathbb E_J\bigl[\chi_{J,p,N}\bigr]$. To separate self-averaging from the growth of the mean response, Fig.~\ref{fig:finite_size_chi_distribution} shows the distribution of the normalized variable $\chi_{J,p,N}/\mathbb E_J[\chi_{J,p,N}]$.

At $p=1$, the normalized distributions concentrate rapidly near $\chi_{J,p,N}/\mathbb E_J[\chi_{J,p,N}]=1$ with increasing $N$, consistent with the susceptibility becoming self-averaging over the accessible sizes. At $p=13$, the distributions remain much broader and strongly asymmetric over the same range of $N$, with substantial weight both below and above the disorder average. The overlap response therefore remains strongly sample-dependent and has not yet become self-averaging over the accessible sizes, consistent with the finite-size crossover discussed in the main text.

\subsection{Finite-size overlap distributions}
\label{subsec:sm_finite_size_overlap_distributions}

\begin{figure}[!t]
    \centering
    \includegraphics[width=0.85\columnwidth]{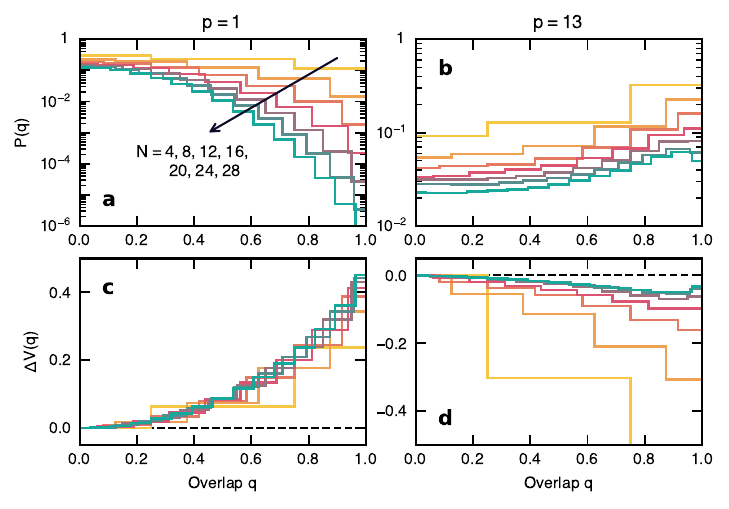}
    \caption{Disorder-averaged positive-overlap distribution $P_{p,N}^{(2)}(q)$ defined in Eq.~\eqref{eq:sm_finite_size_disorder_averaged_histogram} for (a) $p=1$ and (b) $p=13$, and the corresponding finite-size costs $\Delta V_{p,N}(q)$ defined in Eq.~\eqref{eq:sm_finite_size_positive_overlap_cost} for (c) $p=1$ and (d) $p=13$. System sizes $N=4,8,\ldots,28$ are averaged over $10^3$--$10^4$ random problem instances.}
    \label{fig:finite_size_distribution}
\end{figure}

The finite-$N$ QAOA simulations give access to the instance-resolved overlap distribution $P^{(2)}_{J,p,N}(q)$ defined in Eq.~\eqref{eq:sm_overlap_distribution}. By spin-flip symmetry this distribution is even, so we display its restriction to $q\geq0$. Its disorder average is
\begin{equation}
    P_{p,N}^{(2)}(q)=\mathbb E_J\Bigl[P_{J,p,N}^{(2)}(q)\Bigr].
    \label{eq:sm_finite_size_disorder_averaged_histogram}
\end{equation}
For the even system sizes considered here, we define the quenched finite-size histogram cost relative to the zero-overlap sector by
\begin{equation}
    \Delta V_{p,N}(q)=-\frac{1}{N}\mathbb E_J\left[\ln\left(\frac{P_{J,p,N}^{(2)}(q)}{P_{J,p,N}^{(2)}(0)}\right)\right],\qquad q\geq0.
    \label{eq:sm_finite_size_positive_overlap_cost}
\end{equation}
Positive values of $\Delta V_{p,N}(q)$ mean that overlap $q$ is suppressed relative to zero overlap in a typical instance, whereas negative values mean that overlap $q$ has larger weight than zero overlap on the finite-size exponential scale. This finite-$N$ quantity is a direct diagnostic of the exact finite-size overlap distribution. In the thermodynamic limit, if the overlap histogram is controlled by a single large-deviation branch, $\Delta V_{p,N}(q)$ is naturally compared with the overlap-potential difference $V_p(q)-V_p(0)$.

Figure~\ref{fig:finite_size_distribution} compares these quantities at $p=1$ and $p=13$. At $p=1$, the positive-overlap distribution decays rapidly with increasing $q$, and $\Delta V_{p,N}(q)$ is positive and increasing. Thus positive overlap is suppressed relative to zero overlap in typical finite-size instances, consistent with the weak local response derived analytically at depth one. At $p=13$, the histogram is qualitatively different: substantial probability remains at large positive overlaps over the accessible system sizes. The same effect is visible in $\Delta V_{p,N}(q)$, which becomes negative over a broad range of $q$, showing that nonzero positive overlap can be favored relative to zero overlap in the exact finite-$N$ distribution. These data directly illustrate the finite-size crossover discussed in the main text. In the thermodynamic analysis, taking $N\to\infty$ at fixed $p$ gives a thermodynamic potential minimized at zero overlap, whose stiffness decreases with increasing depth. At finite $N$, increasing $p$ can move the system across the corresponding crossover within the accessible size range. For the sizes shown here, $p=1$ remains on the zero-overlap side of the crossover, whereas $p=13$ carries substantial, strongly sample-dependent weight at $O(1)$ overlap. Broad, sample-dependent overlap structures are likewise characteristic of finite-size SK overlap distributions in the spin-glass regime~\cite{AspelmeierBilloireMarinariMoore2008}. Here, however, this high-overlap weight is a finite-size manifestation of the thermodynamic softening.

\subsection{Finite-size scaling at the tilted critical points}
\label{subsec:sm_finite_size_tilted_critical_scaling}

\begin{figure}[!t]
    \centering
    \includegraphics[width=1\textwidth]{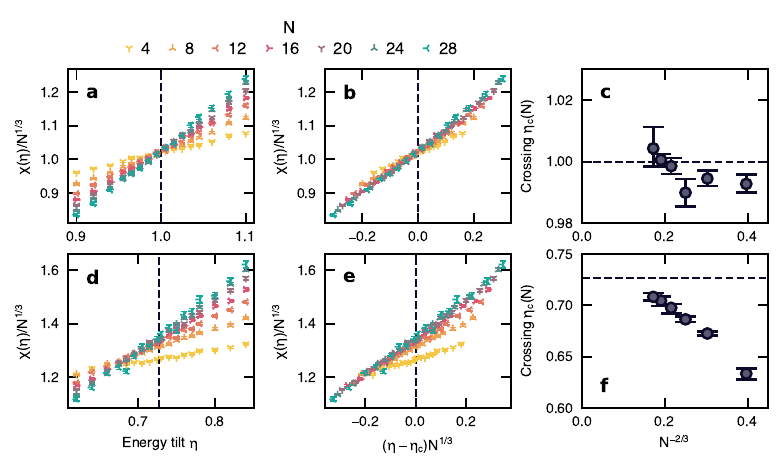}
    \caption{Finite-size scaling of the energy-tilted overlap susceptibility at the $p=0$ Gibbs transition (top row) and the $p=1$ QAOA transition (bottom row). (a, d) Rescaled susceptibility $\chi_{p,\eta,N}/N^{1/3}$ versus the energy tilt $\eta$; vertical dashed lines mark the thermodynamic critical values $\eta_c(0)=1$ and $\eta_c(1)$ from Eqs.~\eqref{eq:sm_p0_tilted_susceptibility} and~\eqref{eq:sm_p1_tilted_eta_c}. (b, e) The same data versus the critical scaling variable $[\eta-\eta_c(p)]N^{1/3}$. (c, f) Size-doubling crossing estimates $\eta_c^{(N,2N)}(p)$ versus $N^{-2/3}$; horizontal dashed lines mark the thermodynamic critical values. Data are averaged over $10^3$--$10^4$ random problem instances. Susceptibility error bars denote standard errors over disorder, while crossing uncertainties are obtained by bootstrap resampling of the local crossing fits.}
    \label{fig:finite_size_tilted_scaling}
\end{figure}

At $p=0$, the energy-tilted distribution is exactly the equilibrium SK Gibbs distribution, and $\eta_c(0)=1$ is the known zero-field spin-glass transition~\cite{SherringtonKirkpatrick1975,Parisi1979,BinderYoung1986,MezardParisiVirasoro1987,CastellaniCavagna2005}. Its finite-size critical behavior provides a direct benchmark for the depth-one QAOA transition. In Sec.~\ref{sec:sm_p1_tilted_parisi_transition}, we found in the thermodynamic limit that the tilted $p=1$ distribution undergoes a continuous Parisi transition at $\eta_c(1)$, with the same critical exponents as the Gibbs SK transition but renormalized nonuniversal coefficients; see Sec.~\ref{subsec:sm_p1_tilted_parisi_sk_comparison}. Here we test this correspondence independently using exact finite-$N$ simulations of the tilted QAOA distribution. Replacing $P_{J,p}$ by $P_{J,p}^{(\eta)}$ in Eq.~\eqref{eq:sm_overlap_distribution} defines $P_{J,p,N}^{(2,\eta)}(q)$, from which we evaluate
\begin{equation}
    \chi_{p,\eta,N}=N\mathbb E_J\sum_q q^2P_{J,p,N}^{(2,\eta)}(q).
    \label{eq:sm_finite_size_tilted_chi}
\end{equation}
At the transition, the mean-field SK critical behavior gives the finite-size scaling form~\cite{BinderYoung1986,AspelmeierKatzgraberLarsonMooreWittmannYeo2016}
\begin{equation}
    \frac{\chi_{p,\eta,N}}{N^{1/3}}=\mathfrak C_p\Bigl(\bigl[\eta-\eta_c(p)\bigr]N^{1/3}\Bigr)+\cdots,
    \label{eq:sm_finite_size_tilted_susceptibility_scaling}
\end{equation}
with $\eta_c(0)=1$ from Eq.~\eqref{eq:sm_p0_tilted_susceptibility} and $\eta_c(1)$ from Eq.~\eqref{eq:sm_p1_tilted_eta_c}, and where $\mathfrak C_p$ denotes the corresponding scaling function.

Figures~\ref{fig:finite_size_tilted_scaling}a and~\ref{fig:finite_size_tilted_scaling}d show the scaled susceptibility around the $p=0$ and $p=1$ transitions. When plotted against $[\eta-\eta_c(p)]N^{1/3}$ in Figs.~\ref{fig:finite_size_tilted_scaling}b and~\ref{fig:finite_size_tilted_scaling}e, the larger sizes approach a common curve corresponding to the scaling function $\mathfrak C_p$ in both cases. The observed $\chi_{p,\eta_c,N}\sim N^{1/3}$ growth and $N^{-1/3}$ critical window are therefore compatible with the thermodynamic susceptibility exponent $\gamma_{\rm crit}=1$ and the cubic SK critical scaling derived in Sec.~\ref{subsec:sm_p1_tilted_parisi_sk_comparison}. Finite-size corrections are visibly stronger at $p=1$ than for the Gibbs benchmark, consistent with the depth-dependent crossover discussed in Sec.~\ref{subsec:sm_finite_size_collapse_diagnostic}, where fixed-$p$ thermodynamic behavior is reached only once $N$ is sufficiently large compared with the corresponding overlap-response scale.

These corrections also shift the crossings of the scaled susceptibility away from the thermodynamic critical point. Since the leading critical theory gives $q=O(N^{-1/3})$, quartic and regular contributions generate $O(N^{-1/3})$ corrections to $\chi_{p,\eta,N}/N^{1/3}$. We therefore write
\begin{equation}
    \frac{\chi_{p,\eta,N}}{N^{1/3}}=\mathfrak C_p(x)+N^{-1/3}\mathfrak G_p(x)+\cdots,\qquad x=[\eta-\eta_c(p)]N^{1/3}.
    \label{eq:sm_finite_size_tilted_corrections}
\end{equation}
where $\mathfrak G_p$ denotes the leading correction-to-scaling function, whose form and amplitude may depend on $p$. Expanding the crossing condition between sizes $N$ and $2N$ around $x=0$ then gives
\begin{equation}
    \eta_c^{(N,2N)}(p)-\eta_c(p)=O\left(N^{-2/3}\right).
    \label{eq:sm_finite_size_tilted_crossing_shift}
\end{equation}
Figures~\ref{fig:finite_size_tilted_scaling}c and~\ref{fig:finite_size_tilted_scaling}f show the corresponding size-doubling crossings, obtained from local fits of the difference between the two scaled susceptibility curves, with statistical uncertainties estimated from disorder fluctuations. The Gibbs crossings remain close to $\eta_c(0)=1$. At $p=1$, the finite-size drift is substantially larger, but the crossing sequence moves toward the independently determined thermodynamic value $\eta_c(1)$ from below as $N$ increases. The accessible sizes are not sufficient to determine the correction exponent independently. Since the depth-dependent finite-size crossover becomes more pronounced with $p$, extending this critical analysis to larger depths would require system sizes beyond those accessible here.

\subsection{Critical overlap distributions under energy tilting}
\label{subsec:sm_finite_size_tilted_overlap_distributions}

\begin{figure*}[!t]
    \centering
    \includegraphics[width=0.85\textwidth]{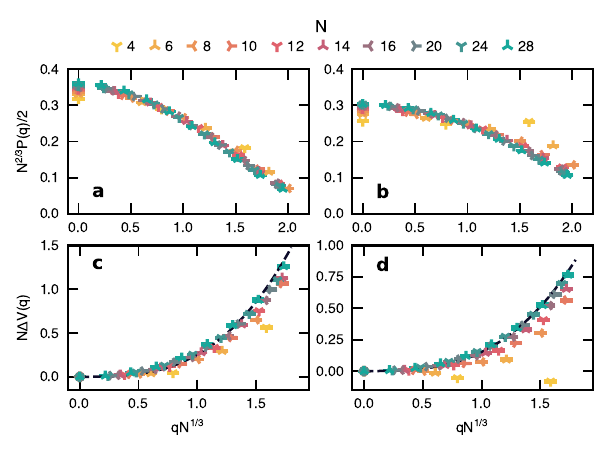}
    \caption{Critical finite-size overlap distributions at the $p=0$ Gibbs transition (left column) and the $p=1$ QAOA transition (right column), evaluated at $\eta_c(0)=1$ and $\eta_c(1)$, respectively. (a,b) Rescaled disorder-averaged positive-overlap distribution $N^{2/3}P_{p,N}^{(2,\eta_c)}(q)/2$ versus $qN^{1/3}$. (c,d) Corresponding quenched finite-size cost $N\Delta V_{p,\eta_c,N}(q)$ versus $qN^{1/3}$. Dashed curves show fits to the predicted cubic form $\sim Nq^3$. Data are averaged over $10^3$--$10^4$ random problem instances, with error bars denoting standard errors over disorder.}
    \label{fig:finite_size_critical_overlap}
\end{figure*}

Beyond the susceptibility, the overlap distribution provides a direct test of the critical overlap scale. We define the disorder-averaged tilted distribution
$P_{p,N}^{(2,\eta)}(q)=\mathbb E_J[P_{J,p,N}^{(2,\eta)}(q)]$ and the corresponding quenched cost $\Delta V_{p,\eta,N}(q)$ by the tilted analogue of Eq.~\eqref{eq:sm_finite_size_positive_overlap_cost}. At $\eta=\eta_c(p)$, the cubic critical theory gives $q=O(N^{-1/3})$, the finite-size scaling of the order parameter associated with the thermodynamic exponent $\beta_{\rm crit}=1$ derived in Sec.~\ref{subsec:sm_p1_tilted_parisi_sk_comparison}. It follows that
\begin{equation}
    \frac{N^{2/3}}{2}P_{p,N}^{(2,\eta_c)}(q)=\mathfrak P_p\left(qN^{1/3}\right)+\cdots,\qquad N\Delta V_{p,\eta_c,N}(q)=\mathfrak V_p\left(qN^{1/3}\right)+\cdots,
    \label{eq:sm_finite_size_critical_overlap_scaling}
\end{equation}
where $\mathfrak P_p$ and $\mathfrak V_p$ denote the corresponding finite-size scaling functions. The factor $1/2$ converts the probability mass on the finite-$N$ overlap grid, with spacing $2/N$, into the corresponding scaled density. The critical Landau theory further predicts $\mathfrak V_p(x)\propto |x|^3$ near the origin.

Figures~\ref{fig:finite_size_critical_overlap}a and~\ref{fig:finite_size_critical_overlap}b show that the larger sizes approach a common scaled distribution at each critical point, supporting the predicted $q\sim N^{-1/3}$ scaling and hence the order-parameter critical behavior. The quenched costs in Figs.~\ref{fig:finite_size_critical_overlap}c and~\ref{fig:finite_size_critical_overlap}d likewise approach a common curve and are consistent with the predicted cubic growth. As for the susceptibility scaling, finite-size corrections are more pronounced at $p=1$. Together, the $N^{-1/3}$ overlap scale and cubic critical cost provide independent finite-size signatures of the shared critical behavior established in Sec.~\ref{sec:sm_p1_tilted_parisi_transition}.

\section{Tabulated QAOA angles and thermodynamic benchmarks}
\label{sec:sm_qaoa_angles_benchmarks}

Table~\ref{tab:sm_qaoa_angles_benchmarks} lists the depth-dependent QAOA angles used in the numerical evaluations of the main text, together with the corresponding thermodynamic energy density $E_p$ and overlap susceptibility $\chi_p$, defined in Eqs.~\eqref{eq:sm_energy_density} and~\eqref{eq:sm_chi_correlation_power_identity}, respectively. The angles are taken from Ref.~\cite{BoulebnaneKhanLiuLarsonHermanShaydulinPistoia2026}. They are reported in the normalized SK convention of Eq.~\eqref{eq:sm_sk_cost}. The use of depth-dependent, instance-independent angles is consistent with concentration results for fixed QAOA parameters on typical random instances~\cite{BrandaoBroughtonFarhiGutmannNeven2018}.

\begin{table*}[!ht]
    \centering
    \scriptsize
    \setlength{\tabcolsep}{3pt}
    \renewcommand{\arraystretch}{1.15}
    \begin{tabular}{c|p{0.35\textwidth}|p{0.35\textwidth}|c|c}
    \hline\hline
    $p$ & $\boldsymbol{\gamma}$ & $\boldsymbol{\beta}$ & $\chi_p$ & $E_p$\\
    \hline
    1  & $1/2$
       & $\pi/8$
       & 1.40171 & -0.30327 \\
    \hline
    2  & 0.38174, 0.66550
       & 0.49597, 0.26904
       & 1.96222 & -0.40755 \\
    \hline
    3  & 0.32987, 0.56894, 0.64085
       & 0.54996, 0.36753, 0.21088
       & 2.65050 & -0.47262 \\
    \hline
    4  & 0.29502, 0.51444, 0.55852, 0.64285
       & 0.57093, 0.41752, 0.30269, 0.17287
       & 3.42573 & -0.51568 \\
    \hline
    5  & 0.27049, 0.48034, 0.50739, 0.56464, 0.63964
       & 0.58991, 0.44921, 0.35594, 0.26428, 0.14856
       & 4.33860 & -0.54765 \\
    \hline
    6  & 0.25281, 0.45310, 0.47495, 0.51461, 0.56500, 0.63920
       & 0.60043, 0.46704, 0.38806, 0.31759, 0.23247, 0.12906
       & 5.34353 & -0.57214 \\
    \hline
    7  & 0.23818, 0.43238, 0.45126, 0.48266, 0.51482, 0.56904, 0.63989
       & 0.60811, 0.48061, 0.40869, 0.35331, 0.28575, 0.20800, 0.11441
       & 6.46003 & -0.59148 \\
    \hline
    8  & 0.22678, 0.41618, 0.43325, 0.46083, 0.48177, 0.51793, 0.57171, 0.63931
       & 0.61510, 0.49056, 0.42438, 0.37801, 0.32237, 0.26056, 0.18842, 0.10300
       & 7.69484 & -0.60727 \\
    \hline
    9  & 0.21699, 0.40219, 0.41882, 0.44415, 0.46036, 0.48504, 0.52295, 0.57791, 0.64379
       & 0.61960, 0.49760, 0.43585, 0.39690, 0.34879, 0.29831, 0.23984, 0.17209, 0.09336
       & 9.03072 & -0.62034 \\
    \hline
    10 & 0.20869, 0.39048, 0.40662, 0.43023, 0.44281, 0.46086, 0.48779, 0.52782, 0.58103, 0.64116
       & 0.62447, 0.50359, 0.44301, 0.40897, 0.36716, 0.32465, 0.27591, 0.22064, 0.15719, 0.08484
       & 10.42714 & -0.63136 \\
    \hline
    11 & 0.20234, 0.38034, 0.39623, 0.41987, 0.42970, 0.44405,\newline
         0.46251, 0.49059, 0.53143, 0.58339, 0.64005
       & 0.62672, 0.50681, 0.44970, 0.41952, 0.38220, 0.34533,\newline
         0.30348, 0.25668, 0.20487, 0.14518, 0.07848
       & 11.96146 & -0.64080 \\
    \hline
    12 & 0.19494, 0.37164, 0.39034, 0.41170, 0.41929, 0.42998,\newline
         0.44551, 0.46764, 0.49562, 0.53532, 0.58631, 0.64225
       & 0.63071, 0.51432, 0.45849, 0.42939, 0.39457, 0.36286,\newline
         0.32806, 0.28639, 0.24087, 0.19082, 0.13489, 0.07260
       & 13.64469 & -0.64893 \\
    \hline
    13 & 0.19061, 0.36331, 0.38036, 0.40346, 0.41110, 0.42092,\newline
         0.43045, 0.44595, 0.46840, 0.49970, 0.54126, 0.59024, 0.64293
       & 0.63171, 0.51340, 0.45981, 0.43491, 0.40369, 0.37478,\newline
         0.34136, 0.30643, 0.26795, 0.22546, 0.17845, 0.12575, 0.06750
       & 15.27232 & -0.65606 \\
    \hline\hline
    \end{tabular}
    \caption{Optimized QAOA angles and thermodynamic benchmarks used in the main text. The angles $(\boldsymbol\gamma,\boldsymbol\beta)$ are taken from Ref.~\cite{BoulebnaneKhanLiuLarsonHermanShaydulinPistoia2026}. The table also reports the thermodynamic QAOA energy density $E_p$ and the overlap susceptibility $\chi_p$.}
    \label{tab:sm_qaoa_angles_benchmarks}
\end{table*}

\section{Numerical implementation}
\label{sec:sm_numerical_implementation}

The numerical results reported here combine two complementary calculations. The thermodynamic data are obtained from the finite-history saddle reduction developed above, whereas the finite-size diagnostics are computed independently from exact QAOA state-vector simulations. We summarize the numerical procedures used for both.

\subsection{Finite-history sums and saddle iteration}
\label{subsec:sm_numerical_saddle_iteration}

A one-site QAOA history defined in Eq.~\eqref{eq:sm_history_vector} contains $2p+1$ Ising variables. The thermodynamic calculation therefore replaces the original $N$-spin problem by an explicit sum over a finite history space whose size grows exponentially with $p$. The collective overlap matrices themselves contain only $O(p^2)$ independent entries, so the practical depth limit is set by the history sums. For the one-sample problem there are $2^{2p+1}$ possible histories when the terminal spin is retained, while the general two-sample trace in Eq.~\eqref{eq:sm_two_replica_local_trace} involves pairs of such histories and is correspondingly more expensive. This exponential dependence on depth is the same finite-depth bottleneck encountered in earlier thermodynamic SK-QAOA calculations~\cite{FarhiGoldstoneGutmannZhou2022,BassoFarhiMarwahaVillalongaZhou2022,BassoGamarnikMeiZhou2022,BoulebnaneKhanLiuLarsonHermanShaydulinPistoia2026}.

We note that the exponential history sum is not the only route to thermodynamic calculations at high depth. A recent spin--boson mapping reformulates the one-sample SK-QAOA iteration as the dynamics of a spin coupled to $p$ bosonic modes, which is amenable to matrix-product-state simulation and has enabled energy calculations to substantially larger depths~\cite{BoulebnaneKhanLiuLarsonHermanShaydulinPistoia2026,BoulebnaneKhanSubrahmanyaHermanFarhiVillalongaShaydulin2026}. This connection suggests that the same representation could also provide a route to the overlap observables considered here, particularly the zero-source linear response. At finite source, however, two physical replicas must be coupled through their output spins, while the energy tilt introduces the nonunitary terminal-slice deformation of Sec.~\ref{subsec:sm_tilted_terminal_slice}. Auxiliary-replica stability and the Parisi continuation further require additional replicated degrees of freedom. Extending the spin--boson construction to these quantities would therefore require a doubled or replicated formulation together with control of the resulting bosonic and matrix-product-state truncations.

For fixed physical parameters, the saddle equations are solved as nonlinear fixed-point equations. Starting from an initial collective matrix, we evaluate the one-site trace, compute the overlaps generated by the corresponding normalized one-site measure, and use them to update Eq.~\eqref{eq:sm_two_replica_saddle_equations}. Successive iterates are linearly damped, and convergence is determined from the largest absolute difference between an overlap variable and its updated value. In the calculations reported here, the one-sample saddles are converged to residuals of order $10^{-13}$--$10^{-14}$ and the finite-source two-sample saddles to $10^{-10}$, with damping factors between $0.35$ and $0.5$ depending on the calculation.

The ordinary $\lambda=0$ solution provides the natural starting point. We first obtain the one-sample SK-QAOA saddle at the chosen depth and initialize the two-sample problem with the product saddle of Eq.~\eqref{eq:sm_zero_source_two_replica_saddle}. Finite overlap source is then reached through the continuation described in Sec.~\ref{subsec:sm_source_connected_continuation}: $\lambda$ is increased in small increments, with each converged saddle initializing the next source value. Continuation thus both facilitates convergence and selects the stationary branch continuously connected to the ordinary QAOA distribution at $\lambda=0$.

On a converged saddle, the quenched free-entropy density follows from Eq.~\eqref{eq:sm_phi_two_replica_branch}, the source-selected overlap from the terminal-spin expectation in Eq.~\eqref{eq:sm_terminal_overlap_from_saddle}, and the overlap potential from the parametric Legendre transform in Eq.~\eqref{eq:sm_parametric_potential}. The curve $V_p(q)$ is therefore obtained by repeating this calculation along the continued source branch.

The energy-tilted calculation follows the same construction. At fixed $\eta$, we first solve the one-sample saddle with the terminal-slice deformation of Sec.~\ref{subsec:sm_tilted_terminal_slice}. The corresponding product saddle in Eq.~\eqref{eq:sm_tilted_product_saddle} then provides the zero-source starting point for the two-sample problem. The selected energy and its large-deviation cost follow from Eqs.~\eqref{eq:sm_tilted_selected_energy} and~\eqref{eq:sm_tilted_energy_rate_parametric}, respectively. The tilt therefore changes the underlying one-sample saddle while leaving the numerical structure of the overlap calculation unchanged.

\subsection{Linear response and stability}
\label{subsec:sm_numerical_linear_response}

The local susceptibilities are obtained directly by linearizing the saddle equations at $\lambda=0$, rather than by fitting finite-source data. At zero source the two physical replicas factorize, and the linearized self-consistency equations give a finite linear system for the source-induced cross-sample overlaps. Solving this system yields $\partial_\lambda\mathsf Q_\star|_{\lambda=0}$ and hence the susceptibility appearing in Eq.~\eqref{eq:sm_q_local_expansion}. For the untilted problem, factorization at $\lambda=0$ allows the coefficients of this linear system to be constructed directly from one-sample history moments. This avoids evaluating the full finite-source two-sample trace and allows the zero-source susceptibility to be computed at larger depths than the complete overlap potential. Accordingly, the main-text data reach $p=13$ for $\chi_p$, compared with $p=8$ for the finite-source potential.

The tilted susceptibility is obtained in the same way around the tilted product saddle. We evaluate the Jacobian of the cross-sample fixed-point map and its explicit response to $\lambda$ numerically from the finite history sums, using a finite source increment of $10^{-6}$, and solve the resulting linear system directly. Its terminal--terminal component gives $\chi_{p,\eta}$ defined in Eq.~\eqref{eq:sm_tilted_overlap_response}. At $p=1$, this construction reduces to the finite response problem derived analytically in Eq.~\eqref{eq:sm_p1_tilted_chi_closed_form} and reproduces the corresponding closed-form result.

For $p\geq2$, the critical tilt entering the main-text lower-tail analysis is located by following the stable branch with increasing $\eta$ until the local stiffness $\chi_{p,\eta}^{-1}$ first changes sign. The critical tilt, selected energy, and one-sample free-entropy density are then obtained by linear interpolation across the bracketing interval, after which $\Delta E_c(p)$ and $I_c(p)$ follow from their definitions in Sec.~\ref{subsec:sm_tilted_susceptibility_interpretation}. The exact $p=0$ Gibbs result of Sec.~\ref{subsec:sm_p0_gibbs_benchmark} and the depth-one value in Eq.~\eqref{eq:sm_p1_tilted_eta_c} provide independent benchmarks for this procedure.

Auxiliary-replica stability is evaluated after the continued saddle has converged. The one-site correlation functions on that saddle are inserted into the linear map of Eq.~\eqref{eq:sm_replicon_linearized_map}, and the resulting finite complex operator is diagonalized directly to obtain the stability gap defined in Sec.~\ref{subsec:sm_replicon_stability_operator}. In the tilted problem, the terminal history coordinate is included among the active coordinates as described in Sec.~\ref{subsec:sm_replicon_results}. Stability is therefore a diagnostic of the already-selected thermodynamic branch rather than an additional condition entering the saddle solution.

\subsection{Branch checks and numerical benchmarks}
\label{subsec:sm_numerical_checks}

Several checks distinguish convergence to a numerical stationary point from the physical selection of a saddle branch. The zero-source normalization is built into the subtractions in Eqs.~\eqref{eq:sm_phi_two_replica_branch} and~\eqref{eq:sm_tilted_stationary_density}. Moreover, although the intermediate Keldysh weights and collective fields can be complex, the physical free-entropy densities, overlaps, energies, and susceptibilities are real on the continued saddle. We therefore monitor residual imaginary components together with the fixed-point residuals.

The analytically controlled cases provide end-to-end benchmarks: at $p=0$, the tilted calculation reproduces the Gibbs free entropy, energy, susceptibility, and stability boundary derived in Sec.~\ref{subsec:sm_p0_gibbs_benchmark}; at $p=1$, the numerical zero-source response reproduces the exact susceptibility in Eq.~\eqref{eq:sm_p1_chi_from_saddle}, while the tilted calculation recovers the depth-one energy, susceptibility, and critical tilt derived in Sec.~\ref{subsec:sm_depth_one_energy_tilted_response}. Together, these comparisons test both the nonlinear saddle solution and the subsequent linear-response and stability calculations.

A separate question is whether continuation can miss stationary branches disconnected from the $\lambda=0$ solution. The multistart analysis of Sec.~\ref{subsec:sm_constrained_overlap_branch_search} addresses this possibility by solving the same fixed-source saddle equations from independent real, $A\leftrightarrow B$-symmetric initial conditions rather than from the continued solution. It therefore provides a complementary test of branch selection without modifying the underlying stationary problem.

\subsection{Exact finite-size state-vector calculations}
\label{subsec:sm_numerical_finite_size}

The finite-size results of Sec.~\ref{sec:sm_finite_size} are obtained independently of the thermodynamic saddle calculation by exact simulation of the QAOA state in Eq.~\eqref{eq:sm_qaoa_state}. For each disorder realization, we draw the SK couplings from the distribution defined in Sec.~\ref{subsec:sm_sk}, construct the corresponding diagonal cost Hamiltonian, and apply the QAOA layers using the same depth-dependent angles as in the thermodynamic calculations. The exact global spin-flip symmetry discussed in Sec.~\ref{subsec:sm_quantum} allows the state evolution to be restricted to the even sector, with one representative for each pair $\{\boldsymbol z,-\boldsymbol z\}$. The simulated state space therefore has dimension $2^{N-1}$.

From the resulting measurement probabilities, we evaluate the full two-replica overlap distribution $P^{(2)}_{J,p,N}(q)$ defined in Eq.~\eqref{eq:sm_overlap_distribution}. Rather than summing explicitly over all pairs of output configurations, we use that the overlap is fixed by their Hamming distance. The corresponding pair probabilities form an autocorrelation of the QAOA measurement distribution, which is evaluated with a fast Walsh--Hadamard transform and then collected according to Hamming distance. Accounting for the two members of each global spin-flip pair restores the physical distribution on the overlap grid $q=-1,-1+2/N,\ldots,1$. This gives the complete overlap distribution exactly for each disorder realization within the state-vector calculation.

The finite-size susceptibility and the overlap-distribution diagnostics then follow directly from this common object. In particular, $\chi_{J,p,N}$ is obtained from the second moment in Eq.~\eqref{eq:sm_finite_size_instance_chi}, while the disorder-averaged distributions and quenched histogram costs are constructed as in Eqs.~\eqref{eq:sm_finite_size_disorder_averaged_histogram} and~\eqref{eq:sm_finite_size_positive_overlap_cost}. No sampling of measured bitstrings is required; the statistical uncertainties reported for these quantities arise solely from averaging over the finite ensemble of disorder realizations.

The energy-tilted finite-size calculations use the same state-vector data. For each instance, the exact QAOA probabilities are reweighted according to Eq.~\eqref{eq:sm_energy_tilted_distribution} and normalized before evaluating the overlap distribution. Thus changing $\eta$ does not require a new quantum-state evolution, but only changes the statistical weights assigned to the already computed output configurations. The resulting $P^{(2,\eta)}_{J,p,N}(q)$ gives the tilted susceptibility in Eq.~\eqref{eq:sm_finite_size_tilted_chi} and the critical overlap distributions analyzed in Secs.~\ref{subsec:sm_finite_size_tilted_critical_scaling} and~\ref{subsec:sm_finite_size_tilted_overlap_distributions}. At $p=0$, the same procedure reduces to the exact finite-size Gibbs distribution, providing the equilibrium benchmark used there.

\end{document}